\documentclass[aps,twocolumn,showpacs,preprintnumbers,amsmath,amssymb,nofootinbib,superscriptaddress,showkeys,floatfix,,superscriptaddress]{revtex4}

\usepackage{epsfig}
\usepackage{graphicx}
\usepackage{dcolumn}

\def\u{{\bf u}} \def\U{{\bf U}} \def\S{{\bf S}} \def\h{{\bf h}}
 \def\E{{\bf 1}}

\begin{document}

\title{Renormalization of NN Interaction with Chiral Two Pion Exchange
  Potential. Non-Central Phases. } \author{M. Pav\'on
  Valderrama}\email{mpavon@ugr.es} \affiliation{Departamento de
  F\'{\i}sica At\'omica, Molecular y Nuclear, Universidad de Granada,
  E-18071 Granada, Spain} \author{E. Ruiz
  Arriola}\email{earriola@ugr.es} \affiliation{Departamento de
  F\'{\i}sica At\'omica, Molecular y Nuclear, Universidad de Granada,
  E-18071 Granada, Spain}

\date{\today}

\begin{abstract} 
\rule{0ex}{3ex} We extend the renormalization of the NN interaction
with Chiral Two Pion Exchange Potential to the calculation of
non-central partial wave phase shifts with total angular momentum $j
\le 5 $. The short distance singularity structure of the potential as
well as the requirement of orthogonality conditions on the wave
functions determines exactly the number of undetermined parameters
after renormalization.
\end{abstract}


\maketitle



\section{Introduction} 

The original proposal by
Weinberg~\cite{Weinberg:1990rz,Weinberg:1991um} and carried out for
the first time by Ray, Ordo\~nez and van Kolck~\cite{Ordonez:1995rz},
of making model independent predictions for NN scattering using Chiral
Perturbation Theory (ChPT) has been followed by a wealth of works
~\cite{Rijken:1995pu,Kaiser:1997mw,Kaiser:1998wa,Epelbaum:1998ka,Epelbaum:1999dj,Rentmeester:1999vw,Friar:1999sj,Richardson:1999hj,Kaiser:1999ff,Kaiser:1999jg,Kaiser:2001at,Kaiser:2001pc,Kaiser:2001dm,Entem:2001cg, Entem:2002sf,Rentmeester:2003mf,Epelbaum:2003gr,Epelbaum:2003xx,Entem:2003cs, Higa:2003jk,Higa:2003sz, Higa:2004cr,Birse:2003nz,Entem:2003ft,Epelbaum:2004fk} 
(for a review see e.g. Ref.~\cite{Bedaque:2002mn}). The renormalized
potential as given in Refs.~\cite{Ordonez:1995rz,Kaiser:1997mw,
Rentmeester:1999vw} in
configuration space is expanded taking $m^2 / 16 \pi^2 f^2 $ and $ m/M
$ as small parameters ($m$ and $M$ are the pion and nucleon masses
respectively and $f$ is the pion weak decay constant), with $ m r $
fixed. In this counting for the potential and in a given partial wave
(coupled) channel with good total angular momentum the reduced
potential can schematically be written as
\begin{eqnarray}
U (r) &=& M m \Big\{ \frac{m^2}{f^2} W^{(0)} (mr ) + \frac{m^4}{f^4}
W^{(2)} (mr ) \nonumber \\ &+& \frac{m^4}{f^4} \frac{m}{M}
W^{(3)} (mr ) + \dots \Big\} \,, 
\label{eq:pot_chpt}
\end{eqnarray}
where $W^{(n)} $ are known dimensionless functions which are
everywhere finite except for the origin and depend on the axial
coupling constant. $W^{(3)} $ depends also on three additional low
energy constants $ \bar c_1 = c_1 M $, $ \bar c_3 = c_3 M $ and $ \bar
c_4 = c_4 M$ which have been determined from $\pi N$ scattering ChPT  
studies in a number of
works~\cite{Fettes:1998ud,Buettiker:1999ap,GomezNicola:2000wk,Nicola:2003zi}.

At the level of approximation of Eq.~(\ref{eq:pot_chpt}) these
potentials are local and energy independent and become singular at the
origin. Thus, non-perturbative renormalization methods must be applied
to give a precise meaning to the scattering amplitude~\cite{Case:1950}
(for a comprehensive review in the one channel case see e.g.
Ref.~\cite{Frank:1971} and Ref.~\cite{Beane:2000wh} for a modern
perspective). Several methods have been proposed to study the LO term
in Eq.~(\ref{eq:pot_chpt}) for
central~\cite{Frederico:1999ps,Beane:2001bc,PavonValderrama:2003np,PavonValderrama:2004nb,PavonValderrama:2005gu}
and non-central~\cite{Nogga:2005hy}
waves. Recently~\cite{PavonValderrama:2005gu,Valderrama:2005wv} we
have shown how a renormalization program can be carried out for the NN
interaction for the One Pion Exchange (OPE) and chiral Two Pion
Exchange (TPE) potentials in the central $^1S_0$ and $^3S_1-^3D_1$
waves and its implications for the deuteron and pion-deuteron
scattering~\cite{Valderrama:2006np}.  In the present work we extend
our analysis to all remaining partial waves with $j \le 5 $ both for
the OPE as well as for the chiral TPE potentials.  As we showed in
Refs.~\cite{PavonValderrama:2005gu,Valderrama:2005wv} the short
distance behaviour of the chiral NN potential,
Eq.~(\ref{eq:pot_chpt}), determines {\it exactly} how many
counterterms are needed in order to generate renormalized and finite,
i.e. cut-off independent, phase shifts. These counterterms can be
determined by fixing some low energy parameters while the cut-off is
removed. It has been {\it assumed } that dimensional power counting in
the counterterms can be made {\it independently} on the short distance
singularity of the potential. This yields conflicts between naive
dimensional power counting and renormalization which have been
reported recently even for low partial waves~\cite{Nogga:2005hy}. So,
one is led to an alternative: either one keeps the power counting and
a finite cut-off or one removes the cut-off at the expense of
modifying the power counting of the short distance interaction. The
finite cut-off route has been explored in great detail in the
past~\cite{Rentmeester:1999vw,Epelbaum:1999dj,Epelbaum:2003gr,Epelbaum:2003xx,Entem:2003ft,Rentmeester:2003mf}. In
this paper we explore further the possibility of taking the
alternative suggested by renormalization and the tight constraints
imposed by finiteness. The analysis becomes rather transparent in
coordinate space, where the counterterms can be mapped into boundary
conditions~\cite{PavonValderrama:2003np,PavonValderrama:2004nb,PavonValderrama:2004td}
at the origin. In practice renormalization may be carried out in
several ways. In coordinate space it seems natural to exploit the
locality of the long distance (renormalized) potentials and then to
renormalize the full scattering problem. In the present work we adhere
to this two-steps renormalization which has the additional advantage
of making possible to determine {\it a priori} and based on simple
analytical arguments the existence of the renormalized limit and how
many independent renormalization conditions (counterterms) are
compatible with this limit. In this regard, let us remind that the
main advantage of renormalization is that {\it identical} finite and
unique results should be obtained regardless of the method of
calculation (coordinate or momentum space) and regularization provided
the same input physical data are used to eliminate the divergencies.
In particular we also expect independence on the way how the limit is
taken.

The origin of the conflict can be traced back to the question whether
for a given energy independent local potential, such as
Eq.~(\ref{eq:pot_chpt}), one can assume any short distance physics
regardless on the form of the long range potential. Renormalization
group invariance, however, requires that any physical parameter sits
on a renormalization trajectory and the corresponding evolution on the
renormalization scale is dictated by the form of the long distance
potential at {\it all} distances. The precise trajectory is uniquely
fixed by a renormalization condition at very long distances. Thus, the
separation between the short and long distance contribution is not
only scale dependent but also potential
dependent~\cite{PavonValderrama:2003np,PavonValderrama:2004nb}.
Renormalization conditions are physical and do not exhibit this
dependence. Finiteness of the scattering amplitude and orthogonality
of scattering (and eventually bound state) wave functions impose very
tight constraints on the allowed number of counterterms and their
possible scale dependence~\cite{PavonValderrama:2005gu,
Valderrama:2005wv}. The discussion becomes rather straightforward in
coordinate space and in terms of boundary conditions for ordinary
differential equations. In addition, unlike momentum space treatments,
a very natural hierarchy of the renormalization problem takes place in
configuration space~~\cite{PavonValderrama:2005gu, Valderrama:2005wv}.
More specifically, orthogonality of different
energy solutions requires an energy independent boundary condition on
the wave function for the long distance local and energy independent
potentials as it is the case for Eq.~(\ref{eq:pot_chpt}) valid to
NNLO, so that in {\it all cases} the effective range, and higher order
threshold parameters cannot be taken as independent input
parameters~\footnote{Actually, the potential in
Eq.~(\ref{eq:pot_chpt}) contains distributional contributions, which
strictly speaking are zero for any {\it finite} distance. See the
discussion in our previous work~\cite{Valderrama:2005wv}}.

The results found in
Refs.~\cite{PavonValderrama:2005gu,Valderrama:2005wv} can be concisely
summarized as follows in the one channel case.  For a regular
potential, i.e., diverging less strongly than the inverse square
potential, $r^2 |U(r)| < \infty $, one may {\it choose} between the
regular and irregular solution. In the first case the scattering
length is predicted while in the second case the scattering length
becomes an input of the calculation. Singular potentials at the
origin, i.e. fulfilling, $r^2 |U(r)| \to \infty $, do not allow this
choice.  If the potential is repulsive, the scattering length depends
on the potential while for an attractive potential the scattering
length must be chosen as an independent parameter. In the coupled
channel situation one must look at the strongest singularity of the
potential eigenvalues at the origin, and apply the single channel
results.

In our formulation of the NN renormalization problem threshold
parameters play an essential role. Unfortunately, scattering threshold
parameters for higher partial waves other than the S-waves have never
been considered in the context of chiral
potentials~\cite{Rentmeester:1999vw,Epelbaum:1999dj,Epelbaum:2003gr,Epelbaum:2003xx,Entem:2003ft,Rentmeester:2003mf}. 
Instead, some calculations adjust their counterterms to fit the phase
shifts in the region above threshold to the Nijmegen
database~\cite{Stoks:1993tb,Stoks:1994wp}. In a recent work we have
filled the gap by carrying out a complete determination of these
threshold parameters for the Reid93 and NijmII
potentials~\cite{PavonValderrama:2004se}. On the light of this new
information it is quite possible that the good fits in the
intermediate energy region imply a somewhat less accurate description
in the threshold region. This issue will become relevant in the
description of some partial waves.

The paper is organized as follows. In Sect.~\ref{sec:form} we review
the formalism for coupled channel scattering in the presence of
singular potentials at the origin. For completeness we list the
potentials in Appendix~\ref{sec:potentials}. Based on the short
distance behaviour of those potentials (see Appendix~\ref{sec:short})
and the requirement of orthogonality we determine the number of
independent parameters for any partial wave with $j \le 5$. In
Sect.~\ref{sec:phases} we present our results for the phase shifts.
Specifically, we make a thorough analysis of cut-off dependence in all
partial waves both for the OPE as well as for the chiral TPE
potential.  We also discuss the perturbative nature of peripheral
waves within the present non-perturbative approach. Finally, in
Sect.~\ref{sec:concl} we present our conclusions. 

\section{Formalism}
\label{sec:form} 

We solve the coupled channel Schr\"odinger equation for the relative
motion which in compact notation reads, 
\begin{eqnarray}
-\u '' (r) + \left[ \U (r) + \frac{{\bf l}^2}{r^2} \right] \u (r) =
 k^2 \u (r) \, , 
\label{eq:sch_cp} 
\end{eqnarray} 
where $\U (r)= 2 \mu_{np} {\bf V}(r)$ is the coupled channel matrix
reduced potential with $\mu_{np}=M_p M_n /(M_p+M_n)$ the reduced
proton-neutron mass which for $j> 0$  can be written as,
\begin{eqnarray}
\U^{0j} (r) &=& U_{jj}^{0j} \, , \nonumber \\ \\   
\U^{1j} (r) &=& \begin{pmatrix}  U_{j-1,j-1}^{1j} (r) & 0 &
U_{j-1,j+1}^{1j} (r) \\ 0 & U_{jj}^{1j} (r) & 0 \\ U_{j-1,j+1}^{1j}
(r) & 0 & U_{j+1,j+1}^{1j} (r)   \end{pmatrix} \, . \nonumber  
\end{eqnarray} 
In Eq.~(\ref{eq:sch_cp}) $ {\bf l}^2 = {\rm diag} ( l_1 (l_1+1),
\dots, l_N (l_N +1) )$ is the orbital angular momentum, $\u(r)$ is the reduced
matrix wave function, $k$ the C.M. momentum and $j$ the total angular momentum.
In our case $N=1$ for the spin singlet channel with $l=j$ and $N=3$ 
for the spin triplet channel with $l_1=j-1$, $l_2=j$ and $l_3=j+1$. 
The potentials used in this paper were obtained 
in Refs.~\cite{Ordonez:1995rz,Kaiser:1997mw,Rentmeester:1999vw} in coordinate 
space and are listed in Appendix~\ref{sec:potentials} for completeness. 

\subsection{Long distance behaviour} 

At long distances, we assume the usual asymptotic normalization
condition
\begin{eqnarray}
\u (r)  \to \hat \h^{(-)} (r) - \hat \h^{(+)} (r) \S \, ,
\label{eq:asym}
\end{eqnarray} 
with $\S$ the  coupled channel unitary S-matrix. The
corresponding out-going and in-going free spherical waves are given by
\begin{eqnarray}
\hat \h^{(\pm)} (r) &=& {\rm diag} ( \hat h^\pm_{l_1} ( k r) , \dots ,
\hat h^\pm_{l_N} (k r) ) \, ,
\end{eqnarray} 
with $ \hat h^{\pm}_l ( x) $ the reduced Hankel functions of order
$l$, $ \hat h_l^{\pm} (x) = x H_{l+1/2}^{\pm} (x) $ ( $ \hat h_0^{\pm}
(x) = e^{ \pm i x}$ ), and satisfy the free Schr\"odinger's equation
for a free particle. 

For the
spin singlet state, $s=0$, one has $l=j$ and hence the state is
uncoupled
\begin{eqnarray}
S_{jj}^{0j} = e^{ 2 i \delta_{j}^{0j} } \, ,
\end{eqnarray}
whereas for the spin triplet state $s=1$, one has the uncoupled $ l=j$
state
\begin{eqnarray}
S_{jj}^{1j} &=& e^{  2 i \delta_{j}^{1j} } \, ,
\end{eqnarray}
and the two channel coupled $l,l'=j \pm 1$ states for which we use
Stapp-Ypsilantis-Metropolis (SYM or Nuclear bar)~\cite{stapp}
parameterization 
\begin{eqnarray}
S^{1j} &=& \left( \begin{array}{cc} S_{j-1 \,
j-1}^{1j} & S_{j-1 \, j+1}^{1j} \\ S_{j+1 \, j-1}^{1j} & S_{j+1 \,
j+1}^{1j}
\end{array} \right) \nonumber  \\ &=& \left( \begin{array}{cc} \cos{(2
\bar \epsilon_j)} e^{2 i \bar \delta^{1j}_{j-1}} & i \sin{(2 \bar
\epsilon_j)} e^{i (\bar \delta^{1j}_{j-1} +\bar \delta^{1j}_{j+1})} \\
i \sin{(2 \bar \epsilon_j)} e^{i (\bar \delta^{1j}_{j-1} + \bar
\delta^{1j}_{j+1})} & \cos{(2 \bar \epsilon_j)} e^{2 i \bar
\delta^{1j}_{j+1}} \end{array} \right) \nonumber
\end{eqnarray}
In the discussion of low energy properties we 
also use the Blatt-Biedenharn (BB or Eigen phase)
parameterization~\cite{Bl52} defined by
\begin{eqnarray}
S^{1j} &=& \begin{pmatrix} \cos \epsilon_j & -\sin \epsilon_j \\ \sin
\epsilon_j & \cos \epsilon_j  \end{pmatrix} \begin{pmatrix} e^{2 {\rm i}
\delta^{1j}_{j-1}} & 0 \\ 0 & e^{2 {\rm i} \delta_{j+1}^{1j}} \end{pmatrix}
\nonumber \\ &\times& \begin{pmatrix} \cos \epsilon_j & \sin
\epsilon_j \\ -\sin \epsilon_j & \cos \epsilon_j \end{pmatrix} 
\, . \label{eq:BB} 
\end{eqnarray} 
The relation between the BB and SYM phase shifts is  
\begin{eqnarray}
\bar \delta_{j+1}^{1j} + \bar \delta_{j-1}^{1j} &=& \delta_{j+1}^{1j}
+ \delta_{j-1}^{1j} \, , \\ \sin( \bar \delta_{j-1}^{1j} - \bar
\delta_{j+1}^{1j}) &=& \frac{\tan( 2\bar \epsilon_j
)}{\tan(2\epsilon_j )} \, .
\end{eqnarray} 
In the present paper zero energy scattering parameters play an
essential role since they are often used (see below) as input
parameters of the calculation of phase shifts. Due to unitarity of the
S-matrix in the low energy limit, $ k\to 0$, we have
\begin{eqnarray}
\left(\S - \E \right)_{l',l}=- 2 {\rm i} \alpha_{l', l}  k^{l'+l+1} +
\dots   \, ,  
\end{eqnarray} 
with $\alpha_{l' l} $ the (hermitian) scattering length
matrix~\footnote{For non S-wave scattering the dimension of
$\alpha_{l,l'}$ is ${\rm fm}^{l+l'+1}$ which is not a length. For
simplicity we will abuse language and call them scattering
lengths.}. The threshold behaviour acquires its simplest form in the
SYM representation,
\begin{eqnarray} 
\delta^{0j}_{j} &\to&  - \alpha^{0j}_{j} k^{2j+1} \, , \\ 
\delta^{1j}_{j} &\to&  - \alpha^{1j}_{j} k^{2j+1} \, , \\ 
\bar \delta^{1j}_{j-1} &\to&  - \bar \alpha^{1j}_{j-1} k^{2j-1} \, , \\ 
\bar \delta^{1j}_{j+1} &\to&  - \bar \alpha^{1j}_{j+1} k^{2j+3} \, , \\ 
\bar \epsilon_j &\to&  - \bar \alpha^{1j}_{j} k^{2j+1} \, . 
\label{eq:phase-thres}
\end{eqnarray} 
In the BB form one has similar behaviours for the $\delta$'s but
for $\epsilon_j$ which behaves as $k^{2j}$ instead of $k^{2j+1}$
\begin{eqnarray} 
\delta^{1j}_{j-1} &\to&  - \bar \alpha^{1j}_{j-1}\,k^{2j-1} \, , \\ 
\delta^{1j}_{j+1} &\to&  - 
(\bar \alpha^{1j}_{j+1} - 
\frac{{({\bar \alpha}^{1j}_{j})}^2}{\bar \alpha^{1j}_{j-1}})\,k^{2j+3} \, , \\ 
\epsilon_j &\to&  \frac{\bar \alpha^{1j}_{j}}{\bar \alpha^{1j}_{j-1}} \,
k^{2j} \, . 
\label{eq:phase-thres-BB}
\end{eqnarray} 

\subsection{Short distance behaviour} 

The form of the wave functions at the origin is uniquely determined by
the form of the potential at short distances (see
e.g.~\cite{Case:1950,Frank:1971} for the case of one channel and
\cite{PavonValderrama:2005gu, Valderrama:2005wv} for coupled
channels). For the chiral NN potential, Eq.~(\ref{eq:pot_chpt}), one
has
\begin{eqnarray}
\U_{\rm LO} (r) &\to& \frac{M {\bf C}_{3,LO}}{r^3} \, , \nonumber \\ \U_{\rm
NLO} (r) &\to& \frac{M {\bf C}_{5,NLO}}{r^5} \, ,\nonumber \\ \U_{\rm NNLO} (r)
&\to& \frac{M {\bf C}_{6,NNLO}}{r^6} \, , \nonumber \\
\label{eq:singLONLONNNLO} 
\end{eqnarray} 
where LO includes the first term in Eq.~(\ref{eq:pot_chpt}), NLO the
first two terms and so on. Note that higher order potentials
become increasingly singular at the origin. For a potential diverging at
the origin as an inverse power law
\begin{eqnarray}
\U (r) \to \frac{M {\bf C}_n}{r^n} \, , 
\label{eq:singular} 
\end{eqnarray} 
with ${\bf C}_n$ a matrix of generalized van der Waals coefficients 
and $n > 2$ an integer. One  diagonalizes the matrix ${\bf C}_n $ by 
a constant unitary transformation, ${\bf G}$, yielding
\begin{eqnarray}
M {\bf C}_n = {\bf G} \, {\rm diag} ( \pm R_1^{n-2}, \dots , \pm R_N^{n-2} ) \,
{\bf G}^{-1} \, , 
\end{eqnarray} 
with $R_i$ constants with length dimension. The plus sign corresponds
to the case with a positive eigenvalue (repulsive) and the minus
sign to the case of a negative eigenvalue (attractive). Then, at short
distances one has the solutions
\begin{eqnarray}
\u (r) \to {\bf G} \begin{pmatrix}  u_{1,\pm} (r) \cr \cdots \\
u_{N,\pm} (r)
\end{pmatrix} \, ,
\label{eq:eigen_wf}
\end{eqnarray} 
where for the attractive and repulsive cases one has 
\begin{eqnarray}
u_{i,-} (r) &\to & C_{i,-} \left(\frac{r}{R_i}\right)^{n/4} \sin\left[
\frac{2}{n-2} \left(\frac{R_i}{r}\right)^{\frac{n}2-1} + \varphi_i
\right] \, ,\nonumber \\ \label{eq:uA} \\ u_{i,+} (r) & \to & C_{i,+}
\left(\frac{r}{R_i}\right)^{n/4} \exp \left[- \frac{2}{n-2}
\left(\frac{R_i}{r}\right)^{\frac{n}2-1} \right] \, ,\label{eq:uR}
\label{eq:short_wf}
\end{eqnarray} 
respectively.  This behaviour of the wave functions near the origin is
valid regardless of the energy, provided the distances are small
enough
\footnote{In fact, the next correction to the near-the-origin wave
functions, which is energy dependent, is suppressed by a relative $ (k R)^2
(r/R)^{n/2+1}$ power with respect to the main term, so it is
negligible in the $r \to 0$ limit.}.  Here, $\varphi_i$ are arbitrary
short distance phases which in general depend on the energy. There are
as many short distance phases as short distance attractive
eigenpotentials. Orthogonality of the wave functions at the origin
yield the relation
\begin{eqnarray}
\sum_{i=1}^N \left[ {u_{k,i}}^* u_{p,i}'- {u_{k,i}'}^* 
u_{p,i} \right]\Big|_{r=0} = 
\sum_{i=1}^A \cos(\varphi_i (k) - \varphi_i(p) ) \, , \nonumber \\ 
\end{eqnarray} 
where $A \le N$ is the number of the short distance attractive
eigenpotentials.

\begin{table}
\caption{\label{tab:table1} Sets of chiral coefficients
considered in this work.}
\begin{ruledtabular}
\begin{tabular}{|c|c|c|c|c|}
\hline  Set & Source & $c_1 ({\rm GeV}^{-1}) $ & $c_3 ({\rm GeV}^{-1}) $ & $c_4
({\rm GeV}^{-1}) $ \\ \hline
Set I & $\pi N$~\cite{Buettiker:1999ap}  & -0.81  & -4.69    & 3.40   \\ 
Set II & $NN $~\cite{Rentmeester:1999vw} 
& -0.76  & -5.08   & 4.70  \\ 
Set III &  $NN $~\cite{Epelbaum:2003xx}  
& -0.81  & -3.40    & 3.40   \\ 
Set IV &  $NN$~\cite{Entem:2003ft} & -0.81  & -3.20   & 5.40  \\ 
\end{tabular}
\end{ruledtabular}
\end{table}

The simplest choice to fix relative phases for a positive energy
scattering state is to take the zero energy state $p=0$ as a reference
state, and the zero energy short distance phase. In the particular
case where only one eigenvalue is negative the short distance phase is
energy independent. This may happen both in the singlet as well as in
the triplet channels with $j=l$. The short distance phase is then
fixed by reproducing the scattering length in the singlet channel and
one of the three scattering lengths in the triplet channel. In the
case where one has two negative, i.e. attractive, eigenvalues (this
can only happen in triplet channels) there are two undetermined short
distance phases which can be fixed by using the corresponding three
scattering lengths.  The case of two positive, i.e. repulsive,
eigenvalues does not allow to fix any scattering length. The case with
two different signs for the eigenvalues fixes one scattering length
only. Note that in this construction and for two coupled channels
there is no intermediate situation where the solution is specified by
just two scattering lengths; one has either zero, one or three.

Although our arguments are entirely based on analytical calculations,
one should mention that our conclusions are in agreement with the
findings of Ref.~\cite{Nogga:2005hy} for the OPE case. There,
counterterms beyond the ones dictated by Weinberg's power counting are
included in the $^3P_0$, $^3P_2-{}^3F_2$ and $^3D_2$ waves to ensure
renormalizability on numerical grounds. As we will see below, our
renormalized phase shifts for special case OPE reproduce essentially
their results, although our TPE non-perturbatively renormalized
amplitudes go beyond these results.

Another issue is that of the establishment of a theoretically
compelling and mathematically consistent power counting which also
provides phenomenological success. This has been the goal of much the
EFT activity in recent years. Despite the fact that our OPE is
mathematically identical to the one in Ref.~\cite{Nogga:2005hy} where
a strong emphasis on power counting has been made, our motivation is
slightly different. Actually, these authors argue that a consistent
scheme for TPE might be achieved within a perturbative framework,
using the non-perturbative OPE distorted amplitudes as the leading
order approximation. This is theoretically appealing and the issue was
thoroughly discussed within the coordinate space approach in our
previous paper on the central waves~\cite{Valderrama:2005wv}. There,
it was pointed out that with enough counterterms such a program could
be pursued although orthogonality was violated and results did not
exhibit a clear improvement as compared to the fully iterated
potentials. The reason was the appearance of non-analytical
dependences on the would-be dimensional power counting parameter, a
situation that has not been foreseen in the standard EFT set up.  This
suggests that discussion on power counting and the systematics of EFT
is not yet over. Therefore, and as we did in our previous work, we
focus more on establishing long range model independent correlations, 
leaving the possible establishment of a satisfactory power counting
for future studies.

\subsection{Regularization methods} 

In principle, it is possible to implement the short distance behaviour
of the wave functions, Eq.~(\ref{eq:short_wf}), if one goes to
sufficiently small distances, or if the short distance behaviour of the
the wave function is improved~\cite{PavonValderrama:2005gu}.
Computationally, the
implementation of short distance regulators is mostly straightforward. 
The attractive or repulsive nature of the potentials
at short distances requires different choices of
regulators~\cite{PavonValderrama:2005gu,Valderrama:2005wv}. For a
one-channel repulsive singular potential we use the regulator
\begin{eqnarray}
\frac{u_k' (a)}{u_k(a)}= \frac{l+1}{a} \, ,
\end{eqnarray}  
This condition ensures orthogonality of wave functions with different energy.
For the attractive singular case, we integrate in from infinity at
zero energy down to a given boundary radius, $a$, impose orthogonality
at the boundary by matching logarithmic derivatives 
\begin{eqnarray}
\frac{u_k' (a)}{u_k(a)}= \frac{u_0' (a)}{u_0(a)} \, ,
\end{eqnarray}  
and then integrate out at finite energy. In the coupled channel case
we extend the method by applying the one channel regularization to the
short distance eigen functions, Eq.~(\ref{eq:eigen_wf}).

\subsection{Fixing of parameters and renormalization conditions} 

Fixing of the short distance phases requires some renormalization
condition. As we have said, an appealing choice is to impose this
condition at zero energy. The way to proceed in practice is quite
straightforward, although tedious given the large number (27) of
partial waves considered in this work. In the singlet channel case and
for an attractive short distance singularity, one starts at zero
energy and integrates in from large distances $\sim 15 {\rm fm}$ with
a given scattering length until a short boundary radius $\sim 0.1 {\rm
fm} $. At finite energy one integrates out matching the wave function
to the zero energy solution at the short distance boundary generating
a phase shift from a given prescribed scattering length. Of course, in
this method one has to check for cut-off independence (taking
$r=0.1-0.2 {\rm fm}$ proves enough).  For the coupled channel case one
proceeds along similar lines and the procedure has been described in
great detail in our previous
works~\cite{PavonValderrama:2005gu,Valderrama:2005wv} for the $j=1$
channel. The method relies heavily on the superposition principle of
boundary conditions and we use here the extension of that method to
higher partial waves. One of the advantages of our approach is that we
rarely have to make a fit to the data; any phase shift has {\it by
construction } the right threshold behaviour in the case where the
potential at short distances is attractive. For the repulsive
potential case the scattering length is predicted entirely from the
potential. In any case, discrepancies with the data can be attributed
to the potential.

\begin{table}
\caption{\label{tab:table2} The number of independent parameters for
different orders of approximation of the potential. The scattering
lengths are in ${\rm fm}^{l+l'+1}$ and are taken from NijmII and
Reid93 potentials~\cite{Stoks:1994wp} in
Ref.~\cite{PavonValderrama:2004se}. We use the (SYM-nuclear bar)
convention, Eq.~(\ref{eq:phase-thres}).}
\begin{ruledtabular}
\begin{tabular}{|c|c|c|c|c|}
\hline 
Wave  & $\alpha$ NijmII (Reid93)& LO & NLO & NNLO \\ 
\hline 
\hline $^1S_0 $ & -23.727(-23.735) & Input & Input & Input  \\
\hline $^3P_0 $ & -2.468(-2.469) &  Input & --- & Input  \\
\hline
\hline 
$^1P_1 $ & 2.797(2.736)  & --- & --- & --- \\
\hline $^3P_1 $ & 1.529(1.530) & --- & Input & Input  \\
\hline 
$^3S_1 $ & 5.418(5.422) & Input  &  ---  & Input \\
$^3D_1 $ & 6.505(6.453) &  --- &  --- &  Input  \\
$E_1 $ & 1.647(1.645)   & ---   & ---  &  Input   \\
\hline 
\hline 
$^1D_2 $ & -1.389(-1.377) & ---  & Input &  Input  \\
\hline $^3D_2 $ & -7.405(-7.411) & Input  &  Input & Input \\
\hline $^3P_2 $ & -0.2844(-0.2892) & Input  &  Input & ---  \\
$^3F_2 $ & -0.9763(-0.9698) & --- & --- & ---  \\
$E_2 $ & 1.609(1.600) & --- & ---  & --- \\
\hline
\hline 
$^1F_3 $ & 8.383(8.365) & ---  & --- & ---  \\
\hline $^3F_3 $ & 2.703(2.686) & --- &  Input & Input \\
\hline 
$^3D_3 $ & -0.1449(-0.1770) & Input  & --- & Input  \\
$^3G_3 $ & 4.880(4.874) & ---  & --- & Input   \\
$E_3 $ & -9.695(-9.683) & --- & --- & Input  \\
\hline 
\hline 
$^1G_4 $ & -3.229(-3.210) & ---  & Input  & Input \\
\hline $^3G_4 $ & -19.17(-19.14) & Input  & Input  & Input  \\
\hline 
$^3F_4 $ & -0.01045(-0.01053) & Input  & Input  & --- \\
$^3H_4 $ & -1.250(-1.240) & --- & --- & --- \\
$E_4 $ & 3.609(3.586) & --- & --- & ---  \\
\hline
\hline 
$^1H_5 $ & 28.61(28.57) & --- & --- & --- \\
\hline 
$^3H_5 $ & 6.128(6.082) & --- & Input  & Input  \\
\hline 
$^3G_5 $ & -0.0090(-0.010) & Input  & --- & Input    \\
$^3I_5 $ & 10.68(10.66) & --- &--- & Input   \\
$E_5 $ & -31.34(-31.29) & --- & --- & Input  
\end{tabular}
\end{ruledtabular}
\end{table}

Inspection of Table~\ref{tab:table2} illustrates the situation for the
LO, NLO, and NNLO approximations to the potential. We show the
scattering lengths in all partial waves as determined in our previous
work~\cite{PavonValderrama:2004se} together with the corresponding
eigenvalues for the leading short distance coefficients in the LO
(OPE), NLO and NNLO approximations to the potential. In the NNLO one
must also specify the values of the chiral constants $c_1$, $c_3$ and
$c_4$. We use for definiteness the values of Ref.~\cite{Entem:2003ft},
since as we saw in Ref.~\cite{Valderrama:2005wv} they provide a
reasonable description of deuteron properties.

\subsection{Details on the numerical procedure} 

The integration of the coupled differential equations requires some
care, particularly in the vicinity of the short distance
singularities. In the case of attractive singularities due to the
increasing oscillations the wave function has to be sampled with great
detail at a rate similar to the size of the oscillations. For the
repulsive case, one must stop at sufficiently large distances due to
the exponential suppression of the wave function. Another important
condition has to do with preservation of in and out reversibility of
the integration. This last requirement guarantees that for attractive
channels, where the scattering length is supplied as an input
parameter, the threshold behaviour of the phase shift is consistent
with that given scattering length.

Another problem one has to face for high partial waves is related to
the practical influence of the scattering length on the calculated
phase shifts. In principle, and for an attractive singular potential,
the scattering length needs to be specified. For the one channel case,
this is done by integrating in the zero energy large distance
solution, valid for $r \gg 2/m_\pi $
\begin{eqnarray}
u (r) \to r^{-l} - \frac{r^{l+1}}{\alpha_l} \, ,
\end{eqnarray} 
The long distance irregular solution dominates, unless $\alpha_l $ is
anomalously large , i.e. $\alpha_l (m_\pi/2)^{(2l+1)} \gg 1 $ , so that when
integrating in much of the regular solution will be lost and the
result will be rather insensitive to the value of $\alpha_l$ provided
it is of normal size. This fact becomes relevant in the numerical
calculations if the long distance cut-off is taken to be exceedingly
large. To avoid this situation we take typically $R_{\rm max} = 15
{\rm fm}$ for large $l$.

\begin{table*}
\caption{\label{tab:deut_prop} Deuteron properties for the OPE and TPE
potentials. OPE(0) refers to the deuteron computed by orthogonality to
the zero energy scattering states, fixing $\alpha_0$ to its
experimental value, while in OPE(B) the computation is made by fixing
$\gamma$ to its experimental value, constructing the corresponding
bound state. Similarly, TPE(0) refers to the deuteron computed by
orthogonality to zero energy states, fixing $\alpha_0$ to its
experimental value and $\alpha_{02}$ and $\alpha_2$ to their Nijmegen
II values, while in TPE(B) the computation is made by fixing $\gamma$,
$\eta$ and $\alpha_0$ to their experimental values.  The errors quoted
in OPE(0) corresponds to the uncertainty in the value of the
scattering length, while in OPE(B) the errors correspond to changing
the cutoff in the $0.1 - 0.2\,{\rm fm}$ range.  The errors quoted in
both TPE computations reflect the uncertainty in the non-potential
parameters $\gamma$, $\eta$ and $\alpha_0$ only.  We take set
IV~\cite{Entem:2003ft} for the LEC's in the TPE calculation.
Experimental values can be traced from Ref.~\cite{deSwart:1995ui}.}
\begin{ruledtabular}
\begin{tabular}{|c|c|c|c|c|c|c|c|c|c|}
\hline
Set & $\gamma\,\,({\rm fm}^{-1})$ & $\eta$ & $A_S\,\,({\rm fm}^{-1/2})$ &
$r_m\,\,({\rm fm})$ & $Q_d ({\rm fm}^2)$ & $P_D$ (\%) & 
$\alpha_0\,({\rm fm})$ & 
$\alpha_{02} ({\rm fm}^3)$ & $\alpha_2 ({\rm fm}^5)$ \\
\hline
OPE(0) & 0.2274(4) & 0.02564(4) & 0.8568(10) & 1.964(3) & 0.2796(3) 
& 7.208(12) & Input & 1.754(7) & 6.770(7) \\
\hline
OPE(B) & Input & 0.02633 & 0.8681(1) & 1.9351(5) & 0.2762(1) 
& 7.31(1) & 5.335(1) & 1.673(1) & 6.693(1) \\
\hline\hline
TPE(0) & 0.2322(3) & 0.02531(9) & 0.8891(4) & 1.968(3) & 0.2723(3) 
& 7.24(13) & Input & NijmII & NijmII \\
TPE(B) & Input & Input & 0.884(4) & 1.967(6) & 0.276(3) & 8(1) 
& Input & 1.67(4) & 6.6(4) \\
\hline\hline
NijmII & 0.231605 & 0.02521 & 0.8845 & 1.9675 & 0.2707 & 5.635
& 5.418 & 1.647 & 6.505 \\
Reid93 & 0.231605 & 0.02514 & 0.8845 & 1.9686 & 0.2703 & 5.699 
& 5.422 & 1.645 & 6.453 \\
\hline
Exp. 
& 0.231605 & 0.0256(4) & 0.8846(9) & 1.971(6) & 0.2859(3) 
& - & 5.419(7) & - & -\\
\hline
\end{tabular}
\end{ruledtabular}
\end{table*}

\begin{figure*}[tbc]
\begin{center}
\epsfig{figure=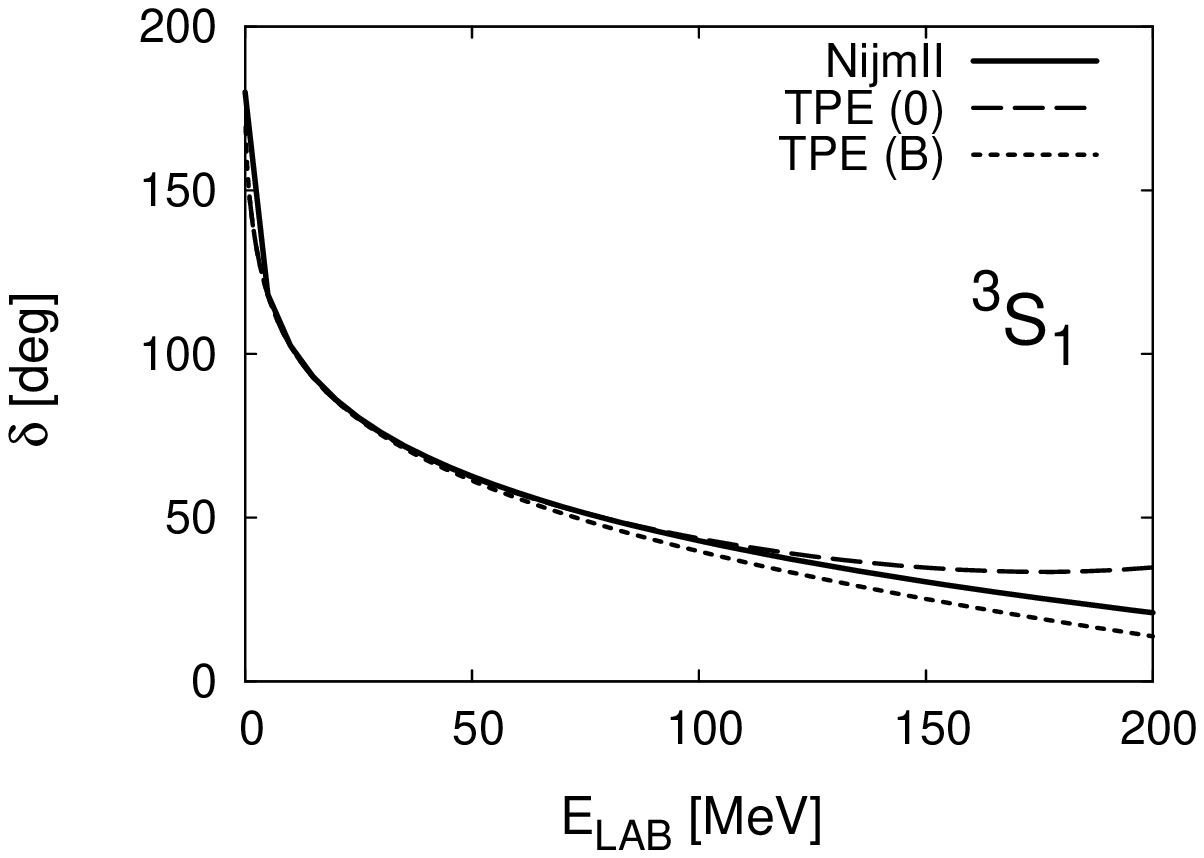,height=4cm,width=5cm}
\epsfig{figure=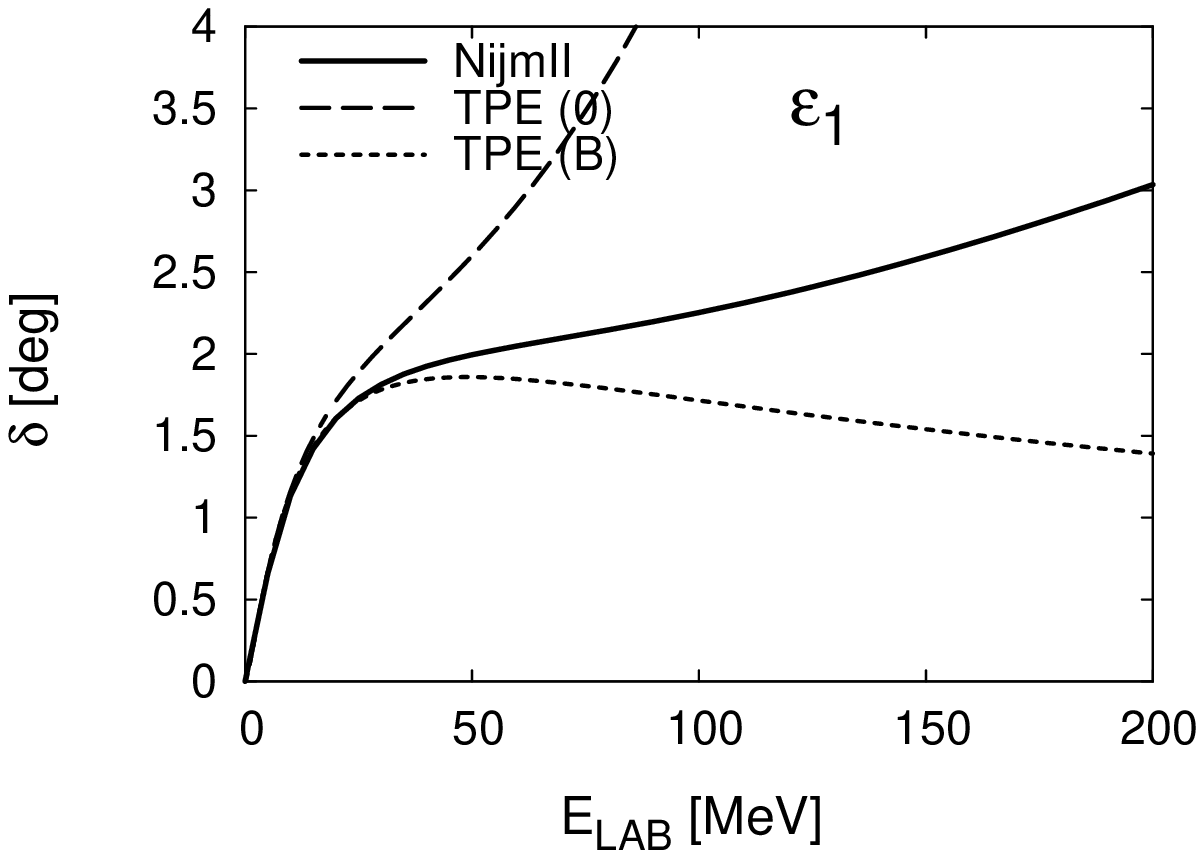,height=4cm,width=5cm}
\epsfig{figure=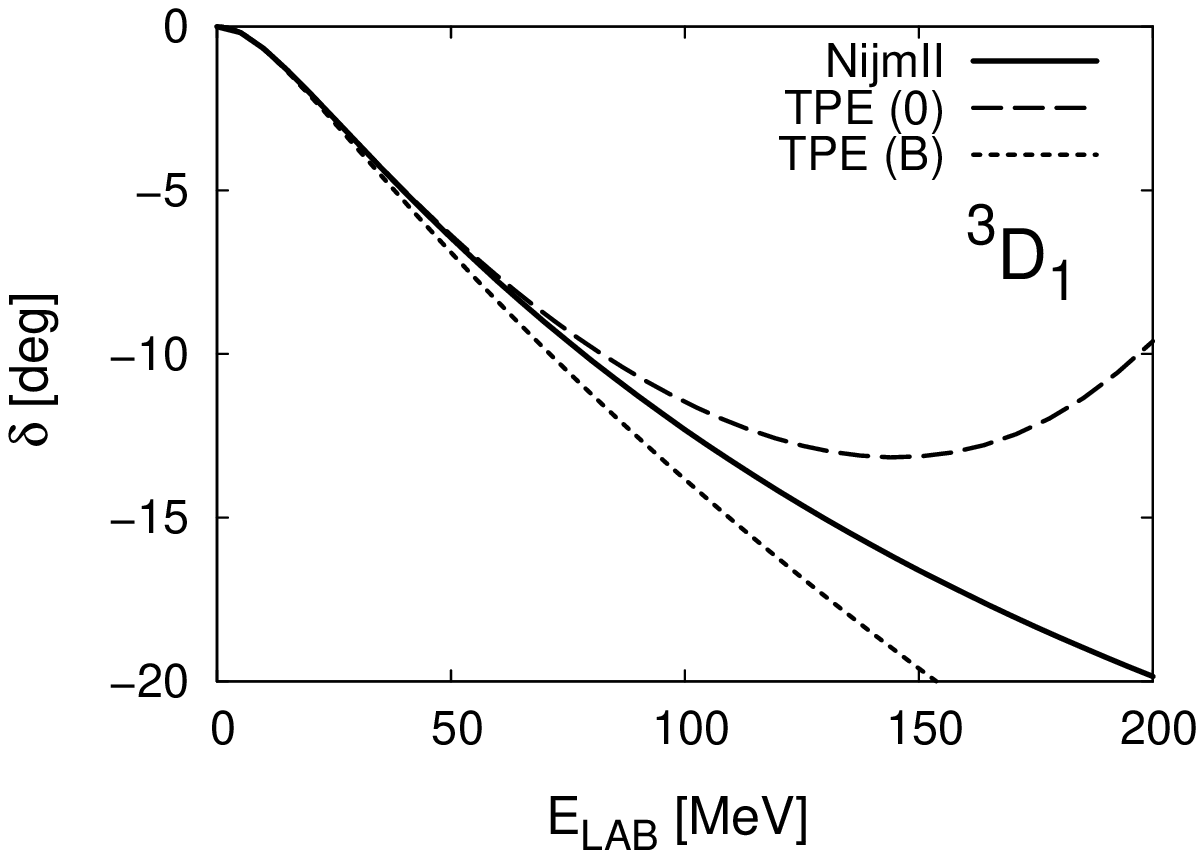,height=4cm,width=5cm} 
\end{center}
\caption{Dependence of the $^3S_1-^3D_1$ channel (SYM-nuclear bar)
phase shifts for the NNLO TPE potential on the reference state used to
orthogonalize the scattering state compared to the corresponding
phases of the database of
Ref.~\cite{Stoks:1993tb,Stoks:1994wp}. Label TPE(0) means  
the zero energy reference state with $\alpha_0 = 5.418 {\rm fm}$,
$\alpha_{02}= 1.647 {\rm fm}^3 $ and $\alpha_{2}=6.505 {\rm fm}^5$. Label TPE(B)stands for the deuteron bound reference state with
the experimental binding energy, asymptotic $D/S $ ratio together
with $\alpha_0 = 5.418 {\rm fm}$ (corresponding to $\alpha_{02}=
1.67\,{\rm fm}^3 $ and $\alpha_{2}=6.6\,{\rm
fm}^5$~\cite{Valderrama:2005wv}.)}
\label{fig:3C1}
\end{figure*}

\begin{figure*}[]
\begin{center}
\epsfig{figure=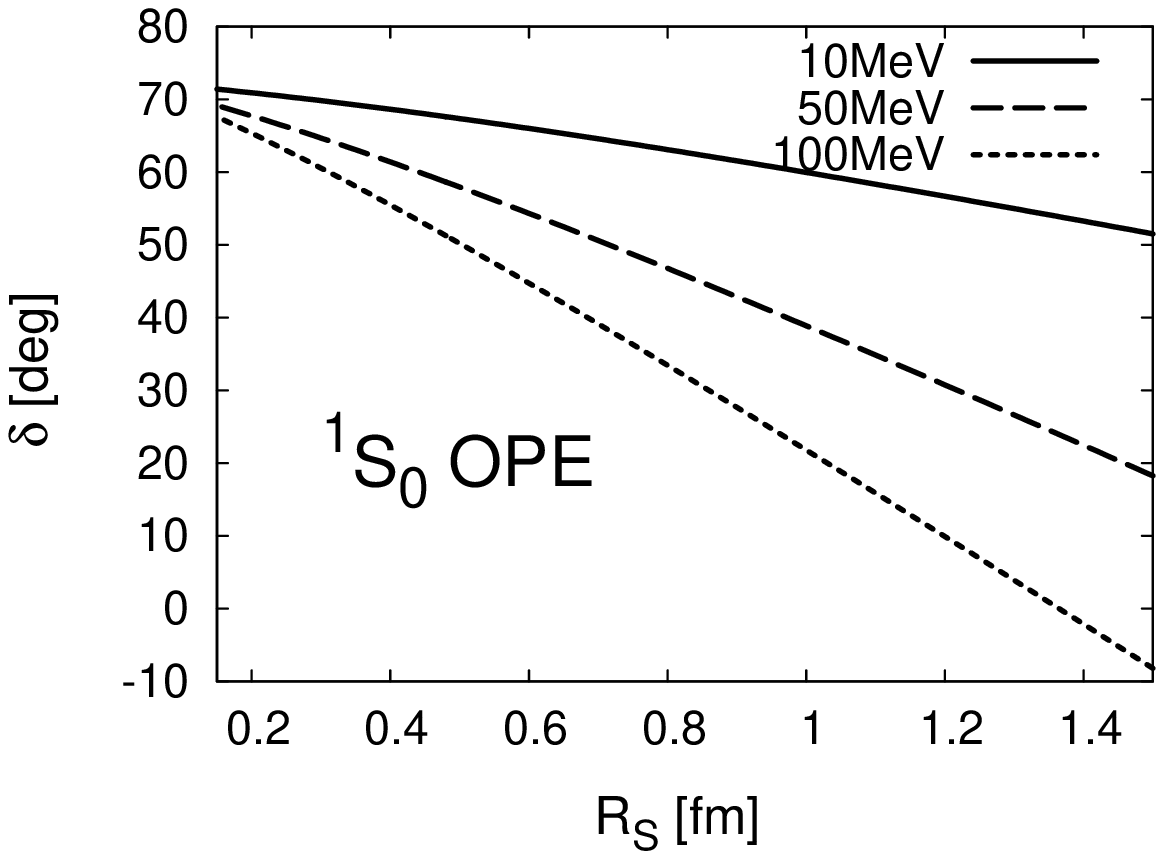,height=4cm,width=5cm}
\epsfig{figure=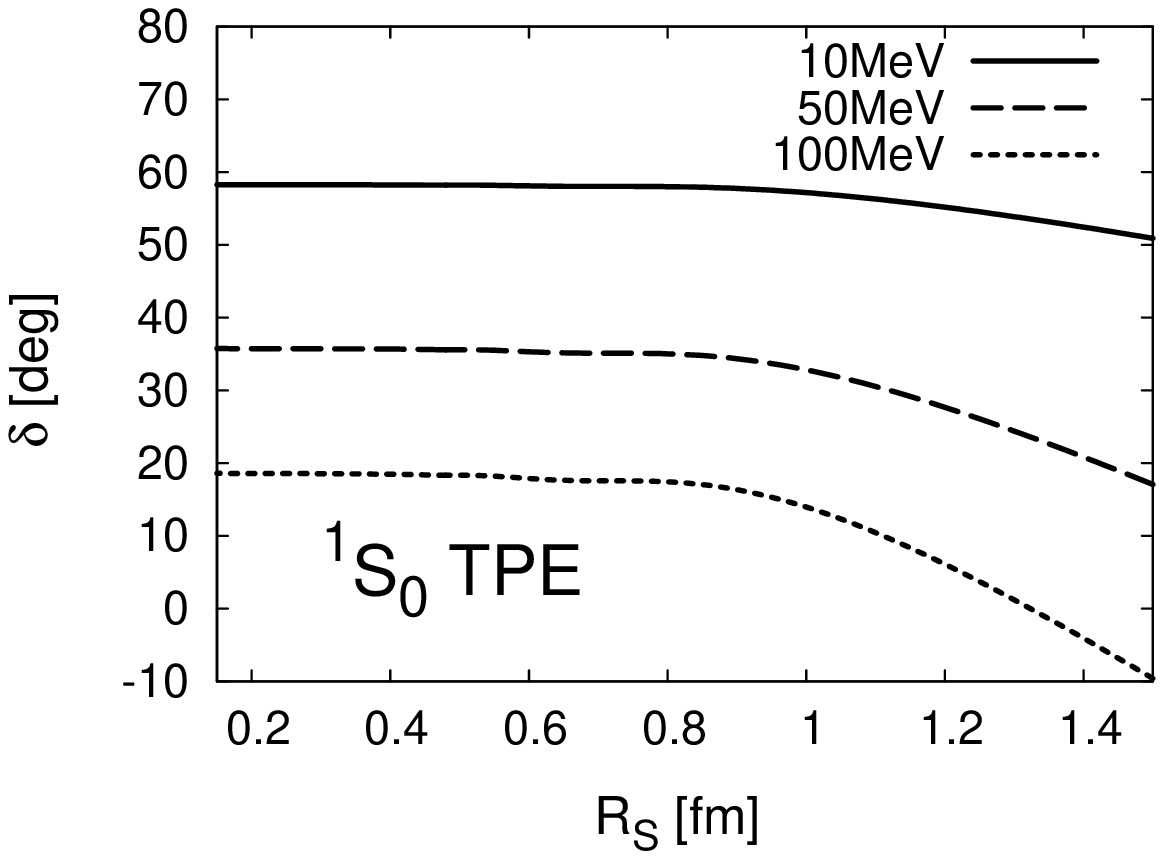,height=4cm,width=5cm}
\epsfig{figure=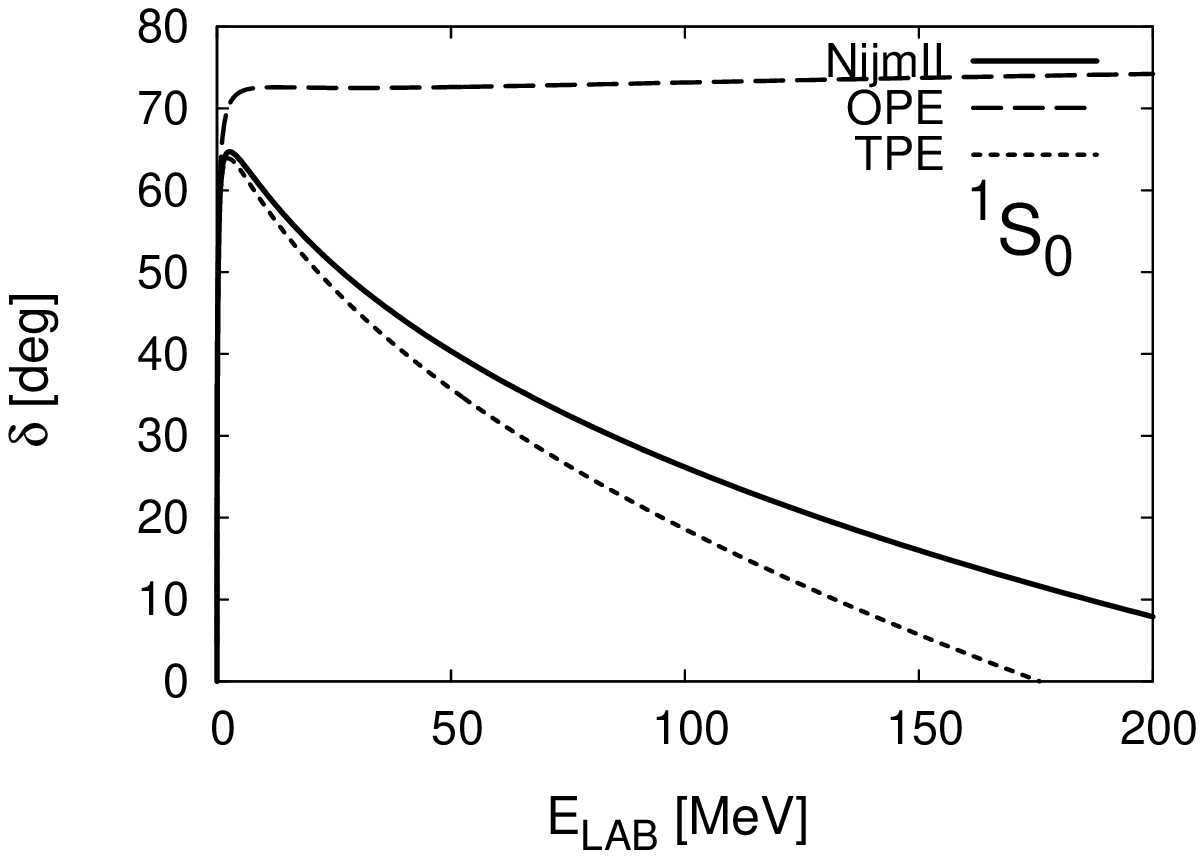,height=4cm,width=5cm} \\ 
\epsfig{figure=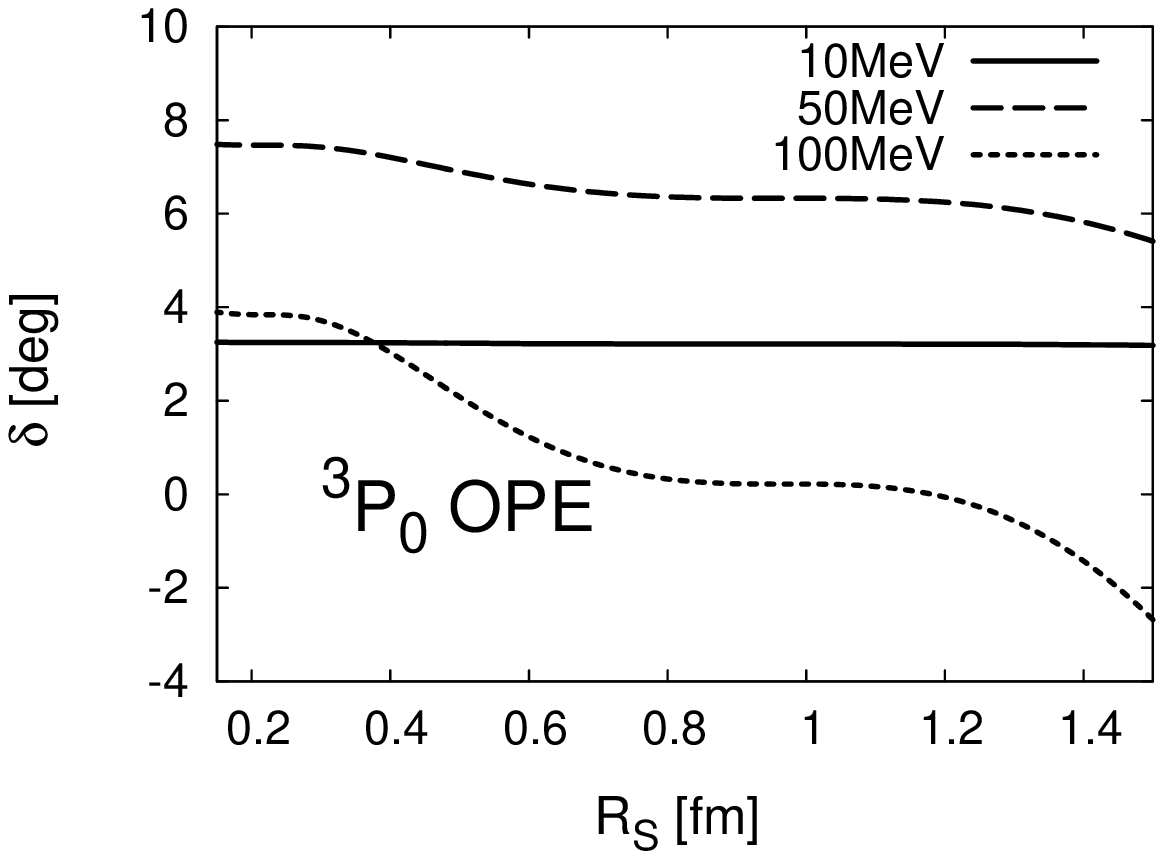,height=4cm,width=5cm}
\epsfig{figure=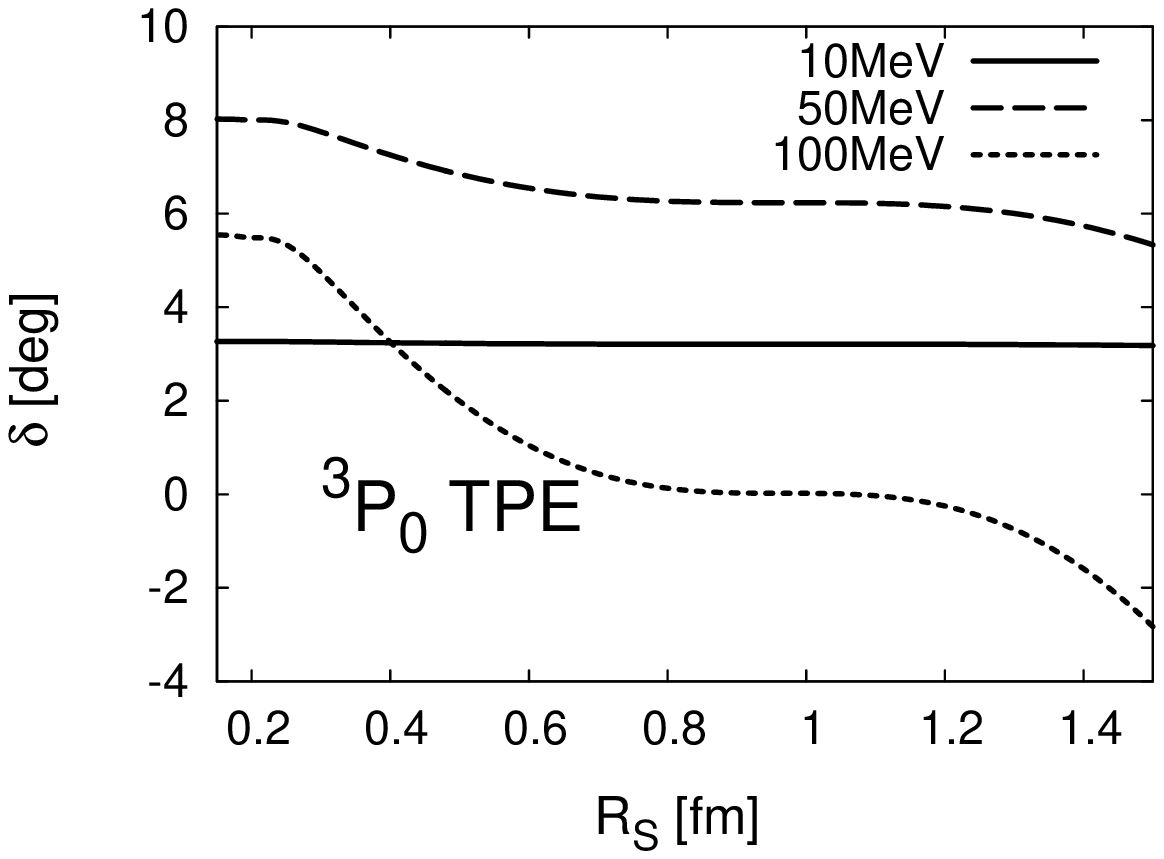,height=4cm,width=5cm}
\epsfig{figure=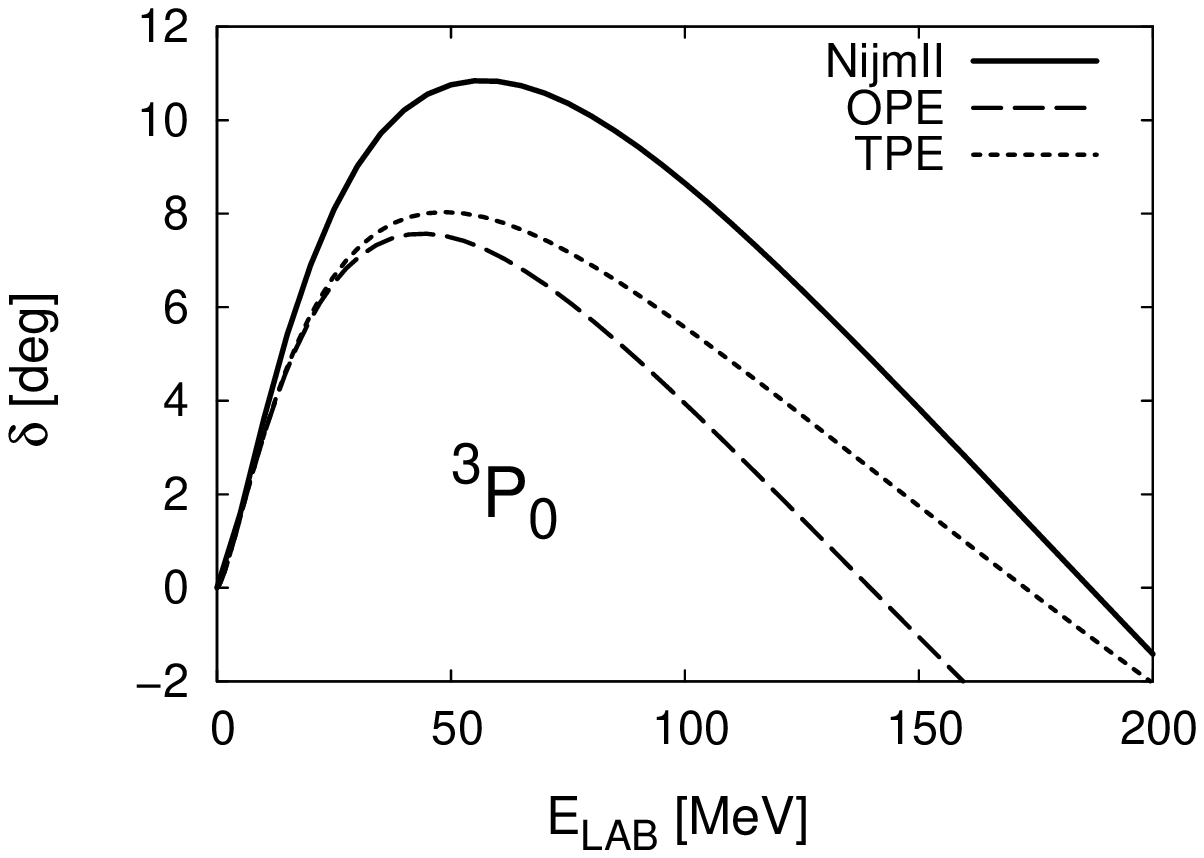,height=4cm,width=5cm} 
\end{center}
\caption{np (SYM-nuclear bar) Phase shifts for the total angular
momentum $j=0$.  OPE (left) and chiral TPE (middle) as a function of
the cut-off radius $R_S$ for fixed LAB energies, $E_{\rm LAB}=10,50,100
{\rm MeV}$. OPE and chiral TPE (left) renormalized (i.e. $R_S \to 0$) 
phase shifts as a function of the LAB energy compared to the Nijmegen
partial wave analysis~\cite{Stoks:1993tb,Stoks:1994wp}.}
\label{fig:fig-j=0}
\end{figure*}

\begin{figure*}[]
\begin{center}
\epsfig{figure=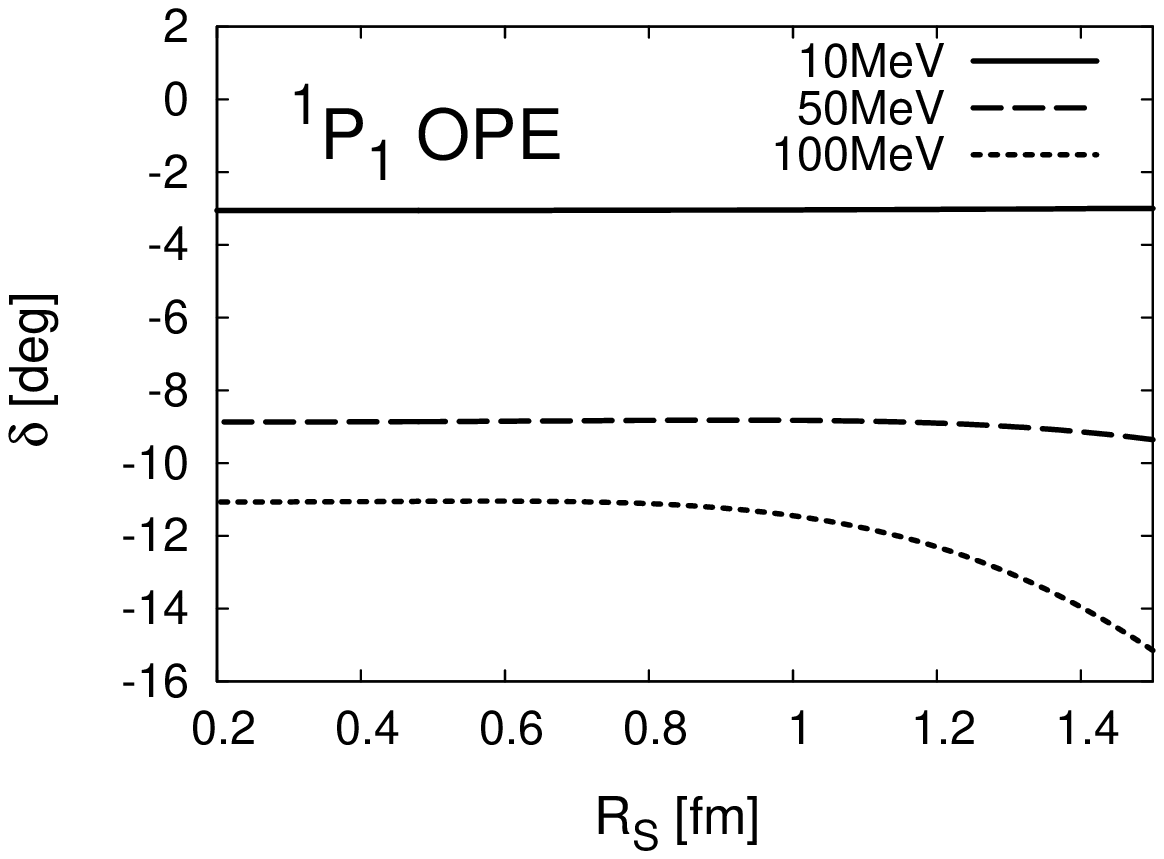,height=4cm,width=5cm}
\epsfig{figure=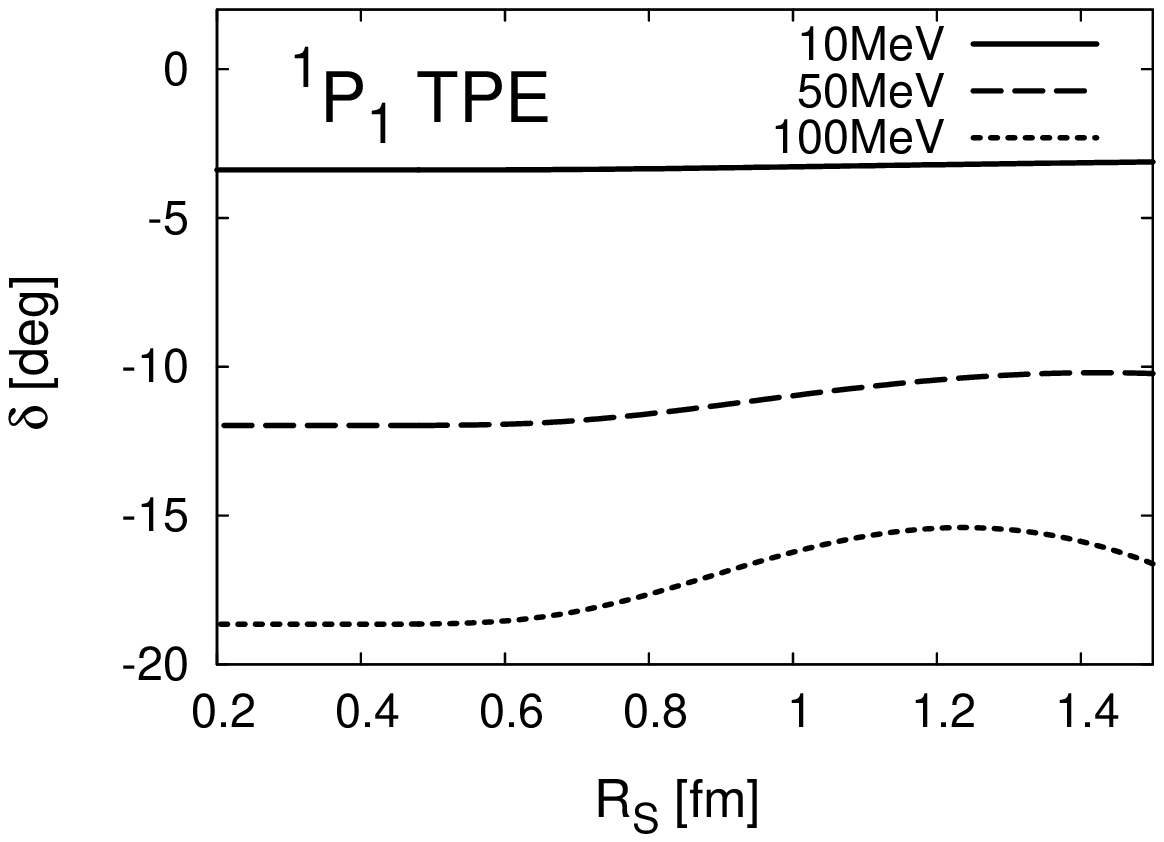,height=4cm,width=5cm}
\epsfig{figure=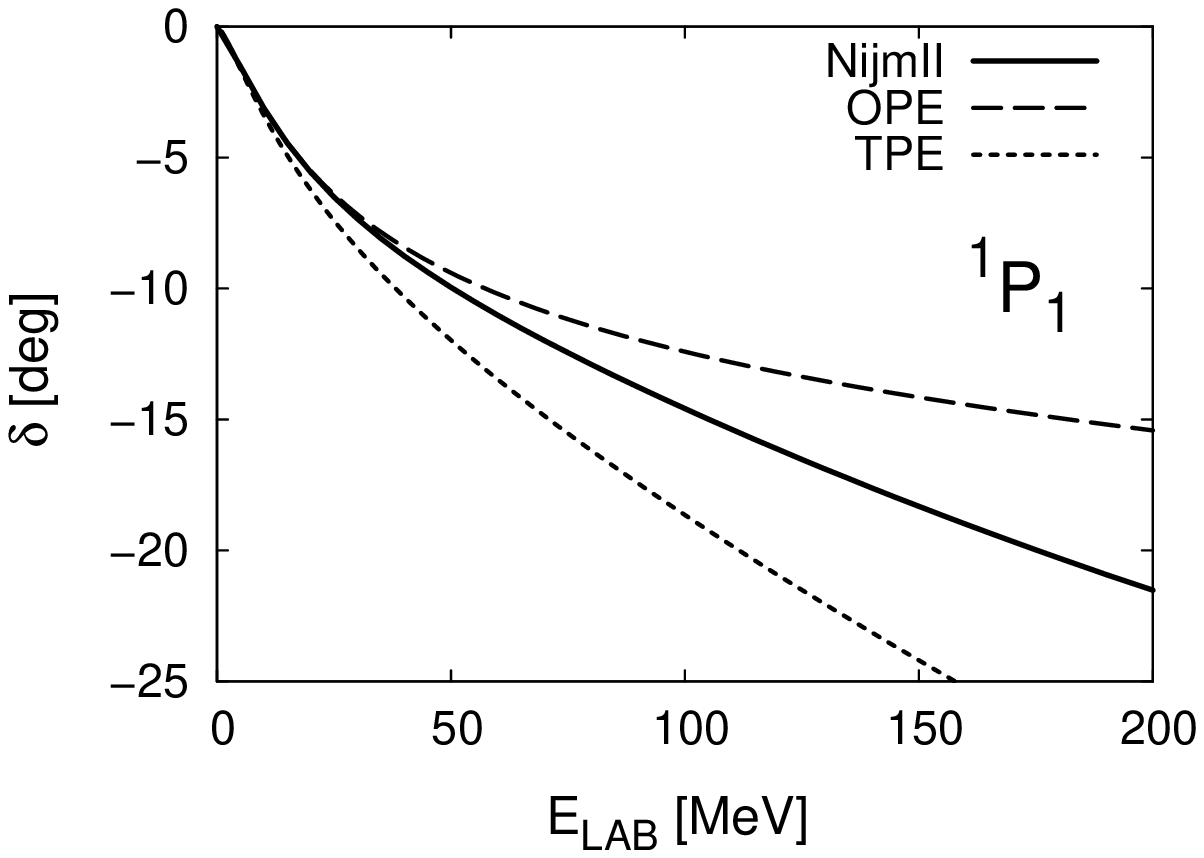,height=4cm,width=5cm} \\ 
\epsfig{figure=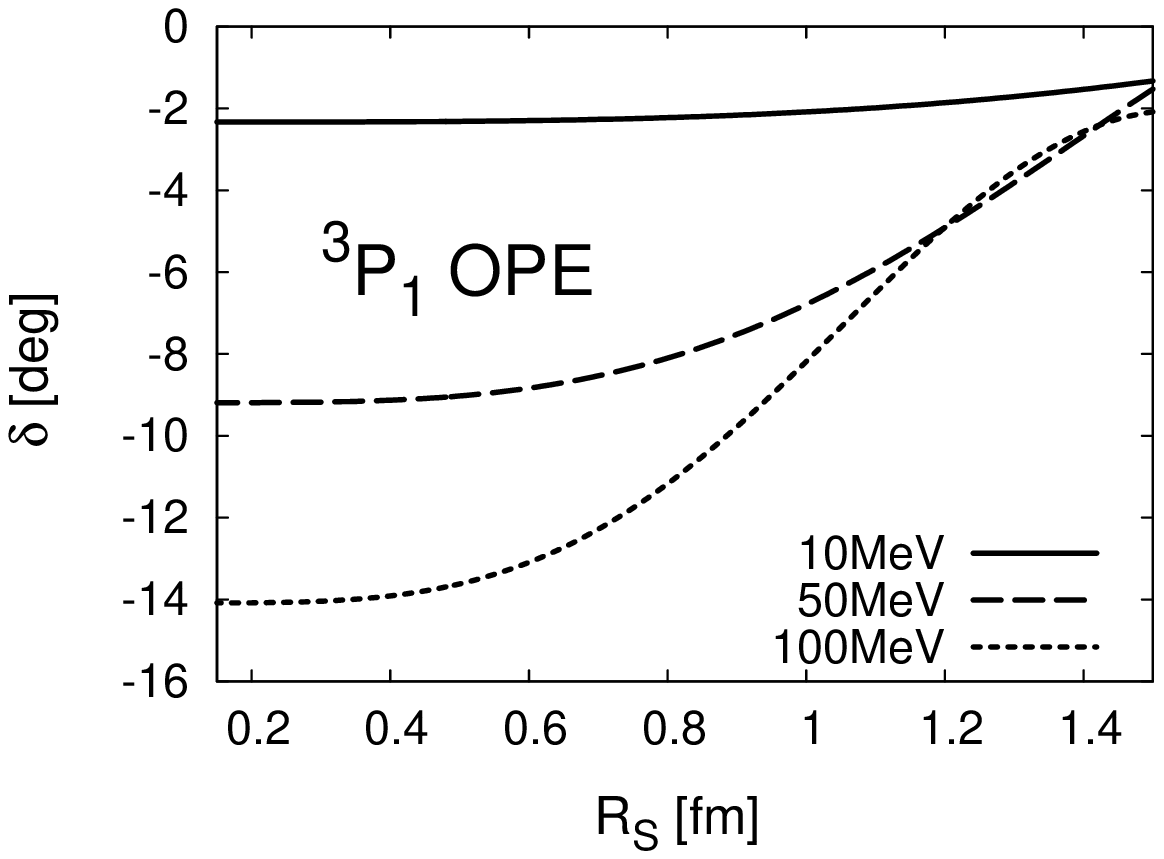,height=4cm,width=5cm}
\epsfig{figure=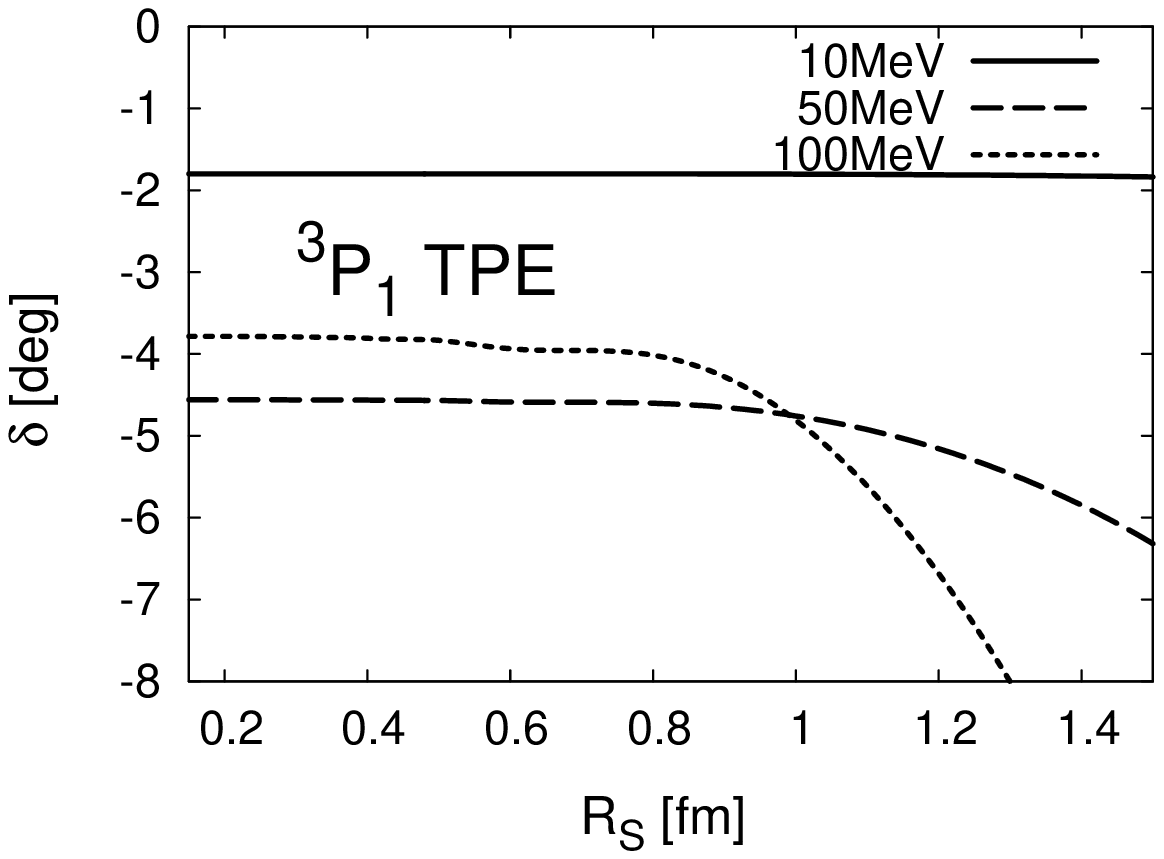,height=4cm,width=5cm}
\epsfig{figure=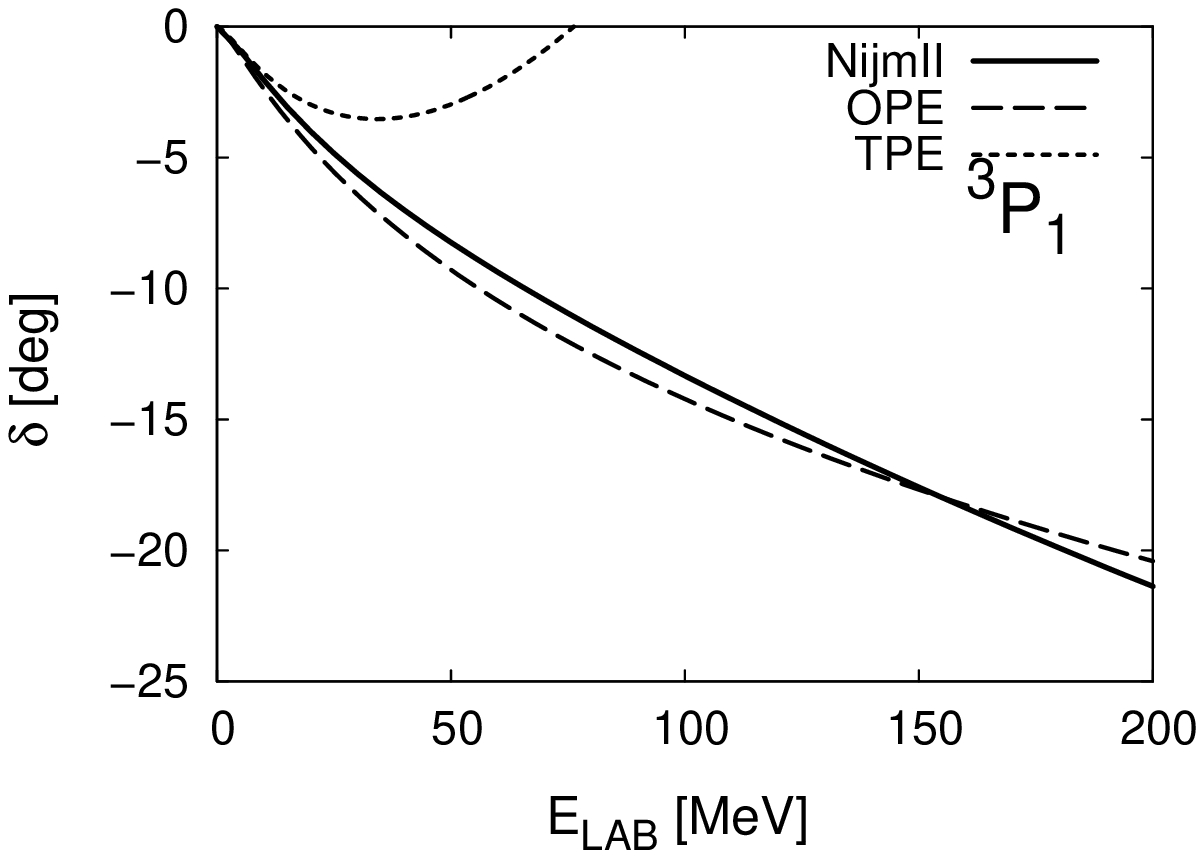,height=4cm,width=5cm} \\
\epsfig{figure=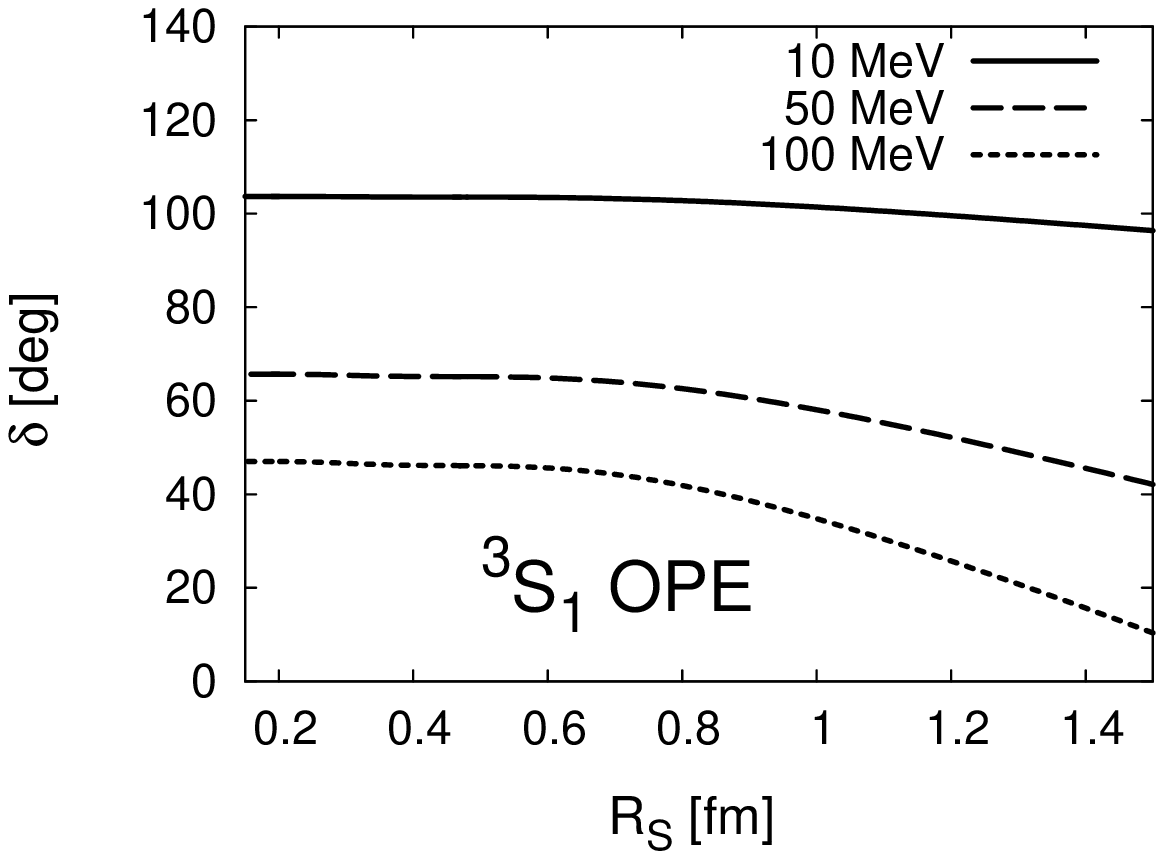,height=4cm,width=5cm}
\epsfig{figure=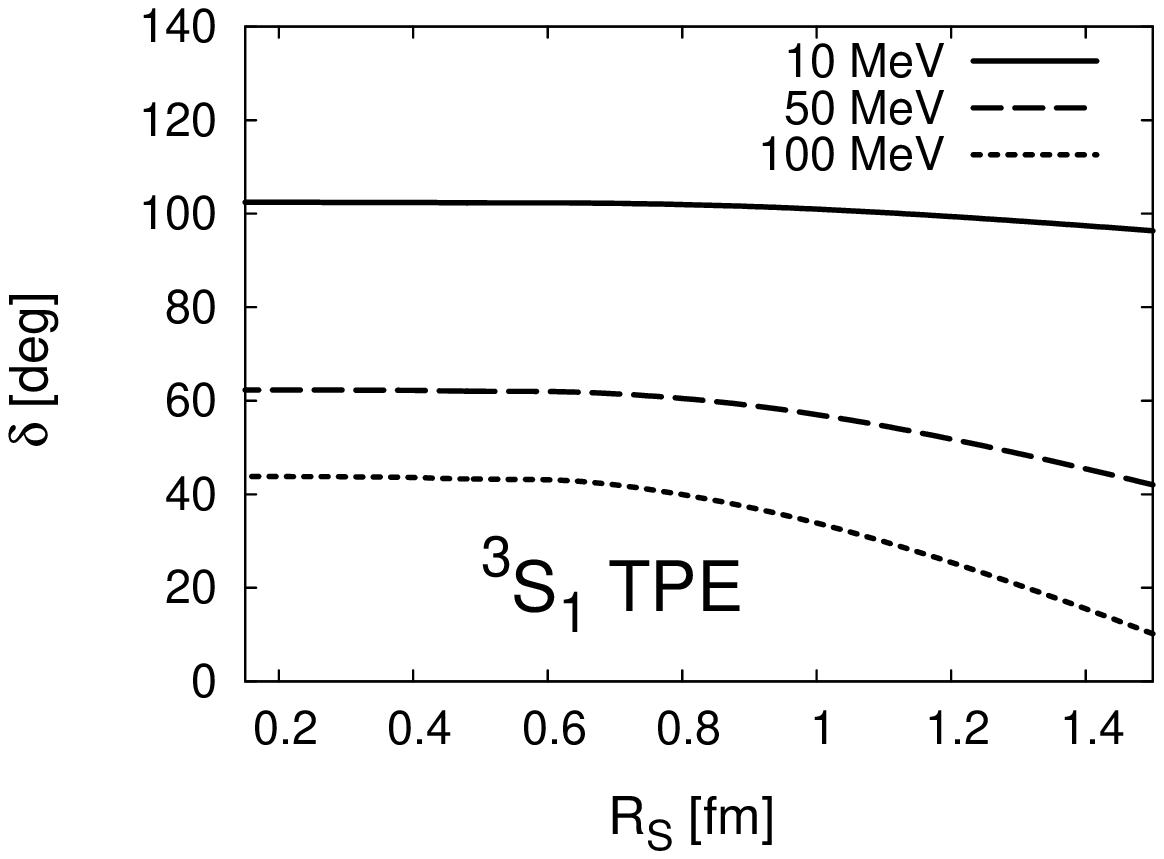,height=4cm,width=5cm}
\epsfig{figure=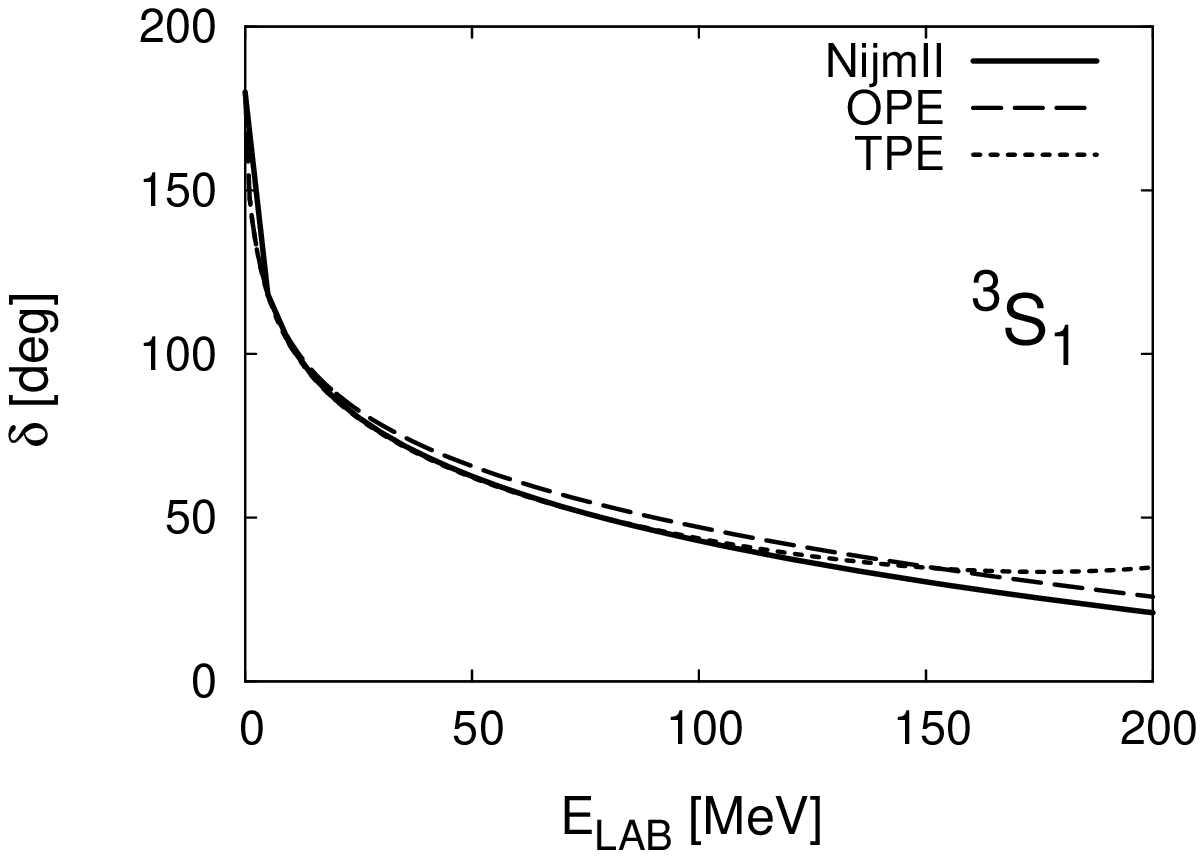,height=4cm,width=5cm} \\ 
\epsfig{figure=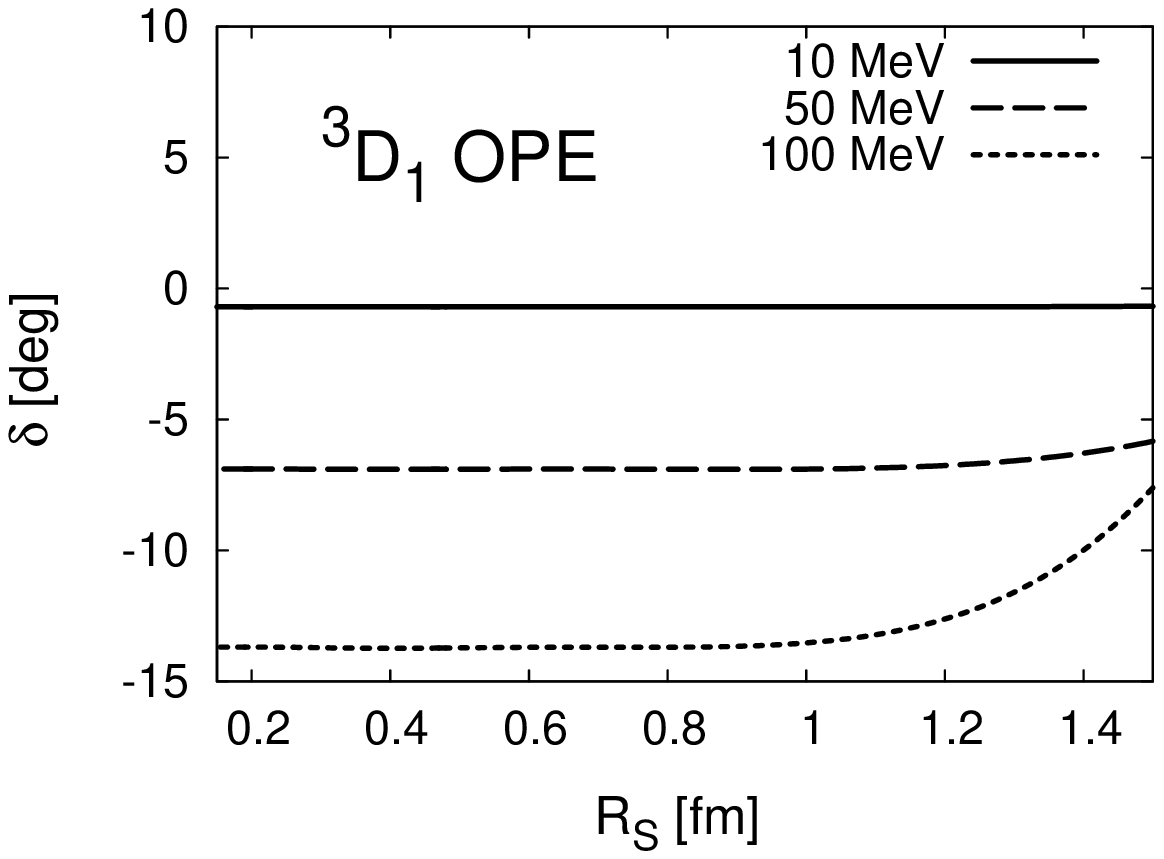,height=4cm,width=5cm}
\epsfig{figure=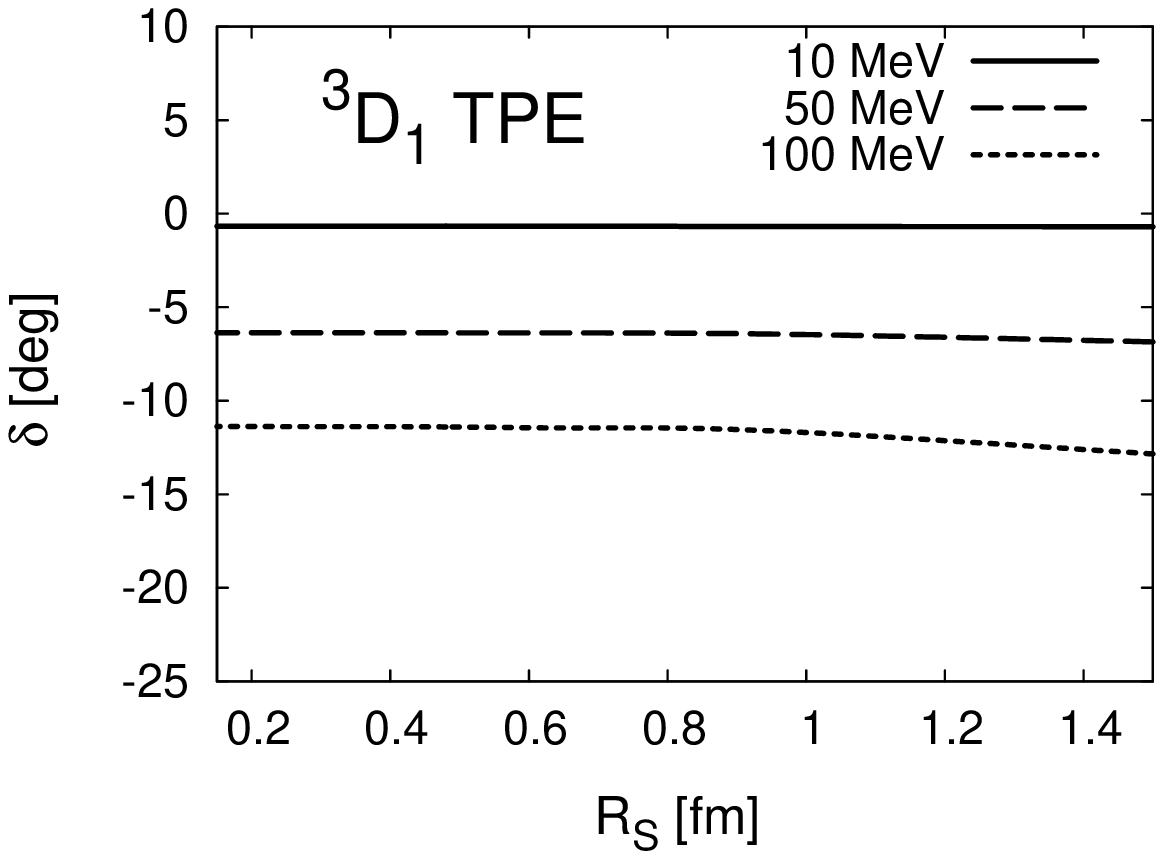,height=4cm,width=5cm}
\epsfig{figure=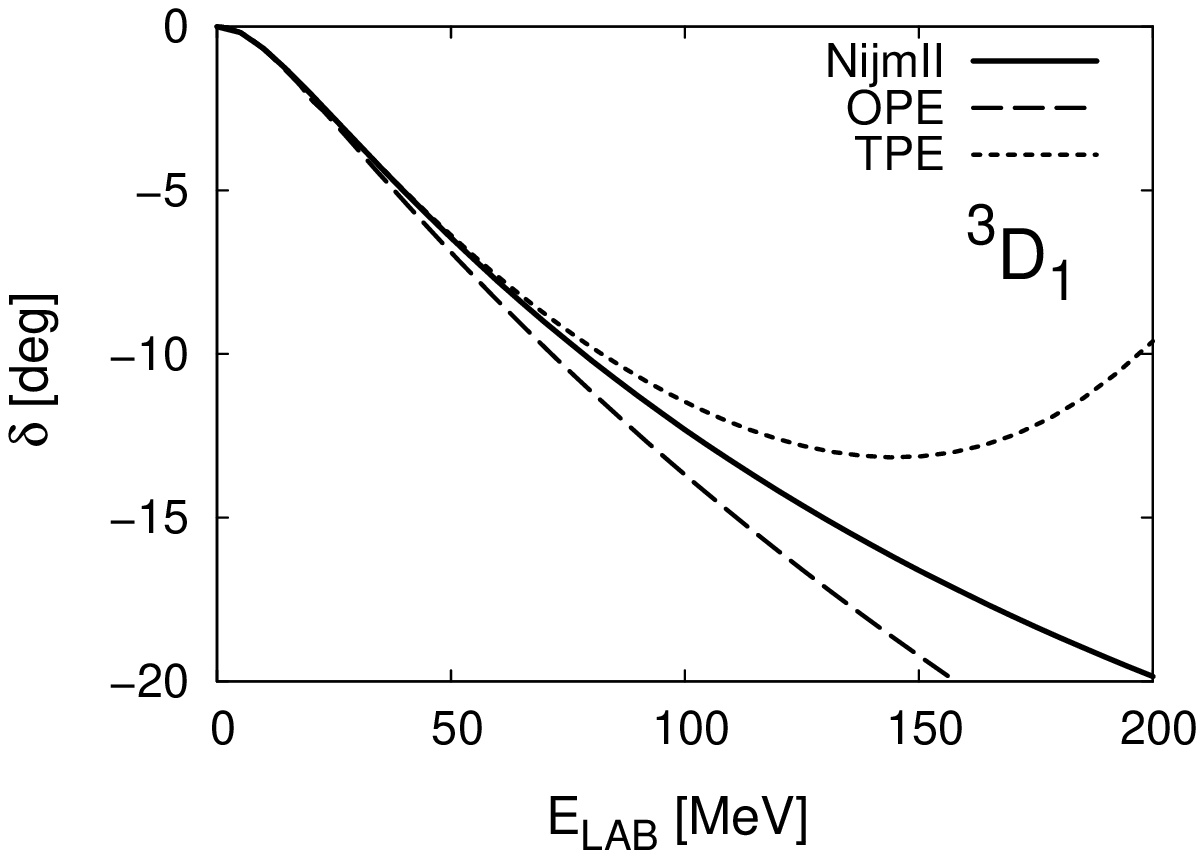,height=4cm,width=5cm} \\ 
\epsfig{figure=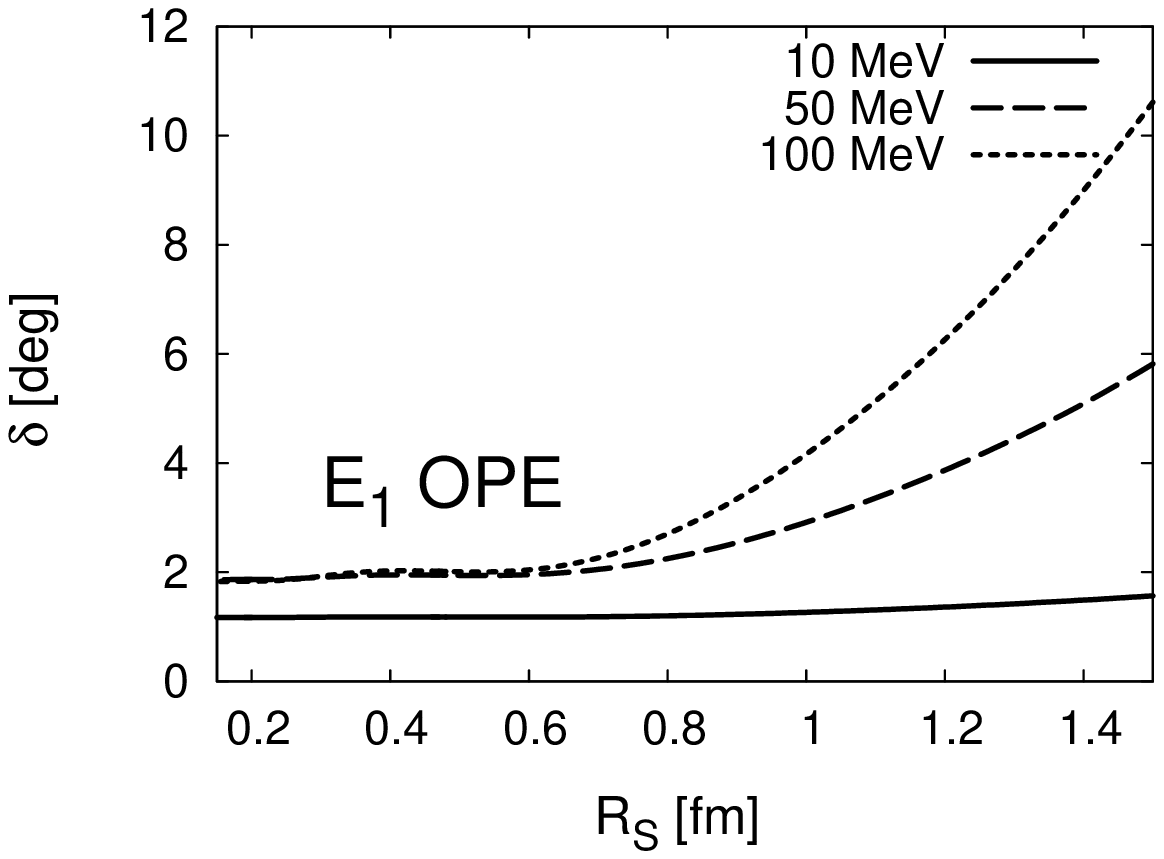,height=4cm,width=5cm}
\epsfig{figure=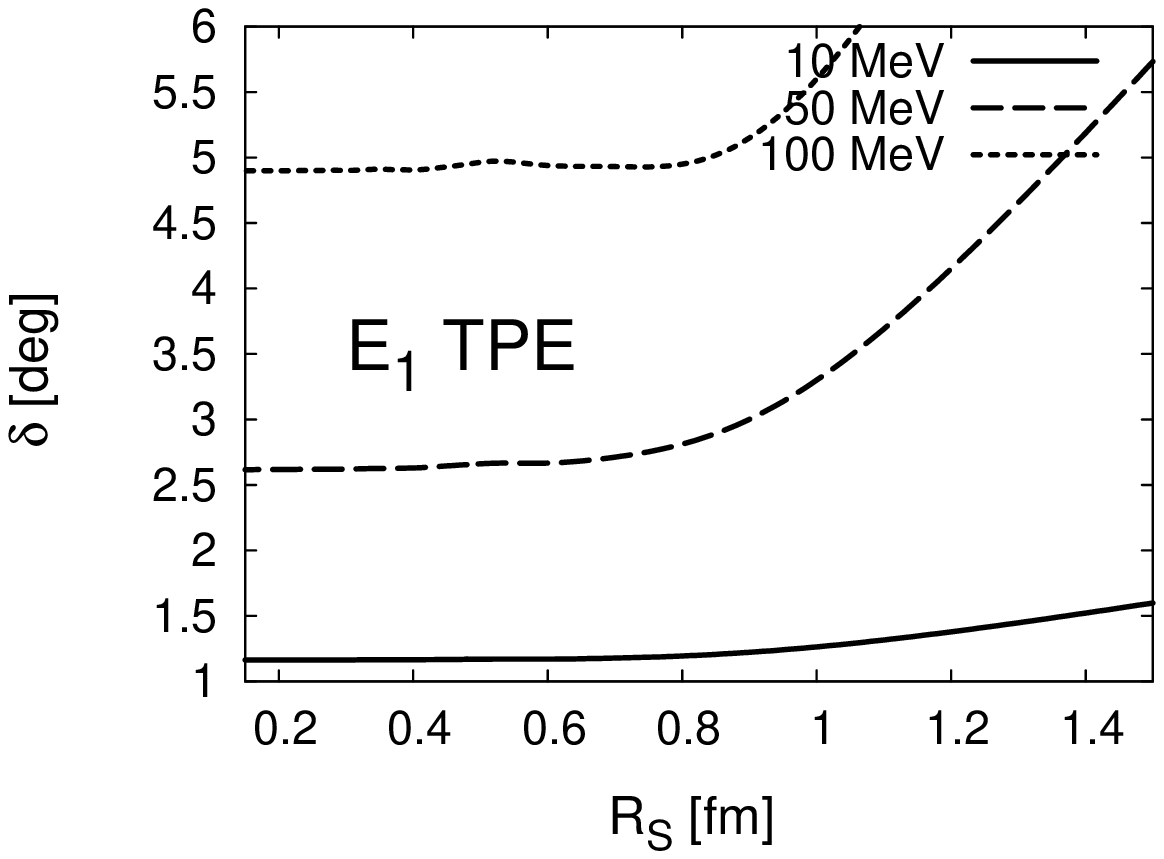,height=4cm,width=5cm}
\epsfig{figure=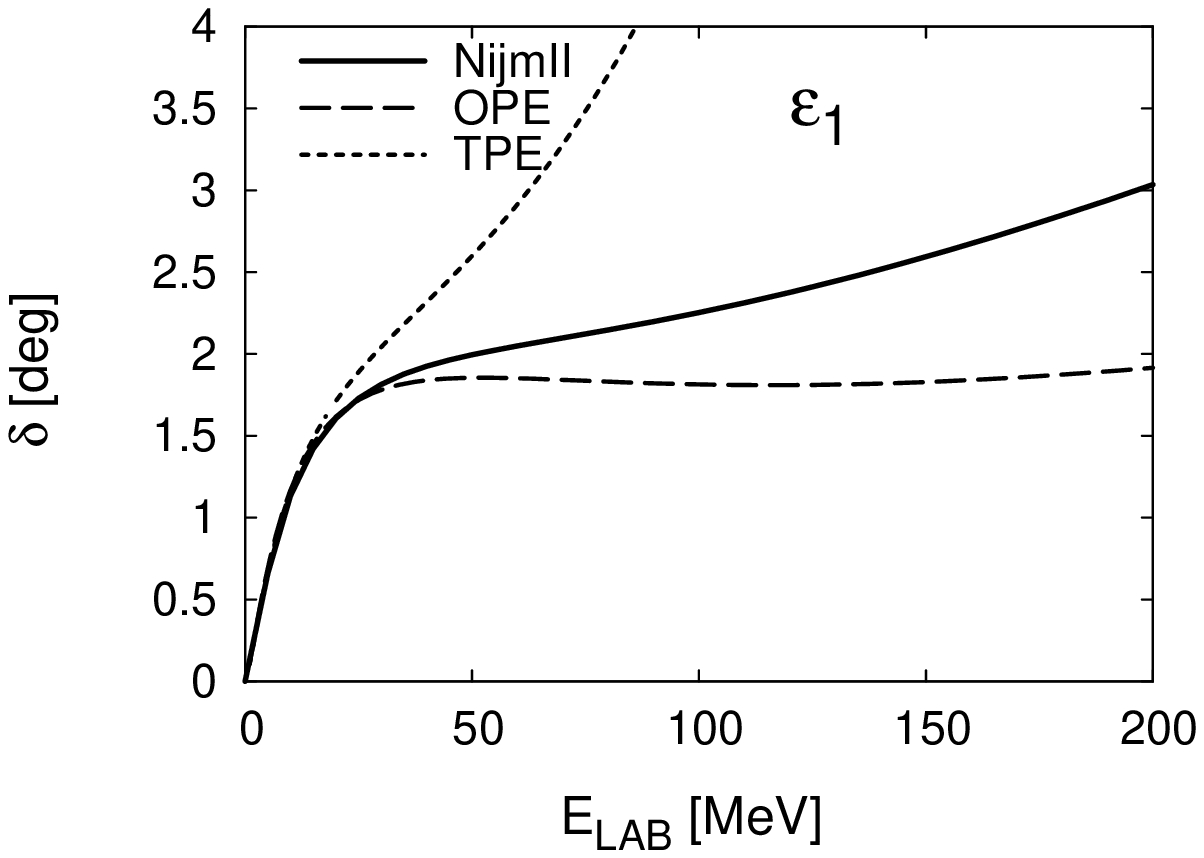,height=4cm,width=5cm} 
\end{center}
\caption{Same as Fig.~\ref{fig:fig-j=0} but for $j=1$.}
\label{fig:fig-j=1}
\end{figure*}

\begin{figure*}[]
\begin{center}
\epsfig{figure=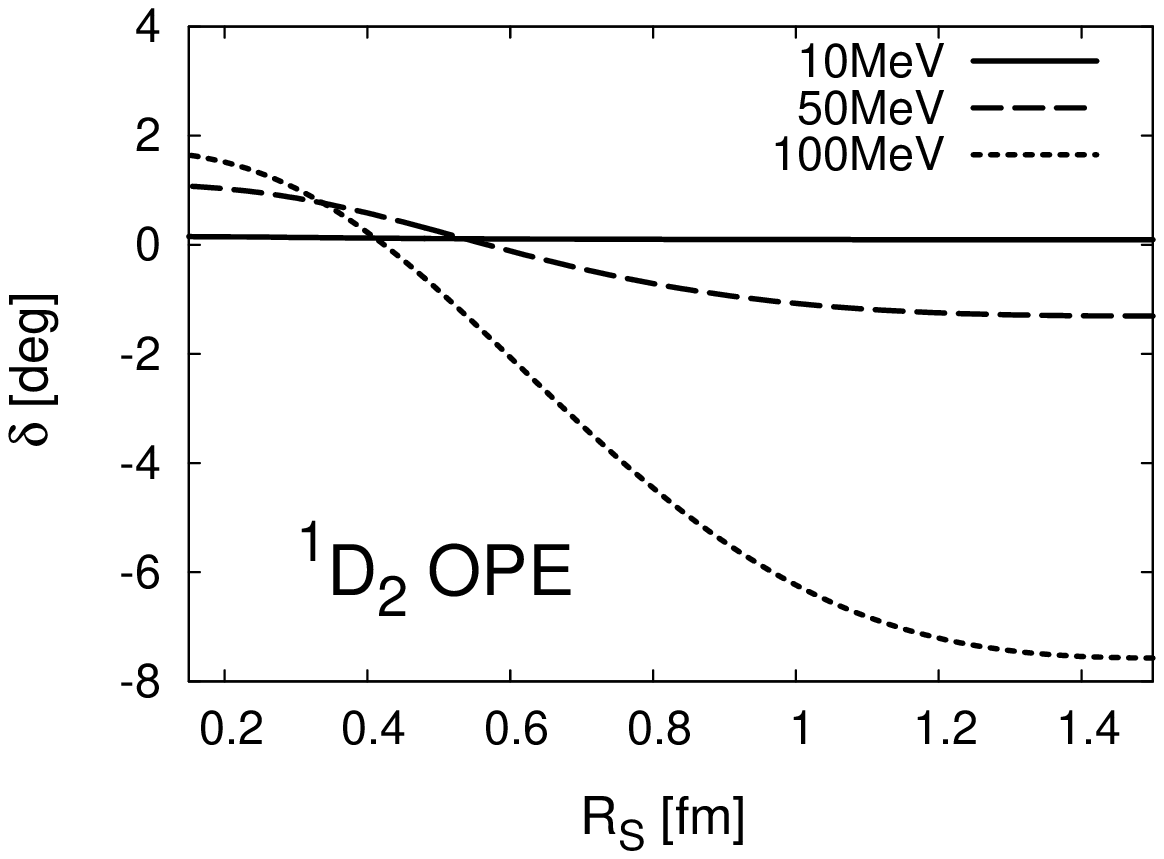,height=4cm,width=5cm}
\epsfig{figure=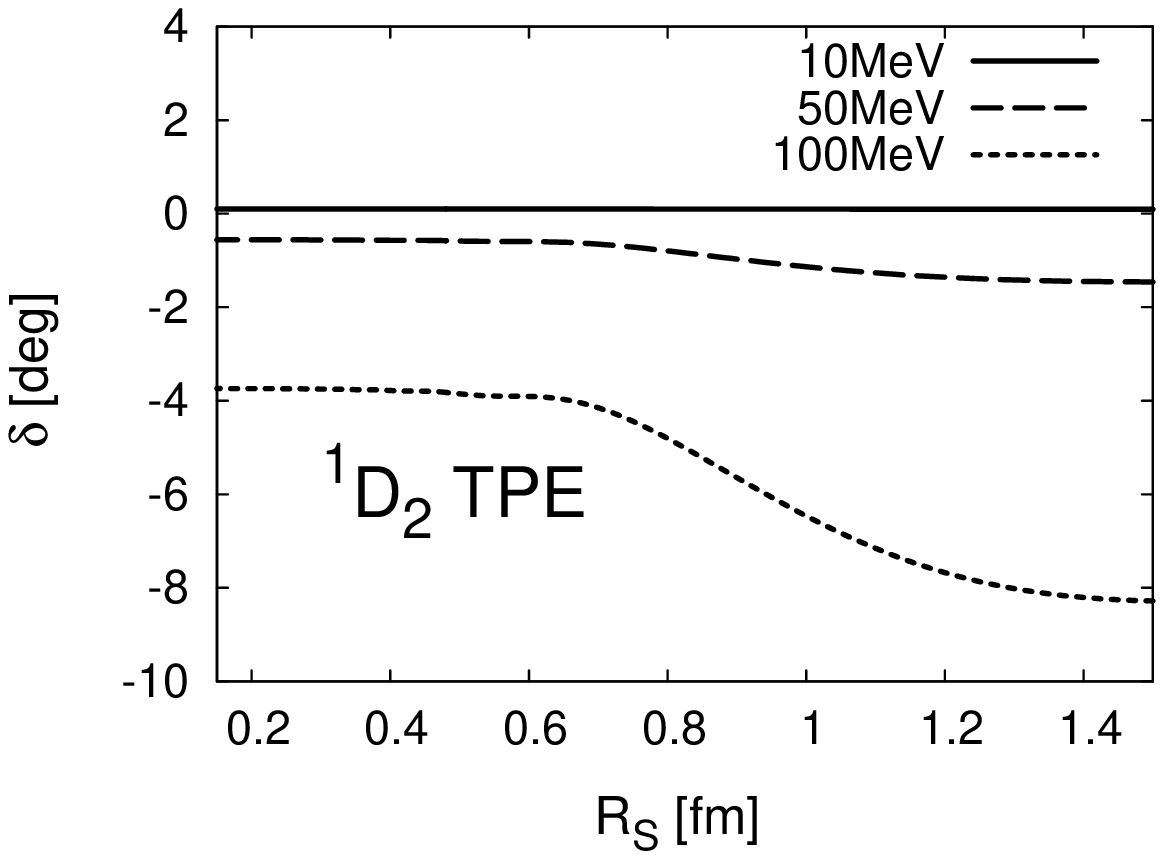,height=4cm,width=5cm}
\epsfig{figure=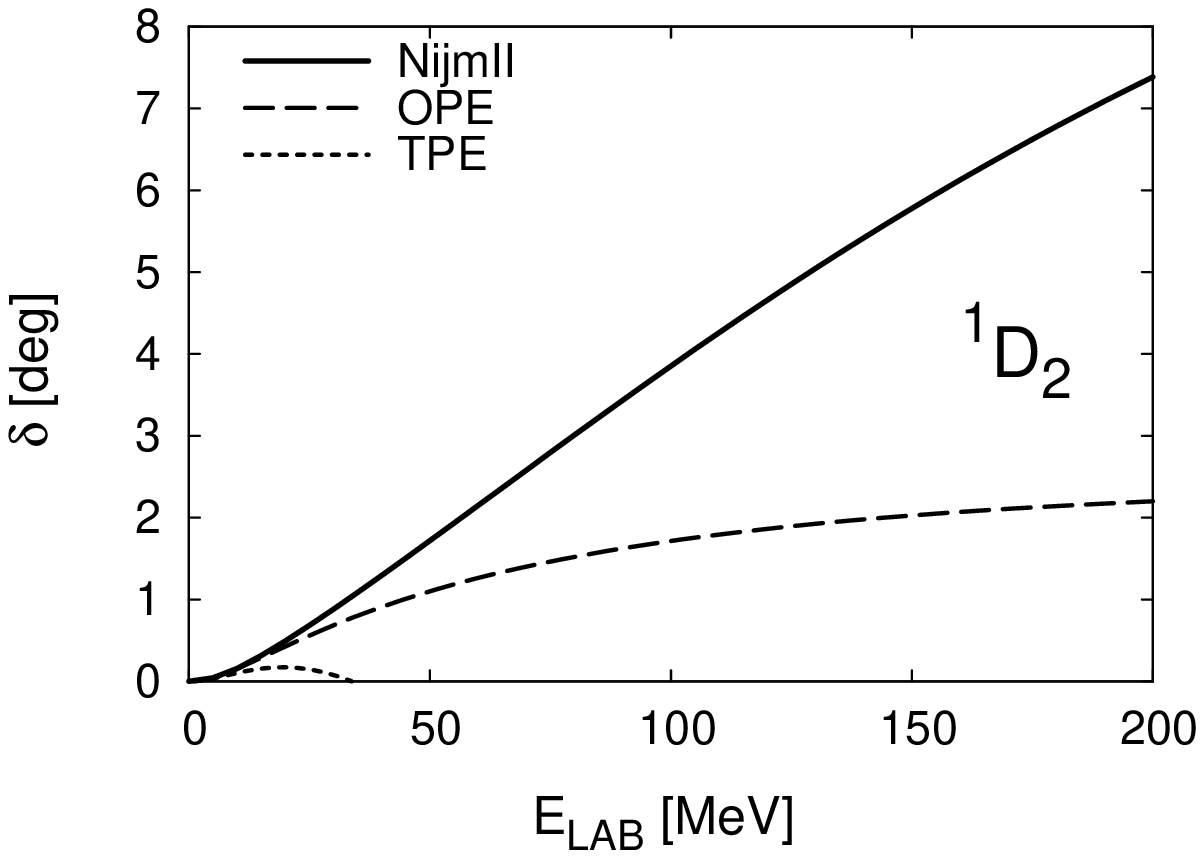,height=4cm,width=5cm} \\ 
\epsfig{figure=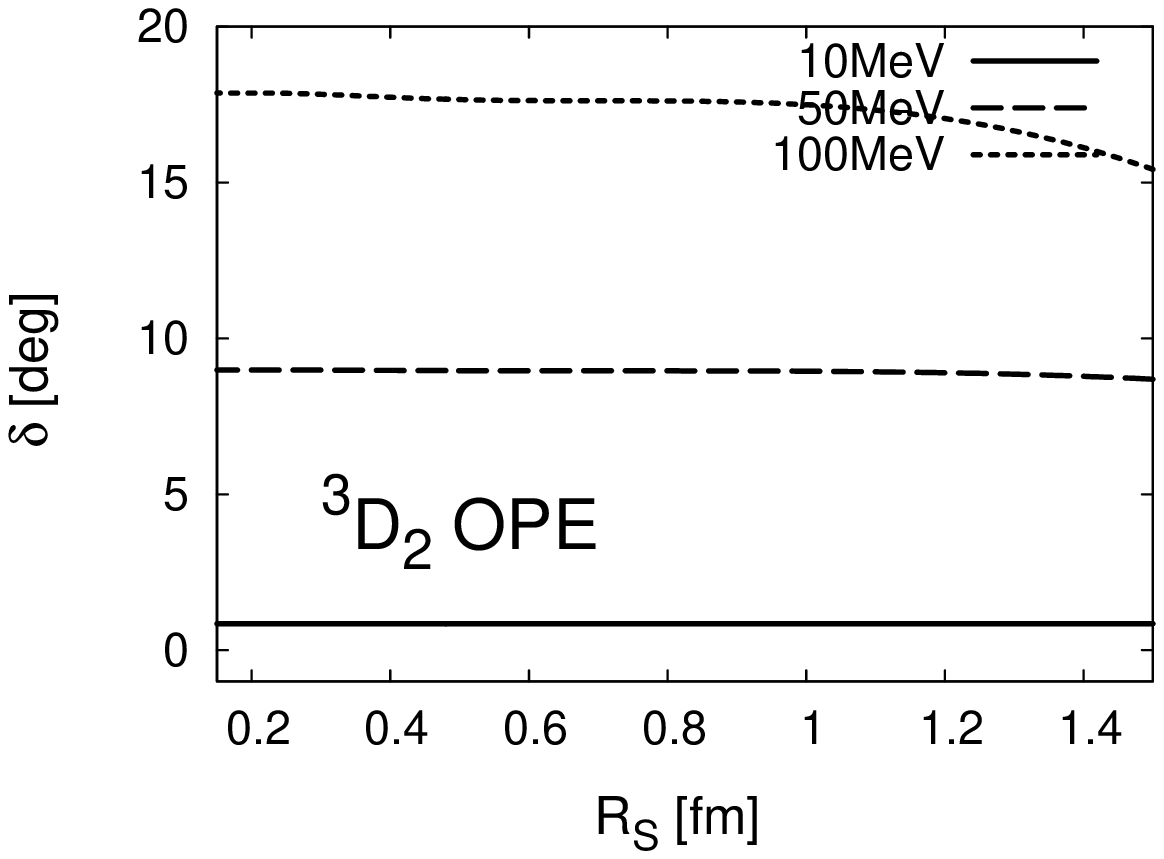,height=4cm,width=5cm}
\epsfig{figure=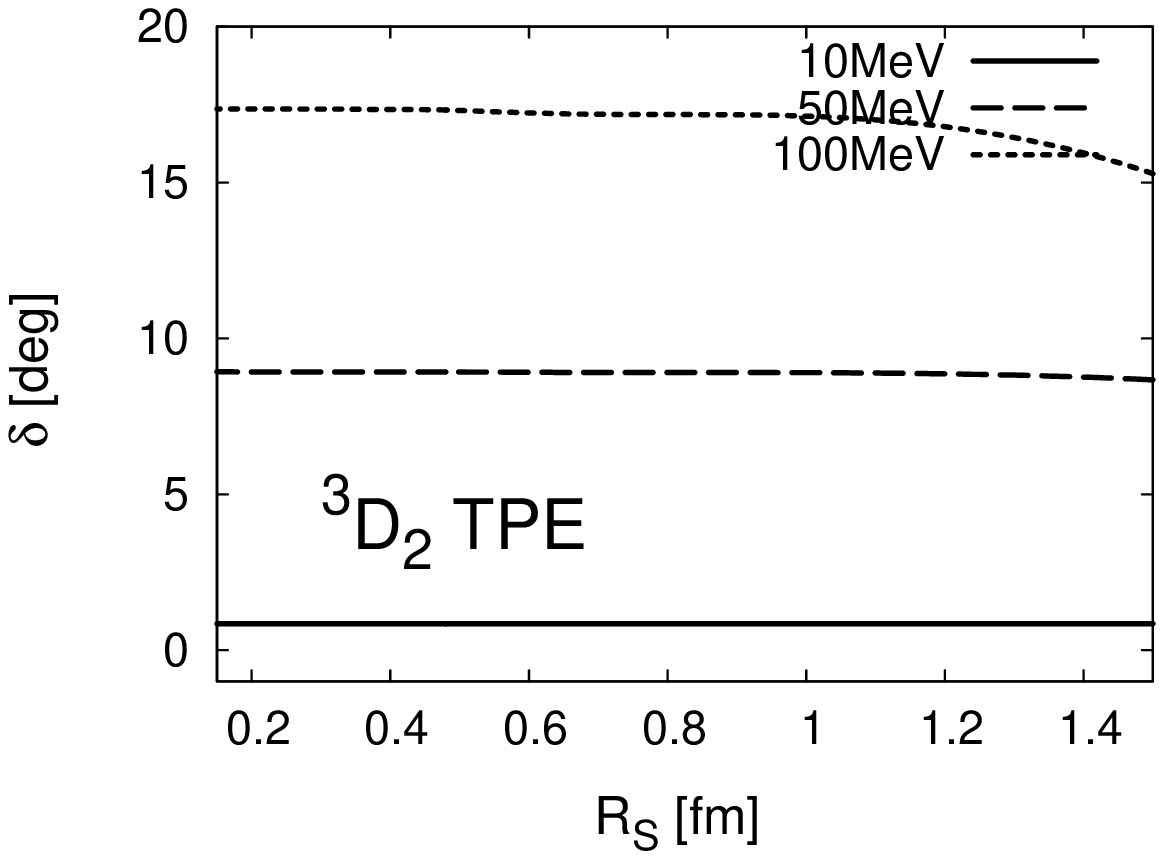,height=4cm,width=5cm}
\epsfig{figure=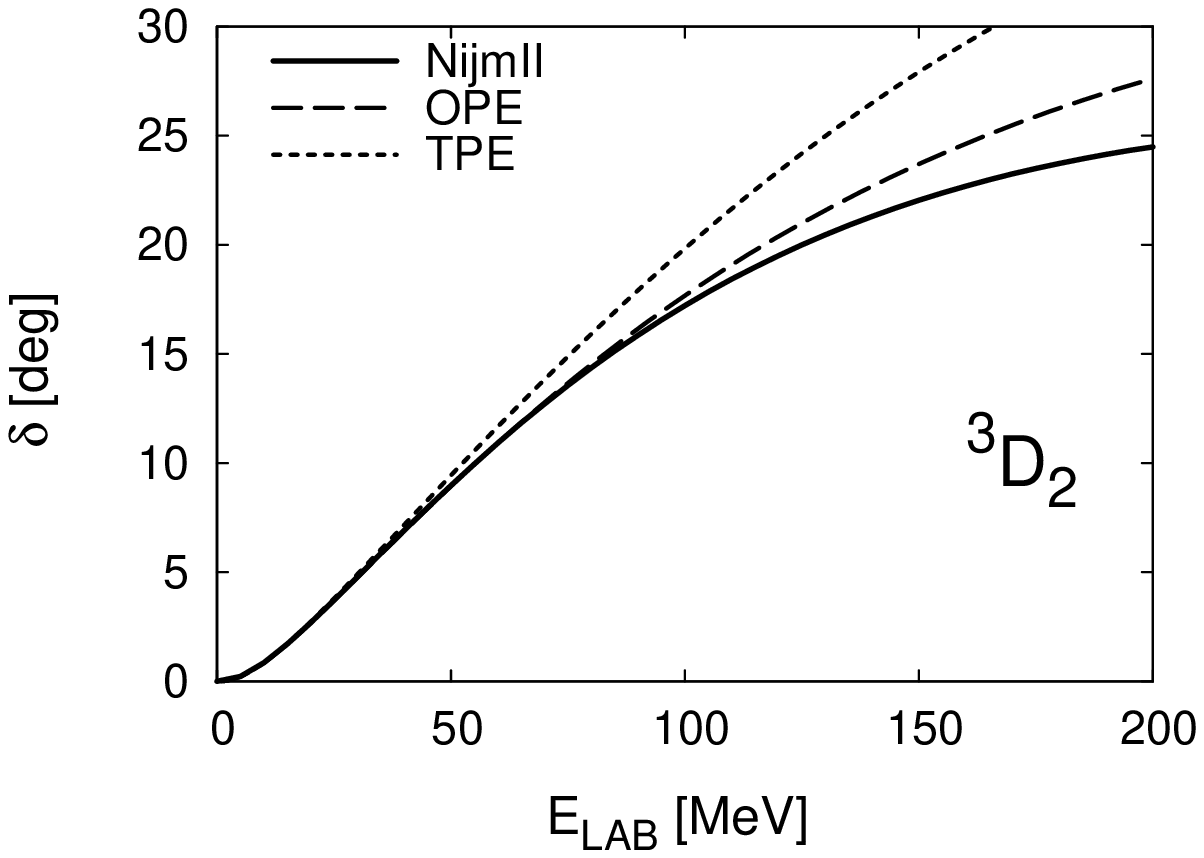,height=4cm,width=5cm} \\
\epsfig{figure=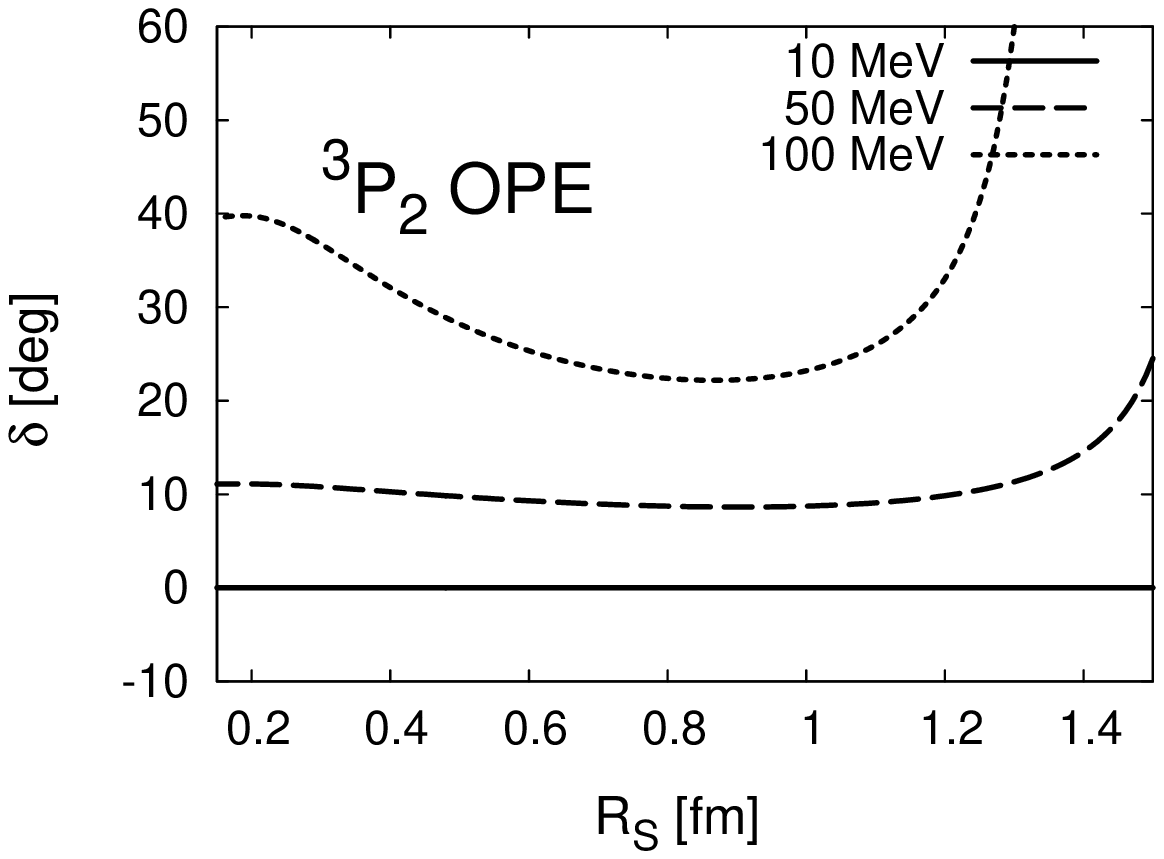,height=4cm,width=5cm}
\epsfig{figure=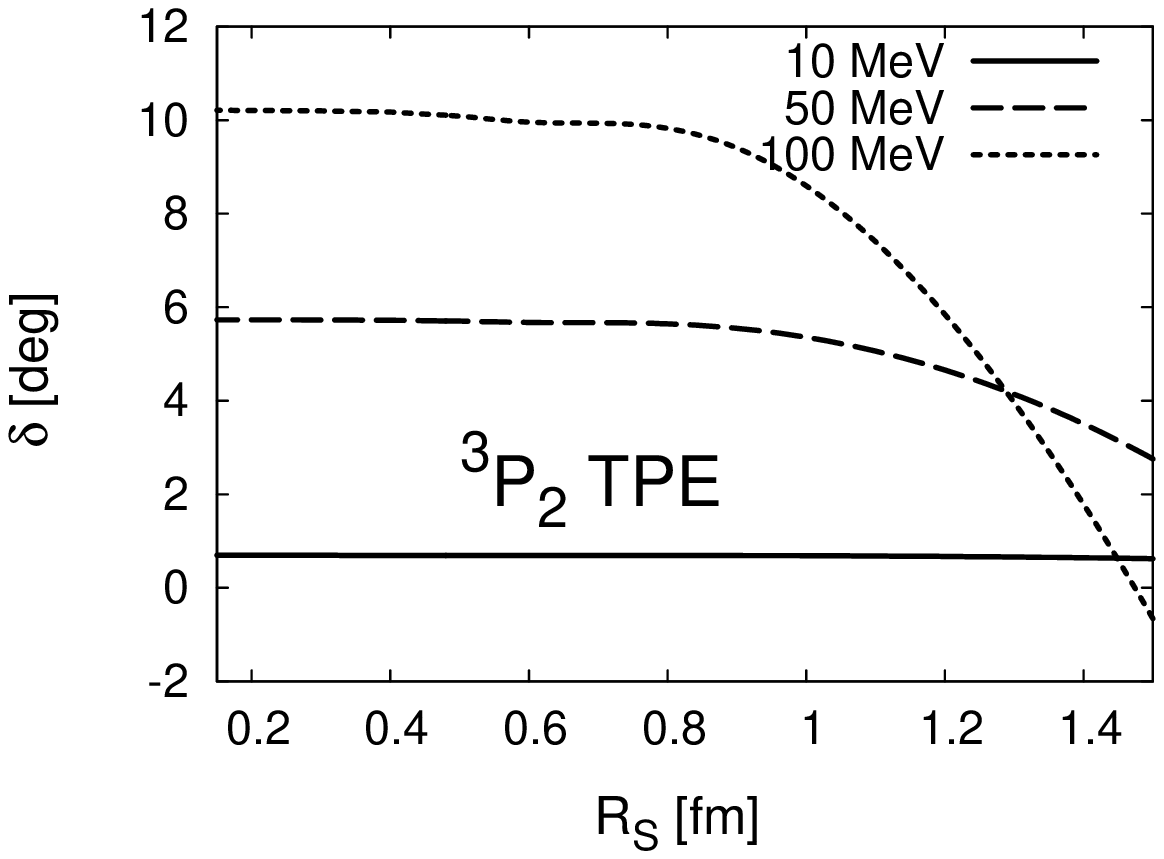,height=4cm,width=5cm}
\epsfig{figure=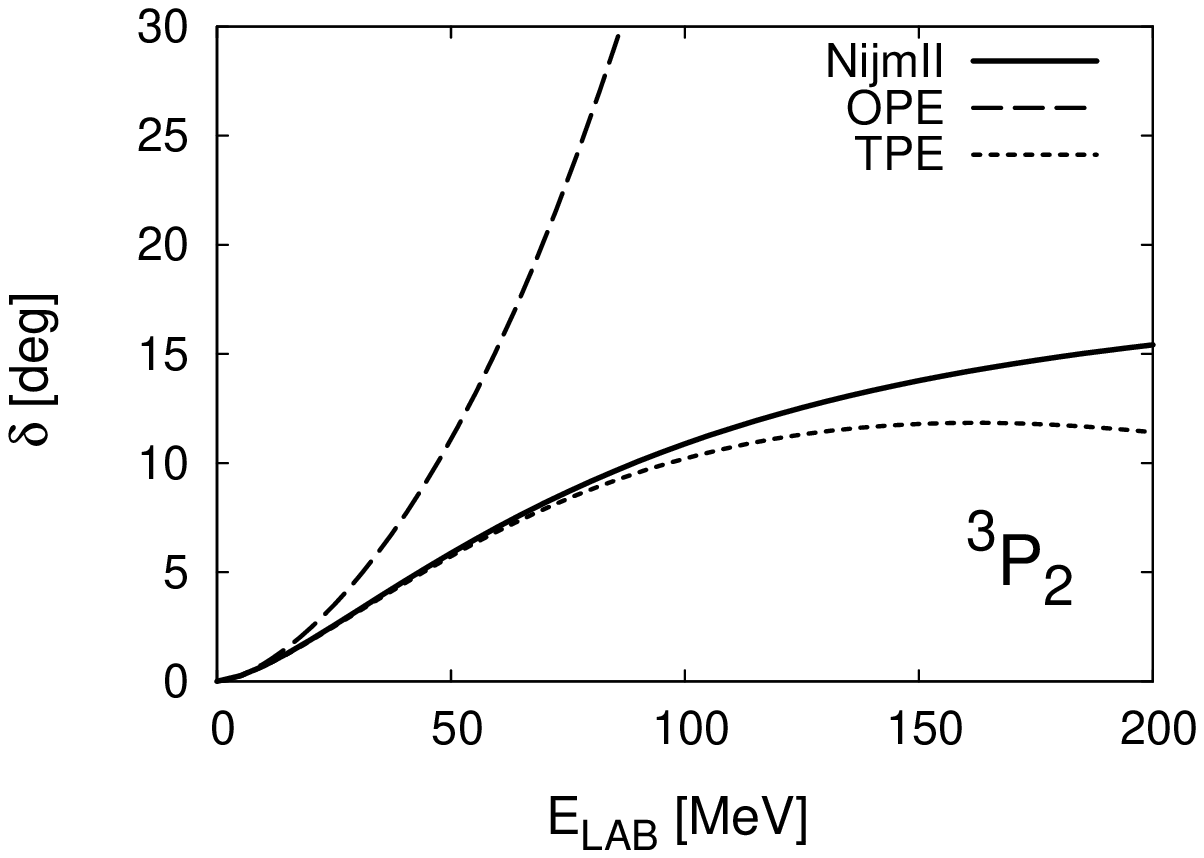,height=4cm,width=5cm} \\ 
\epsfig{figure=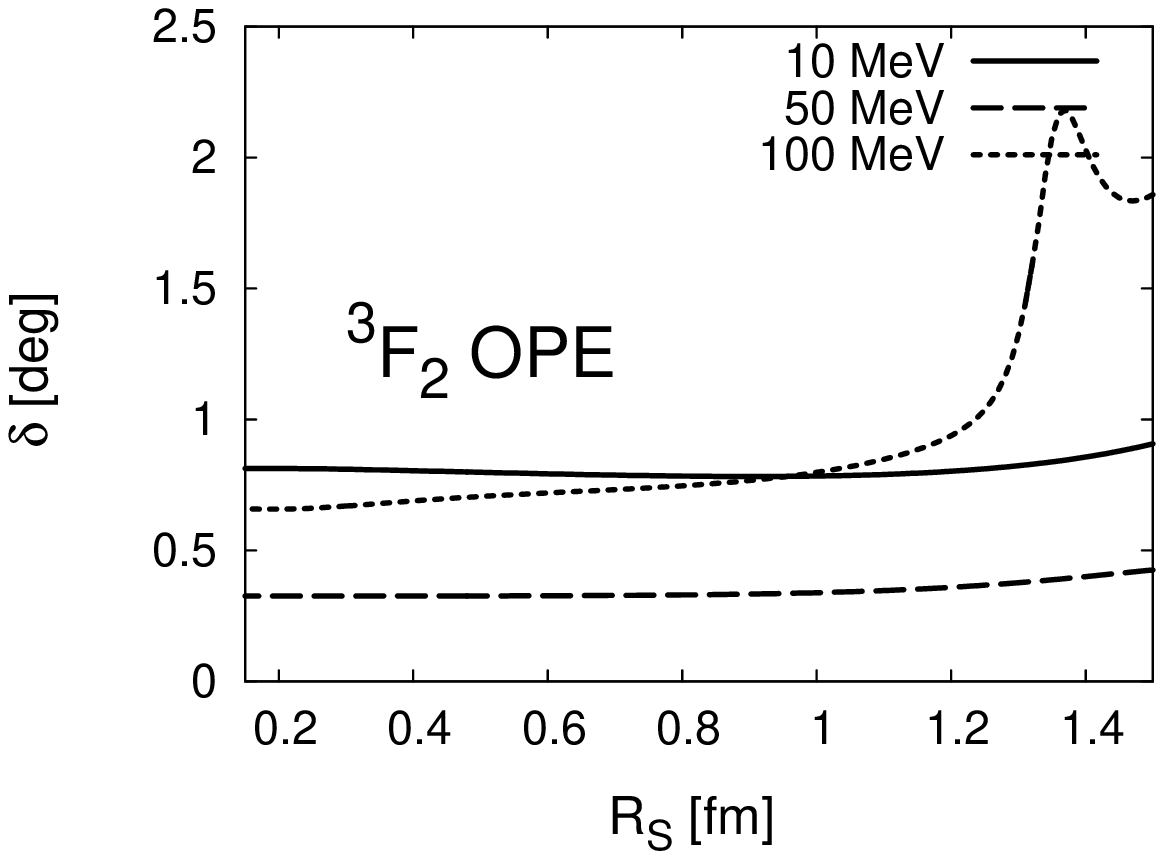,height=4cm,width=5cm}
\epsfig{figure=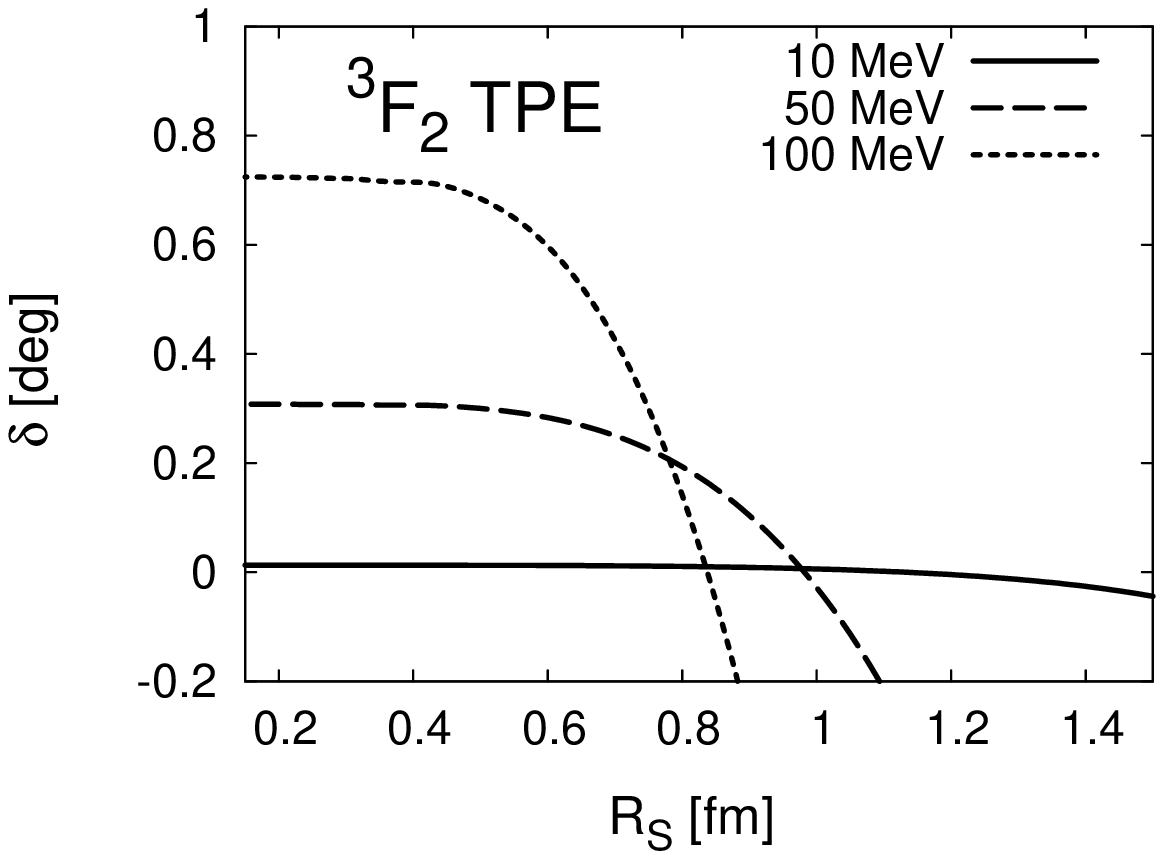,height=4cm,width=5cm}
\epsfig{figure=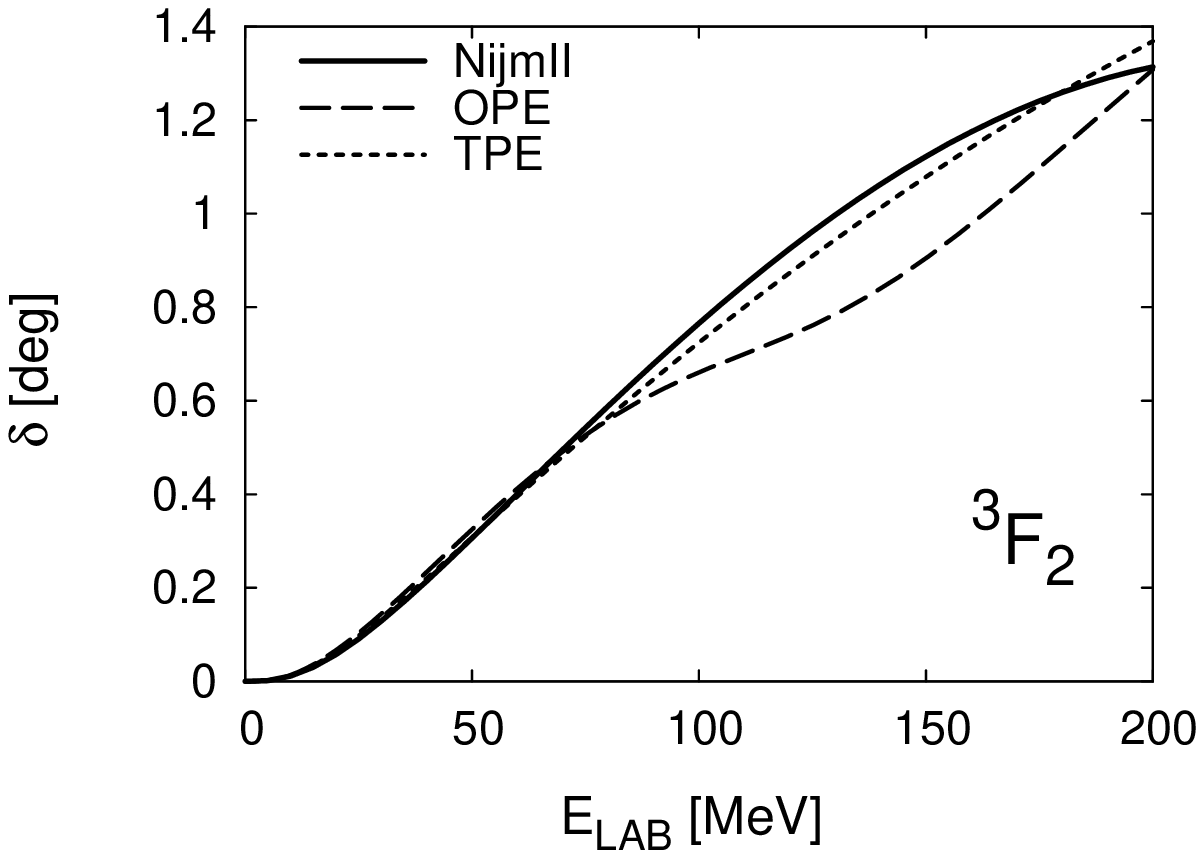,height=4cm,width=5cm} \\ 
\epsfig{figure=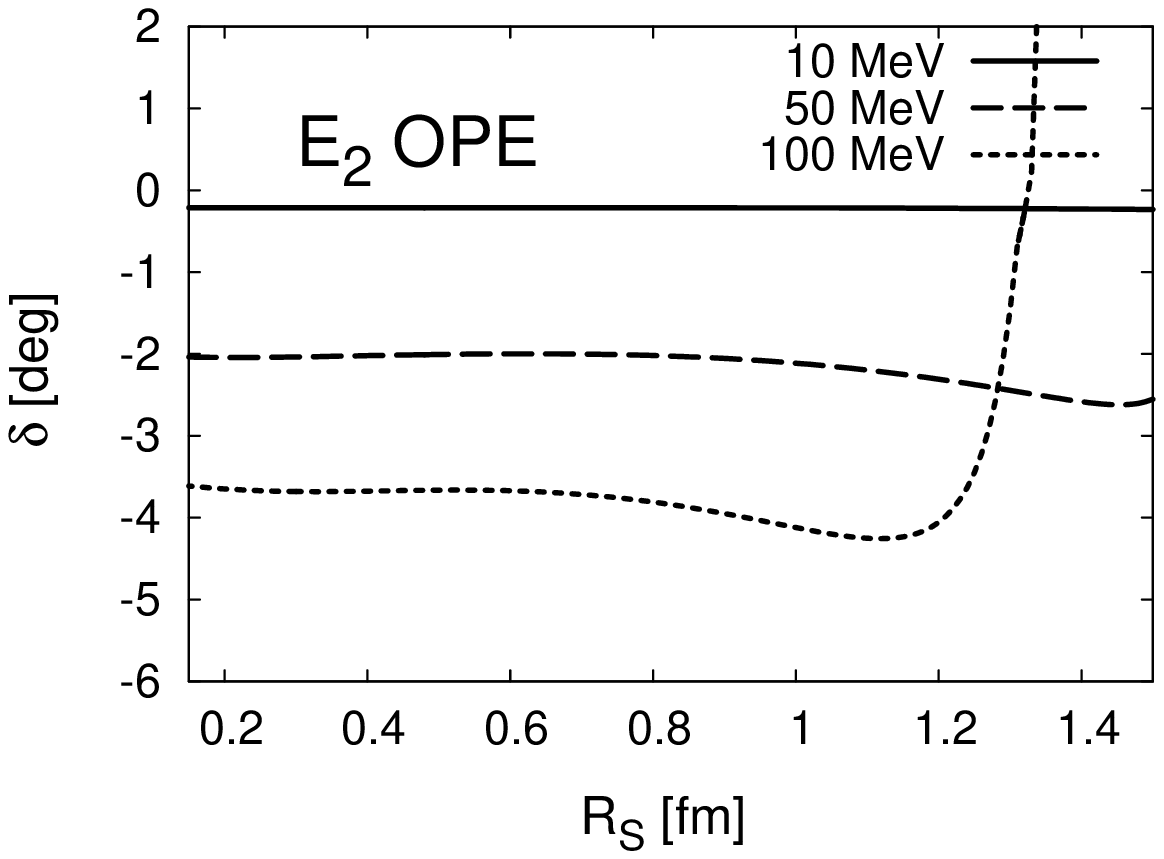,height=4cm,width=5cm}
\epsfig{figure=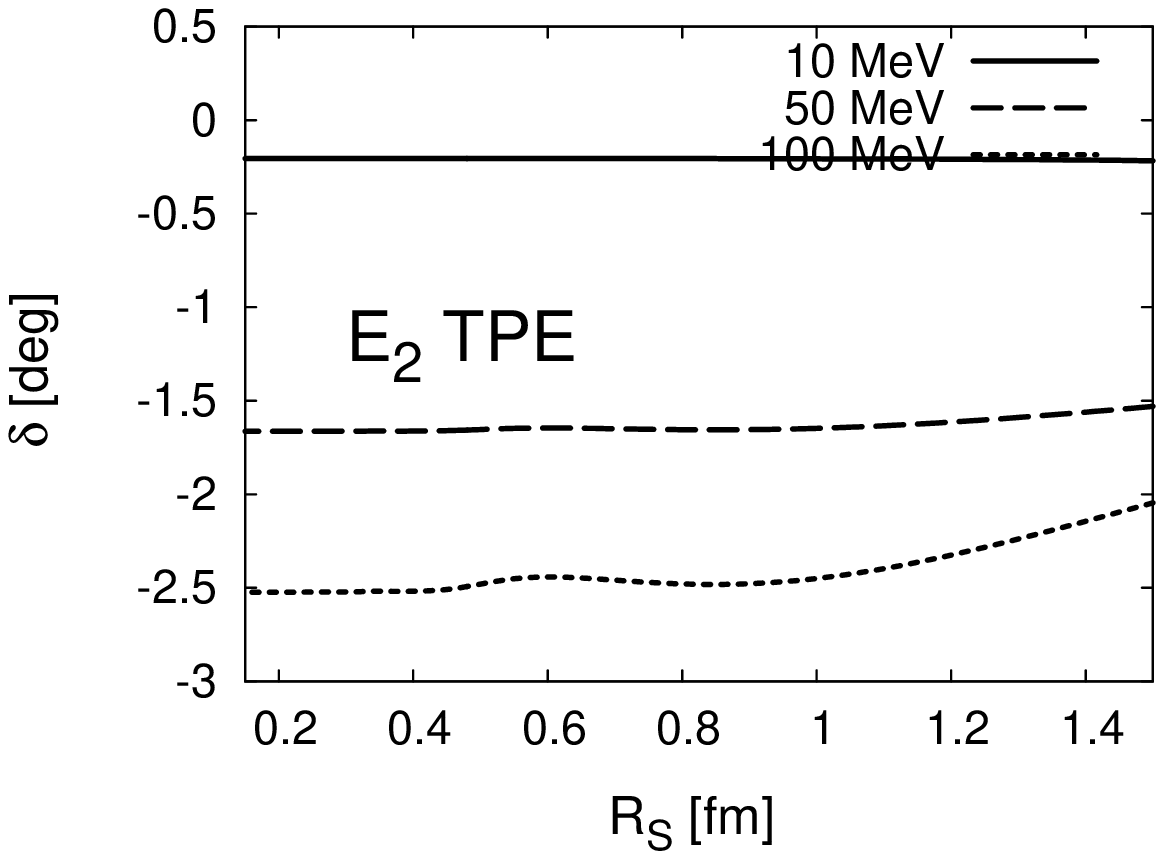,height=4cm,width=5cm}
\epsfig{figure=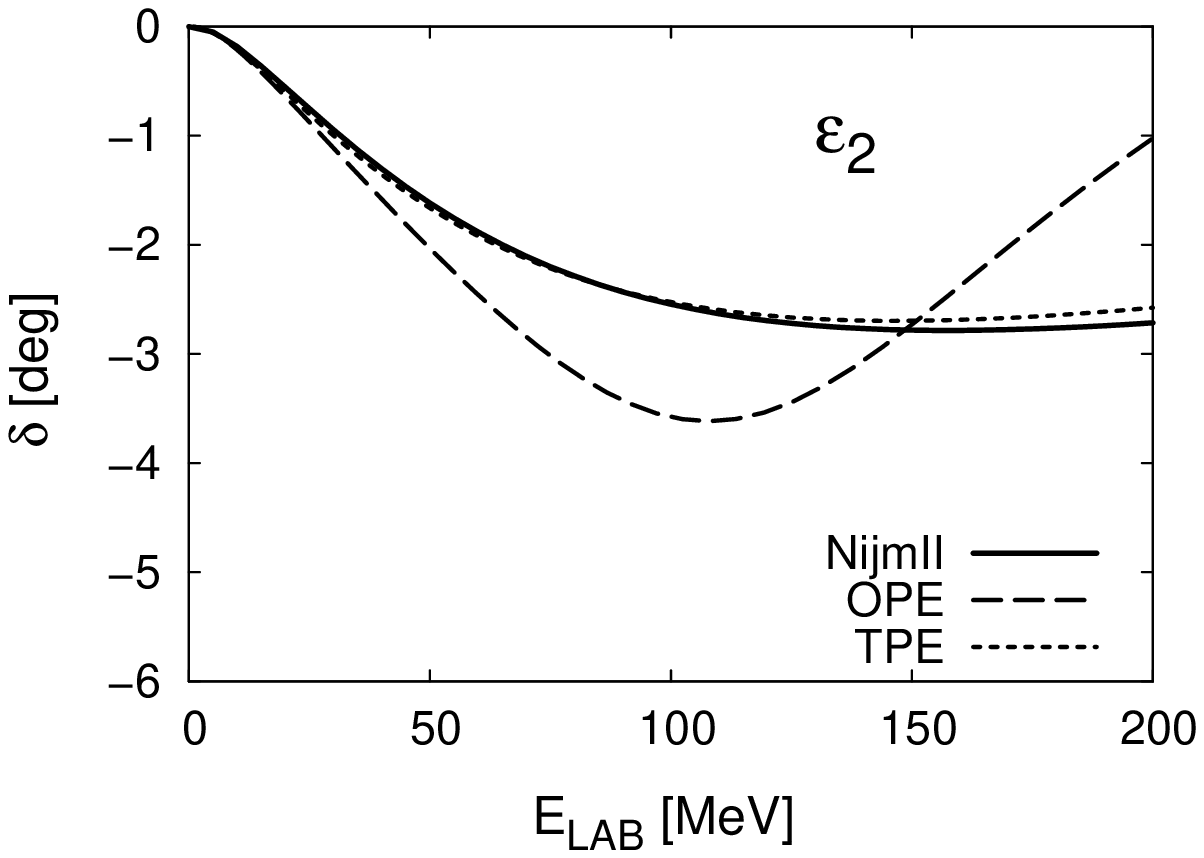,height=4cm,width=5cm} 
\end{center}
\caption{Same as Fig.~\ref{fig:fig-j=0} but for $j=2$.}
\label{fig:fig-j=2}
\end{figure*}

\begin{figure*}[]
\begin{center}
\epsfig{figure=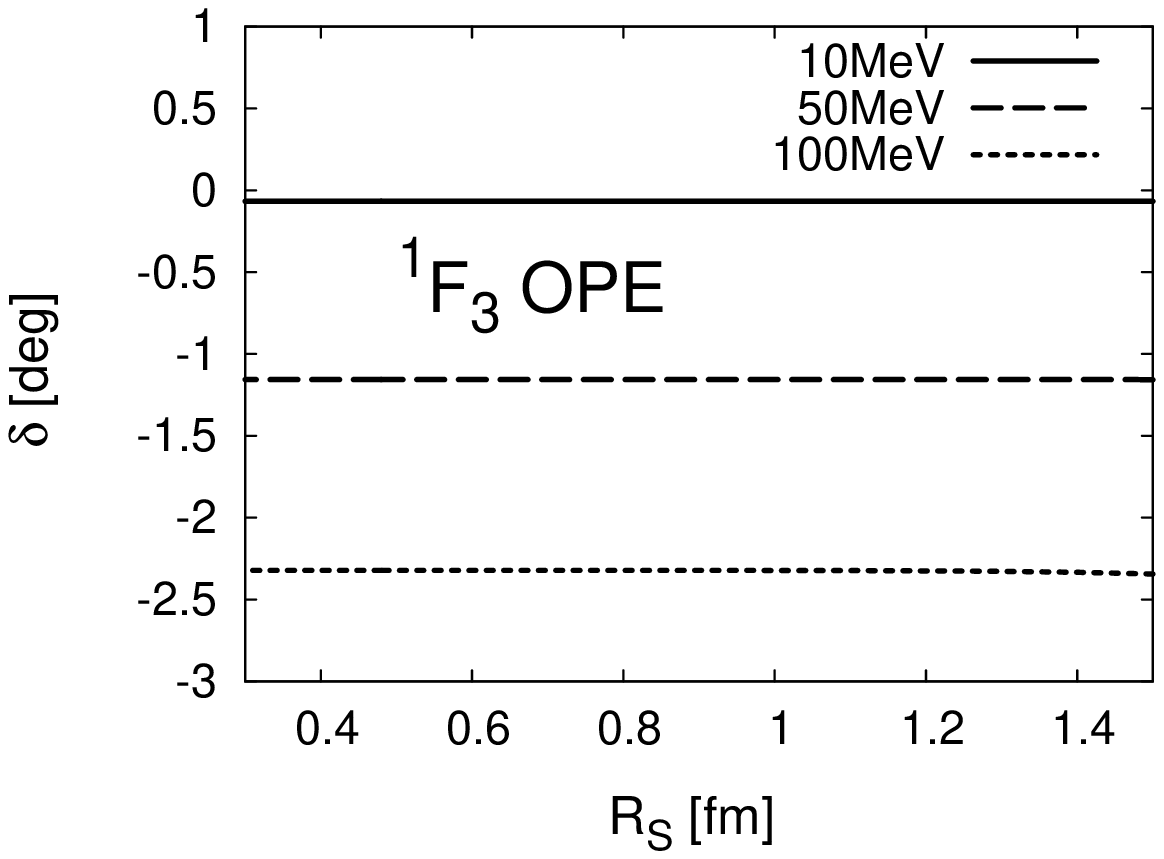,height=4cm,width=5cm}
\epsfig{figure=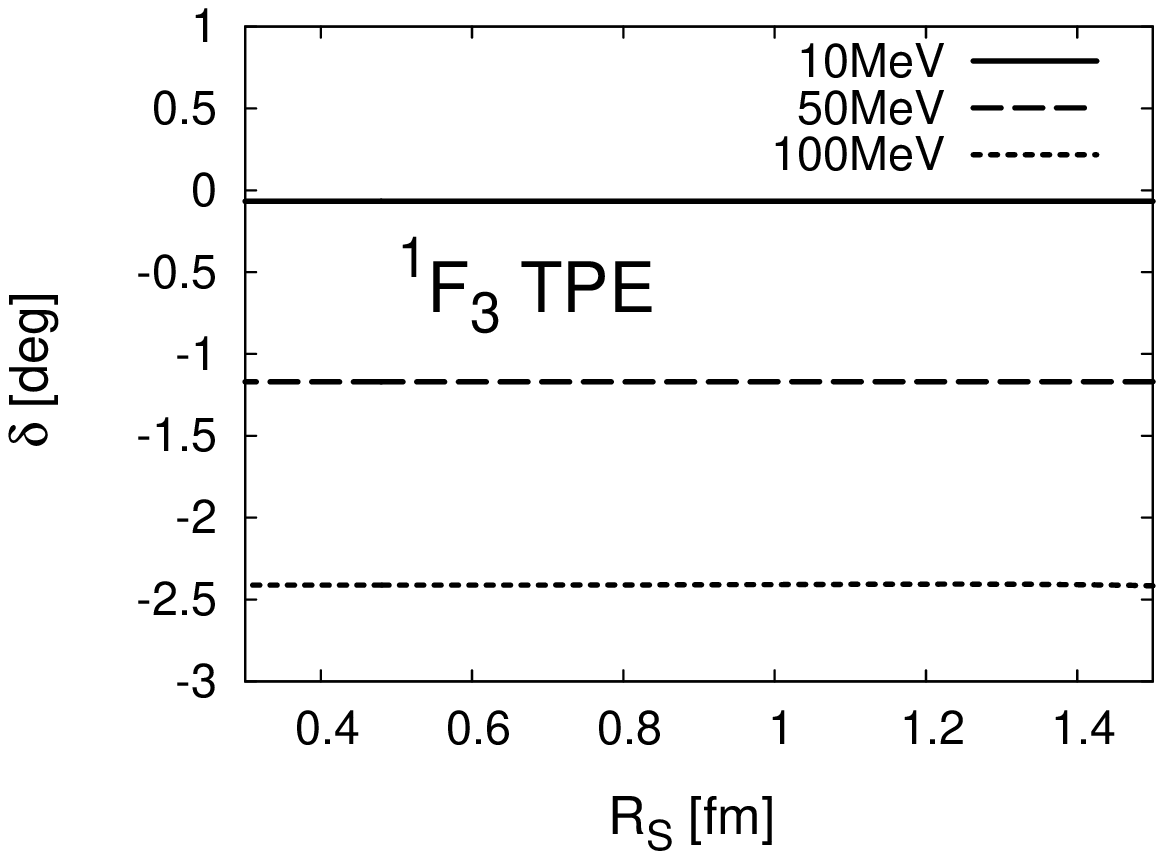,height=4cm,width=5cm}
\epsfig{figure=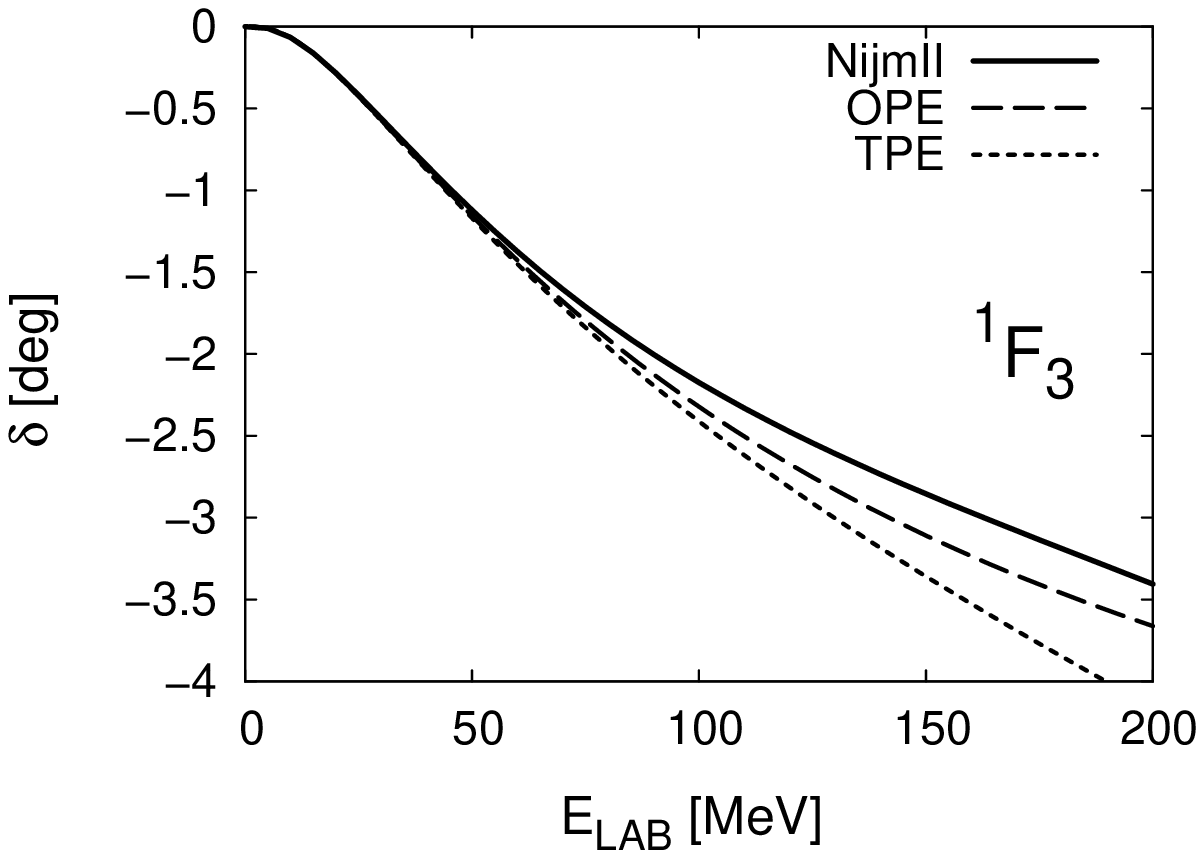,height=4cm,width=5cm} \\ 
\epsfig{figure=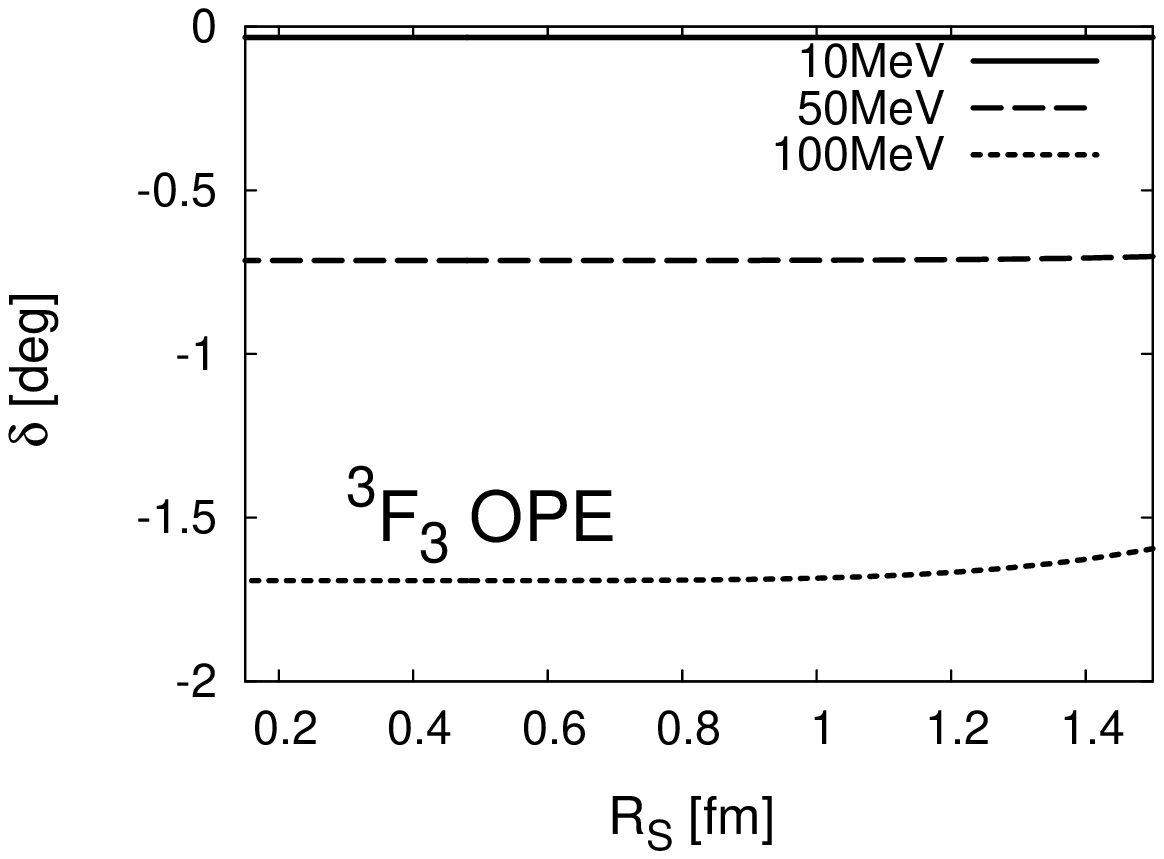,height=4cm,width=5cm}
\epsfig{figure=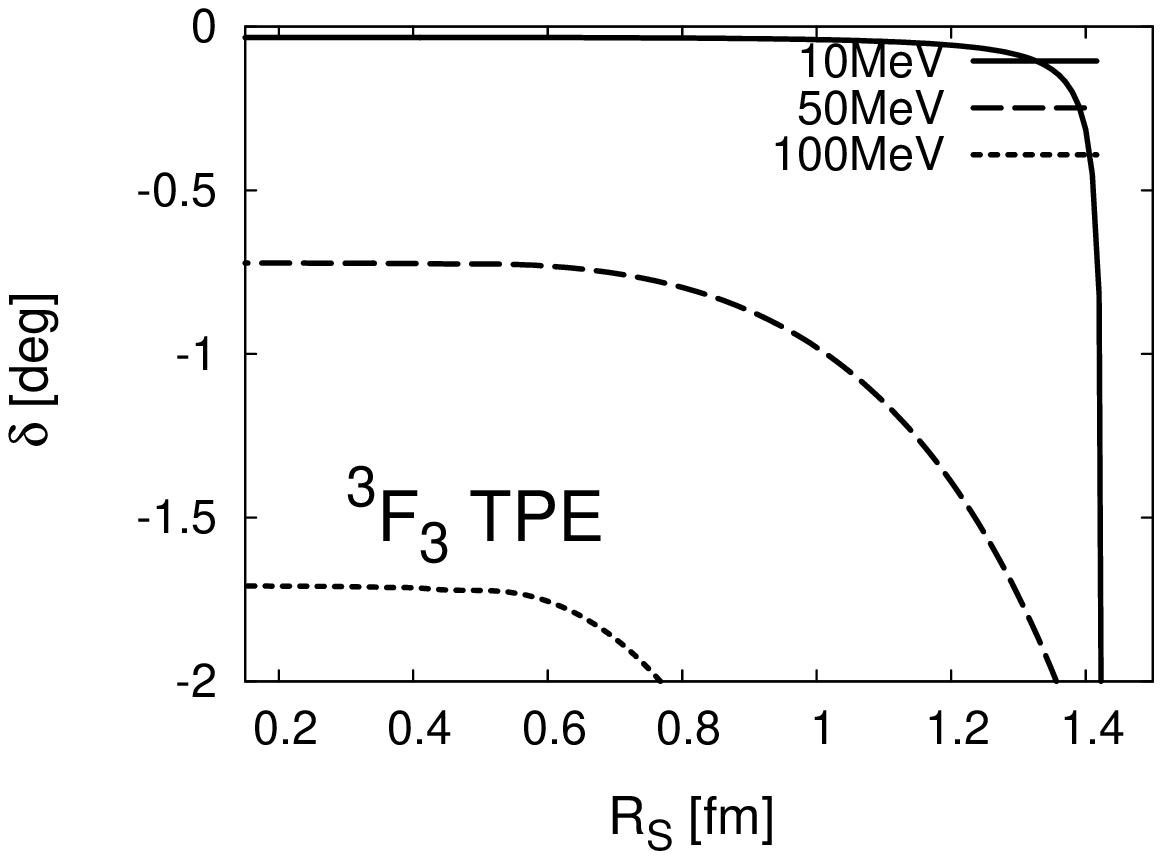,height=4cm,width=5cm}
\epsfig{figure=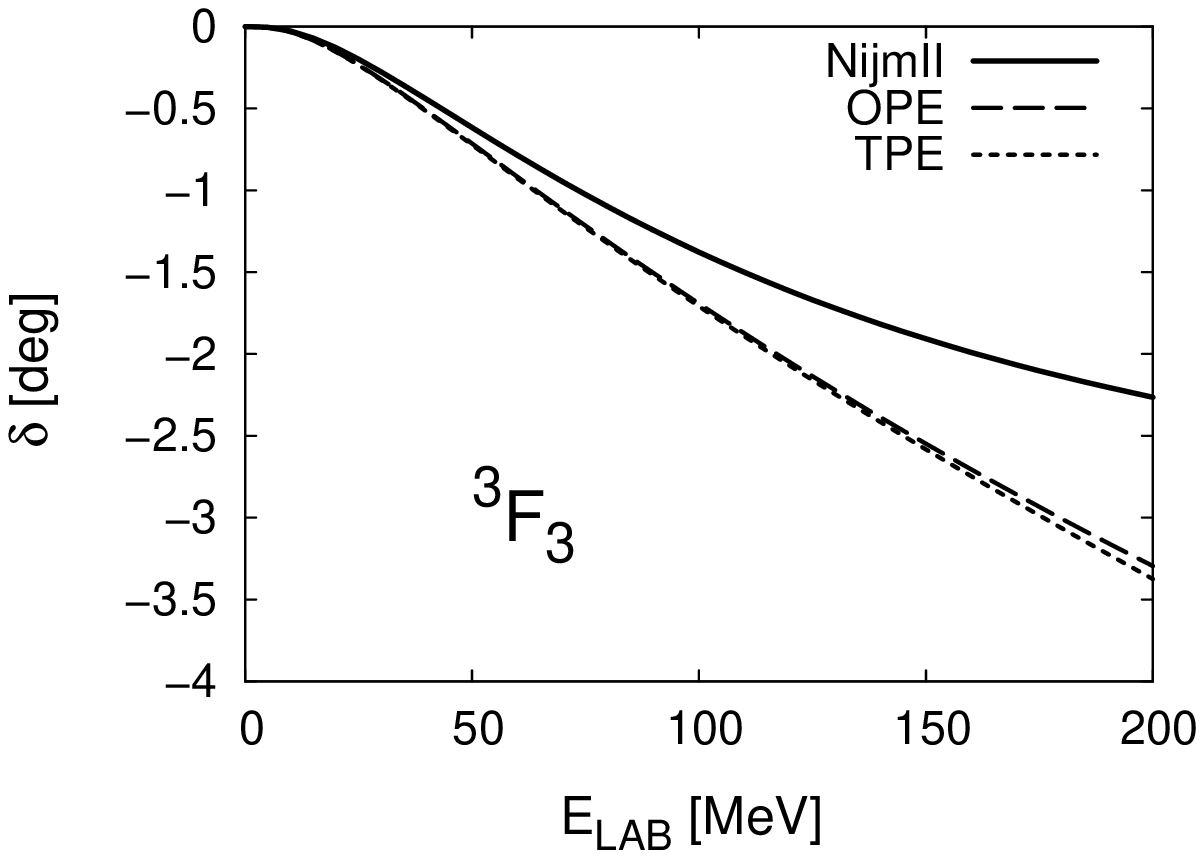,height=4cm,width=5cm} \\
\epsfig{figure=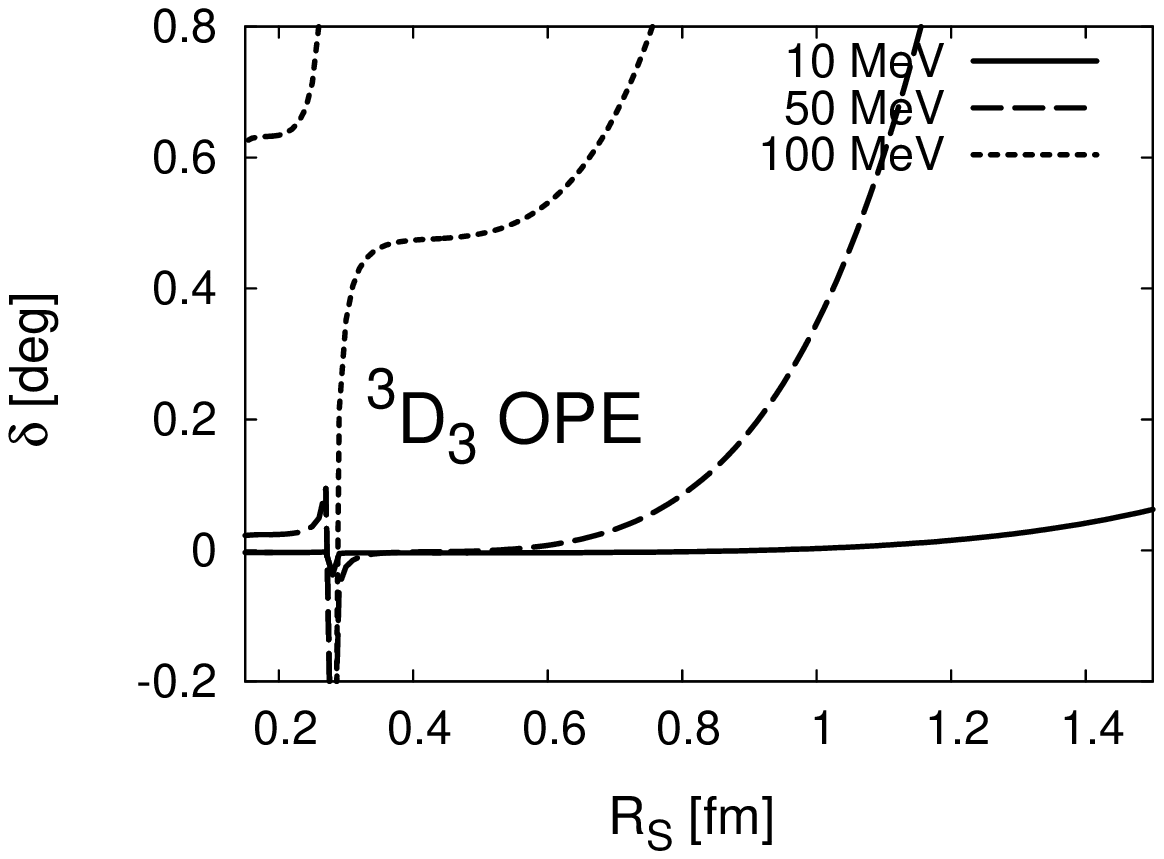,height=4cm,width=5cm}
\epsfig{figure=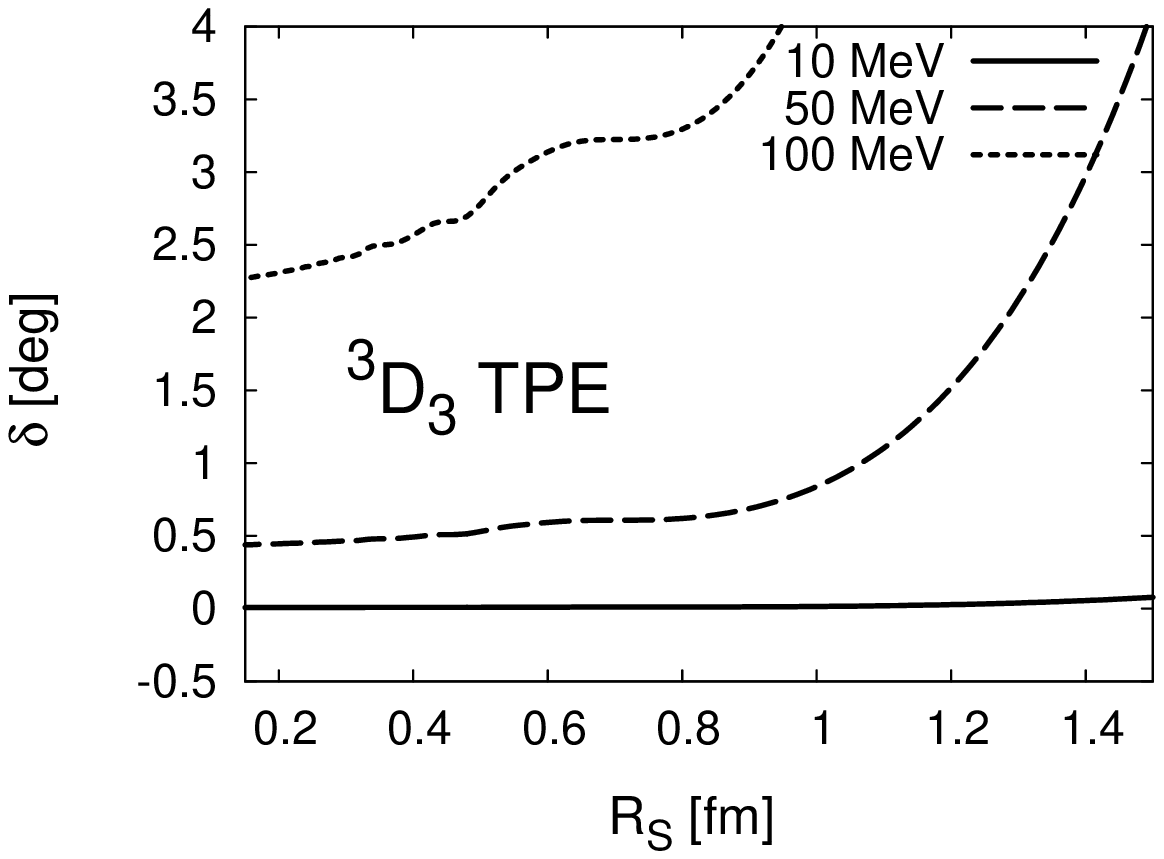,height=4cm,width=5cm}
\epsfig{figure=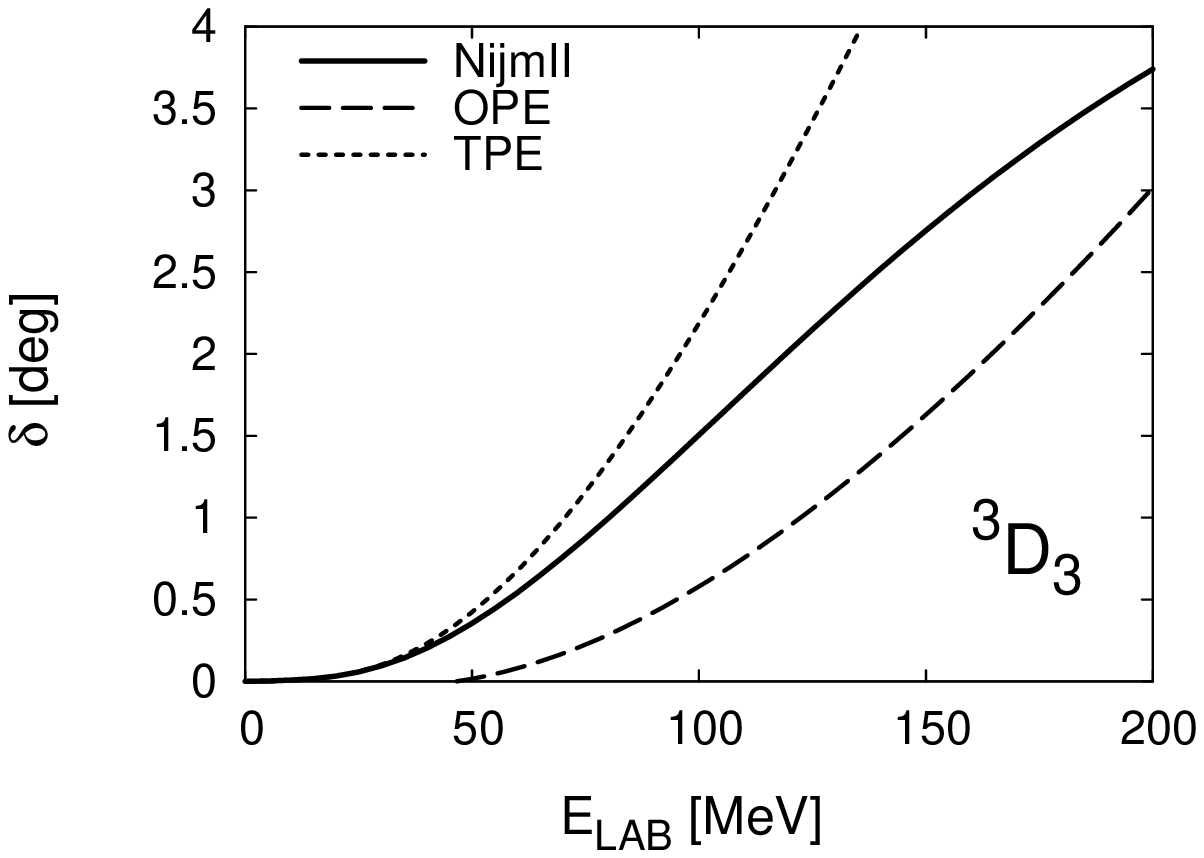,height=4cm,width=5cm} \\ 
\epsfig{figure=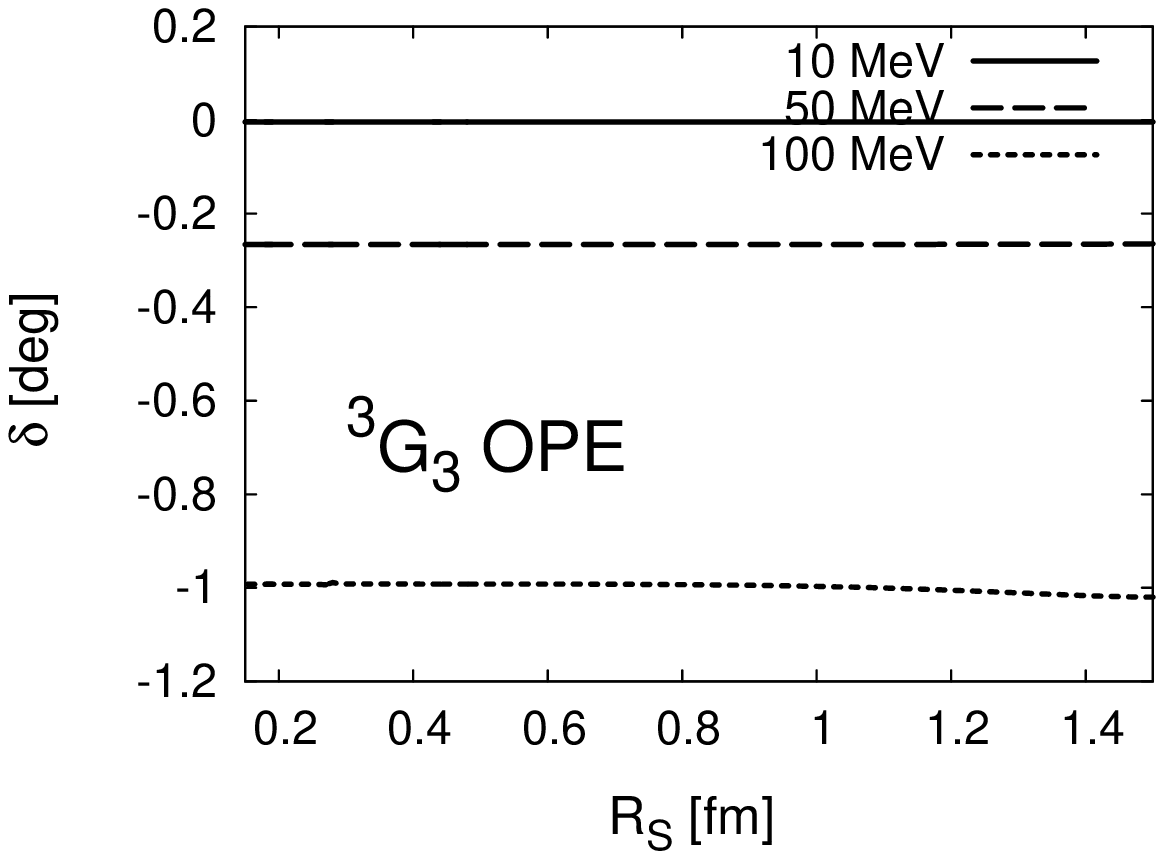,height=4cm,width=5cm}
\epsfig{figure=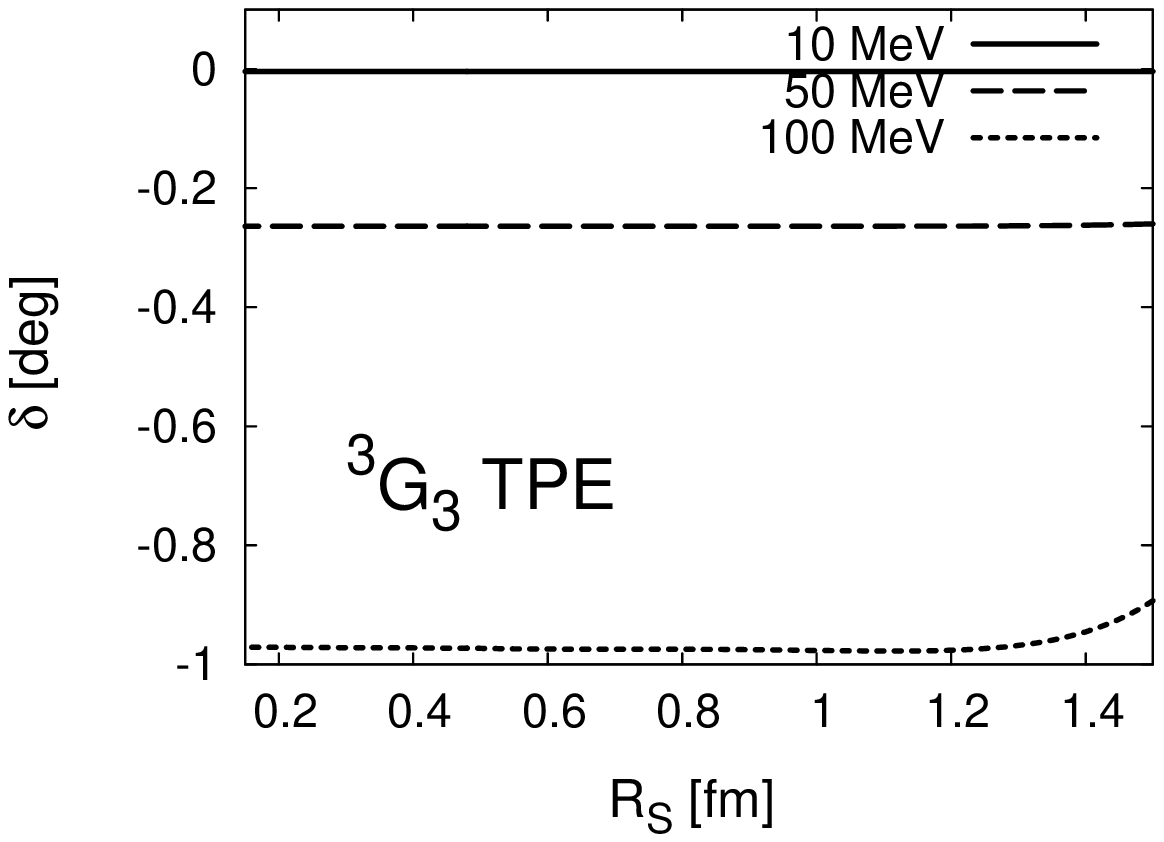,height=4cm,width=5cm}
\epsfig{figure=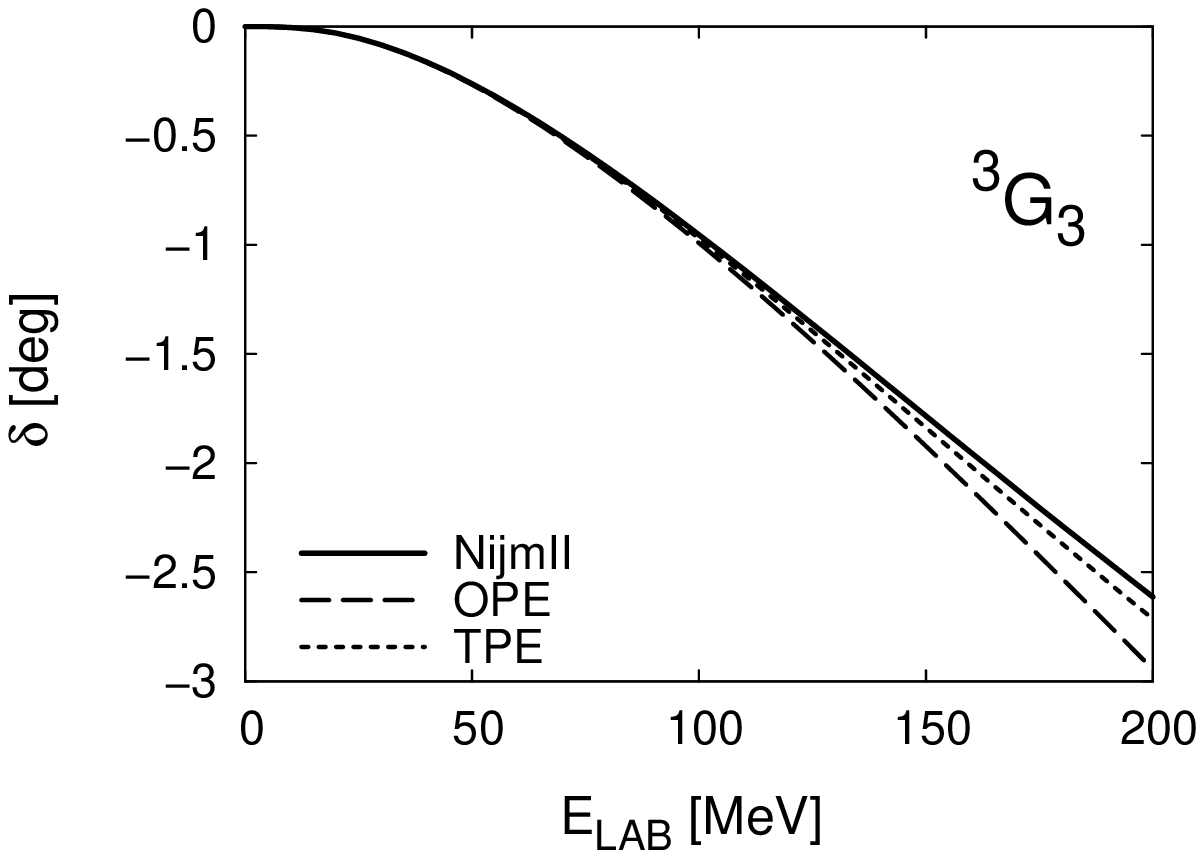,height=4cm,width=5cm} \\ 
\epsfig{figure=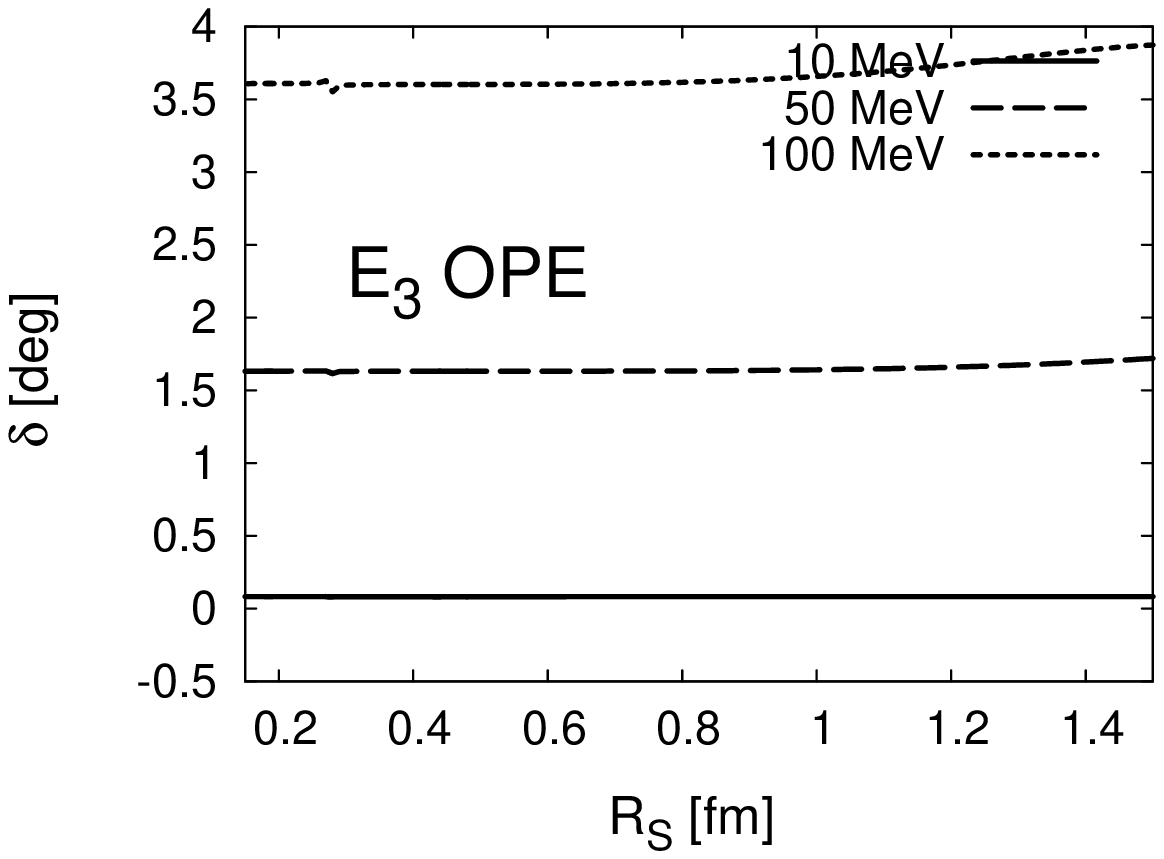,height=4cm,width=5cm}
\epsfig{figure=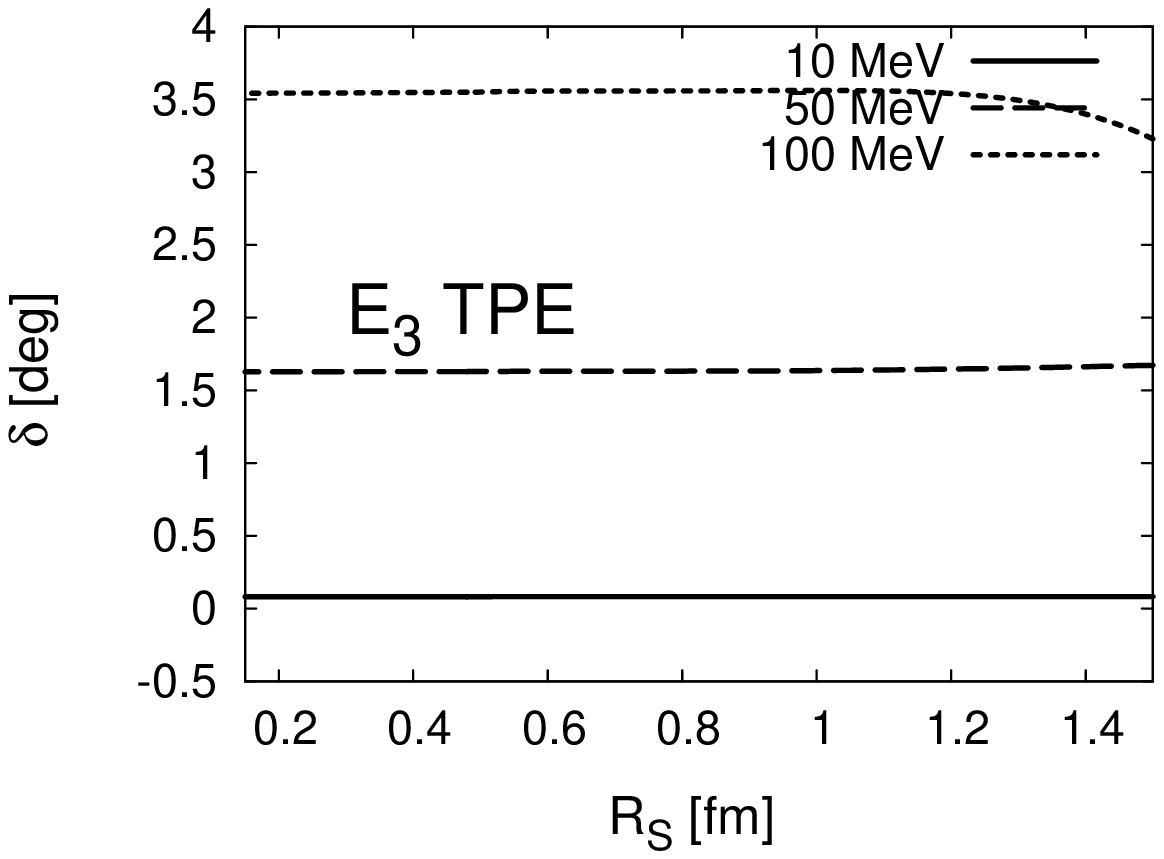,height=4cm,width=5cm}
\epsfig{figure=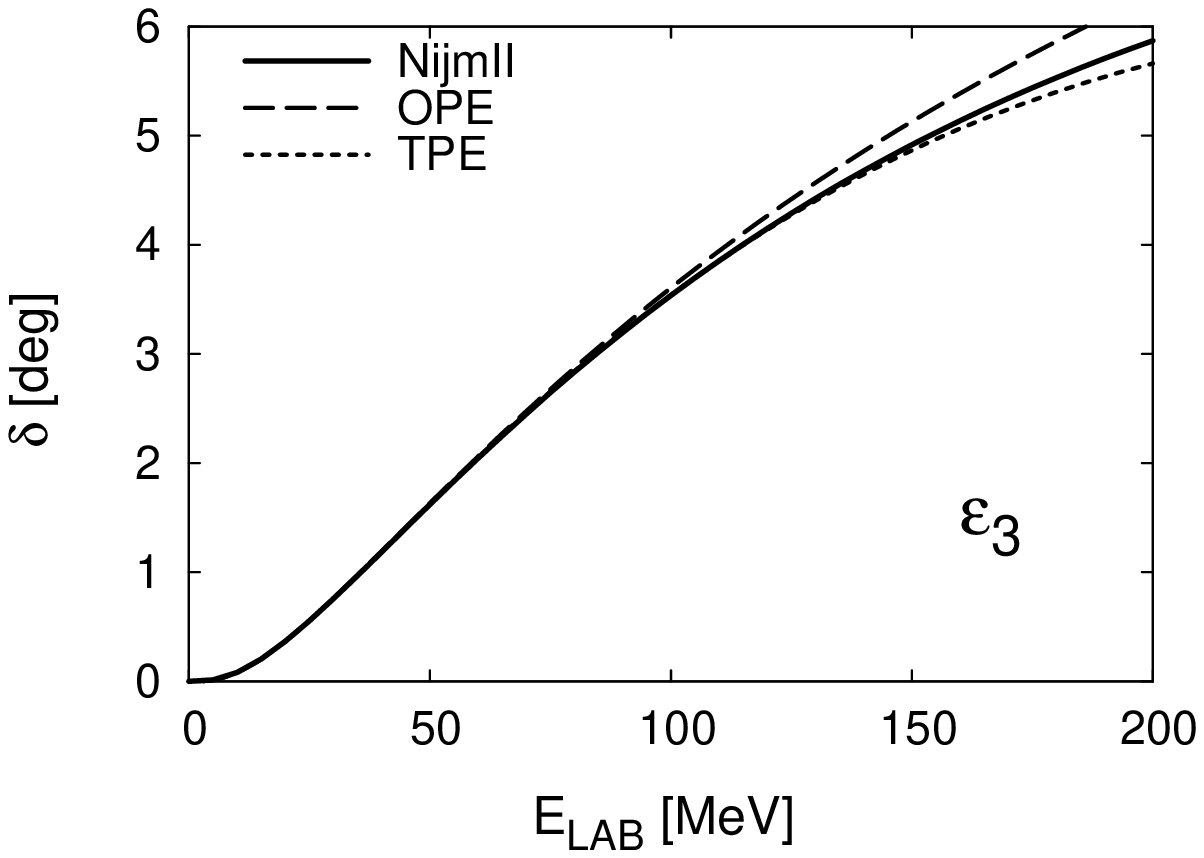,height=4cm,width=5cm} 
\end{center}
\caption{Same as Fig.~\ref{fig:fig-j=0} but for $j=3$.}
\label{fig:fig-j=3}
\end{figure*}

\begin{figure*}[]
\begin{center}
\epsfig{figure=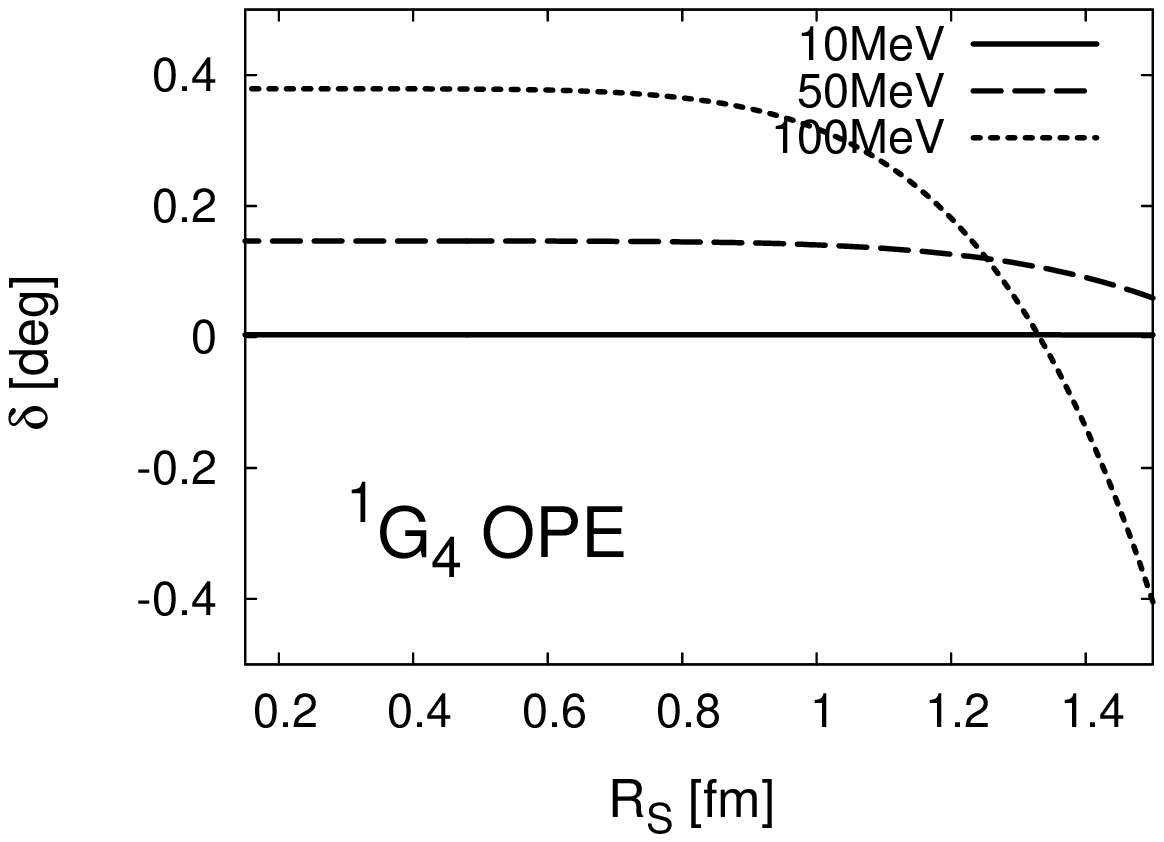,height=4cm,width=5cm}
\epsfig{figure=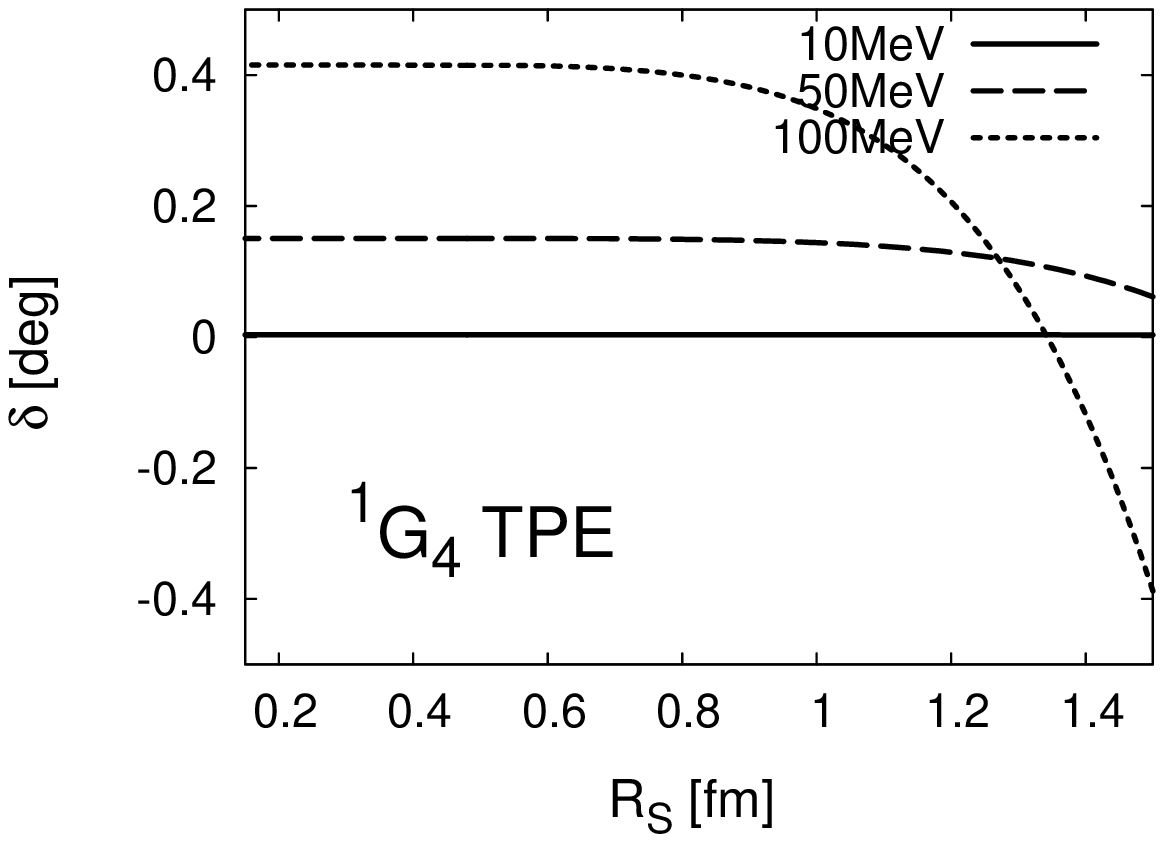,height=4cm,width=5cm}
\epsfig{figure=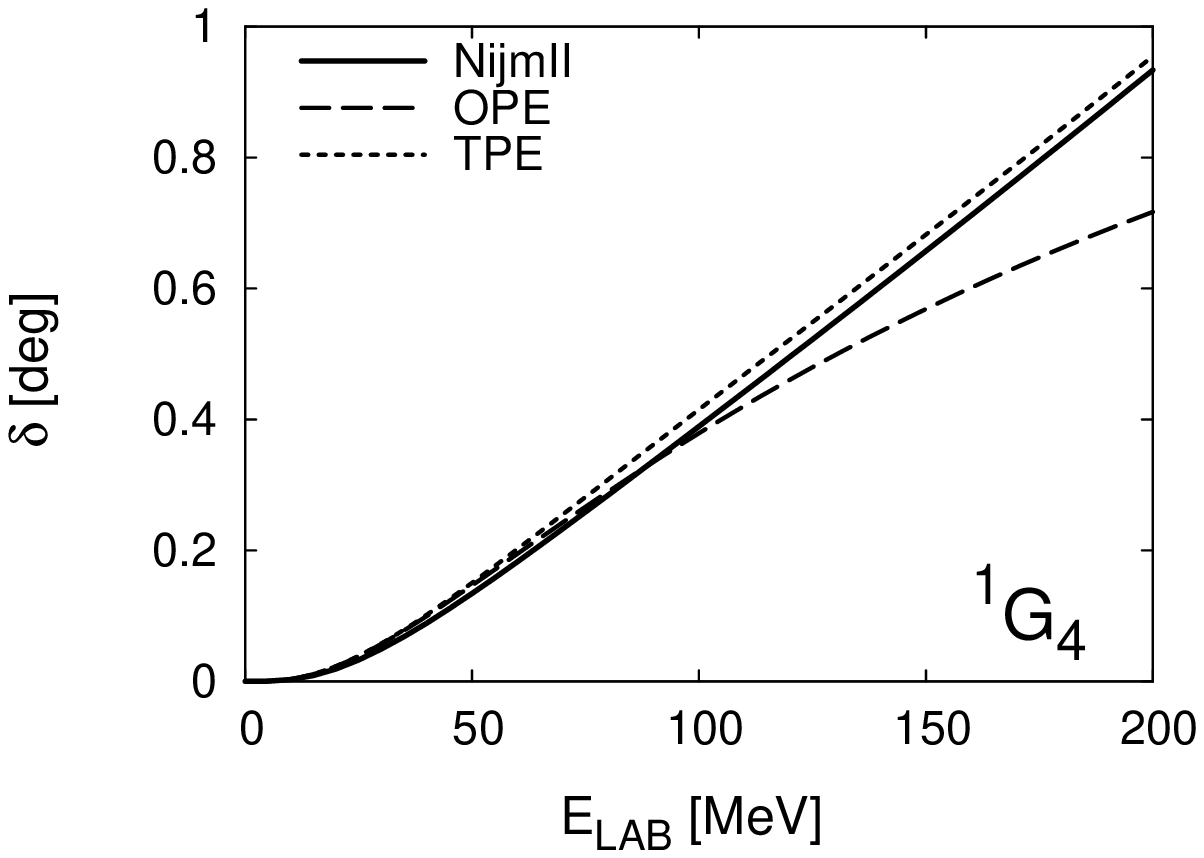,height=4cm,width=5cm} \\ 
\epsfig{figure=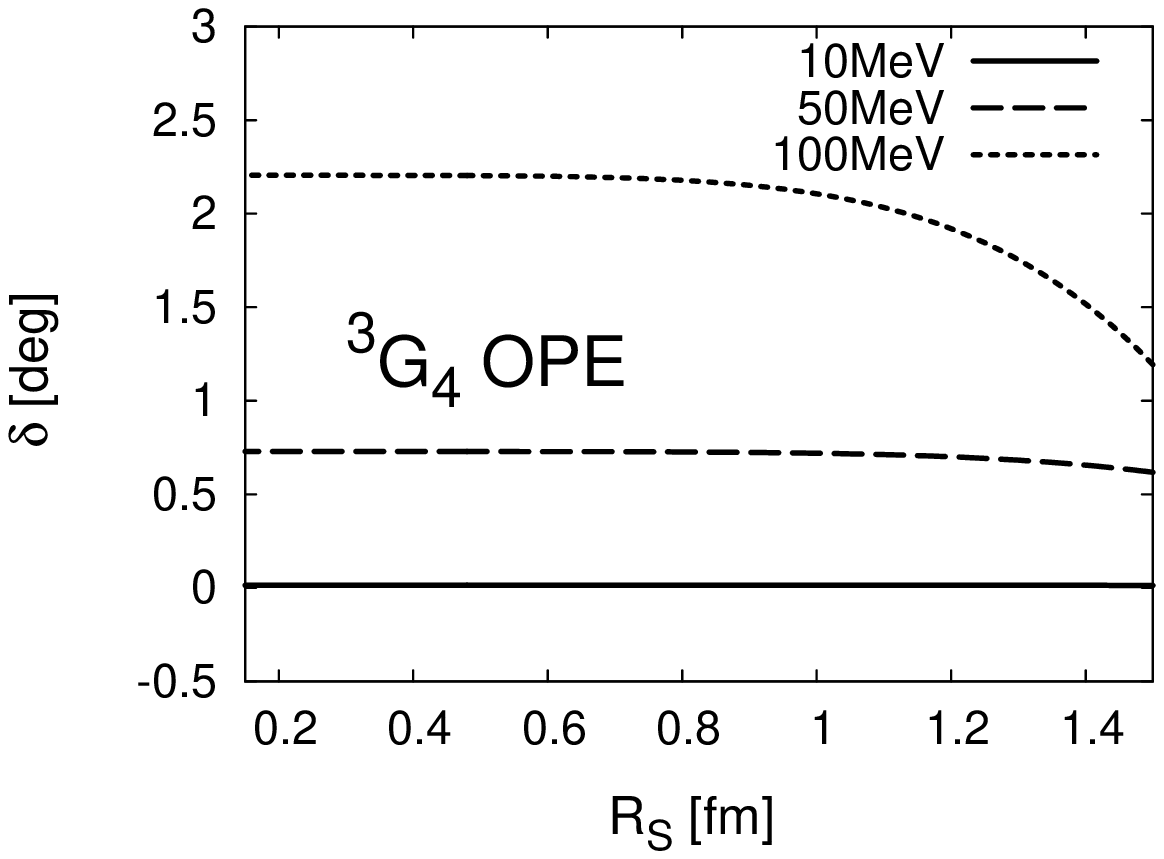,height=4cm,width=5cm}
\epsfig{figure=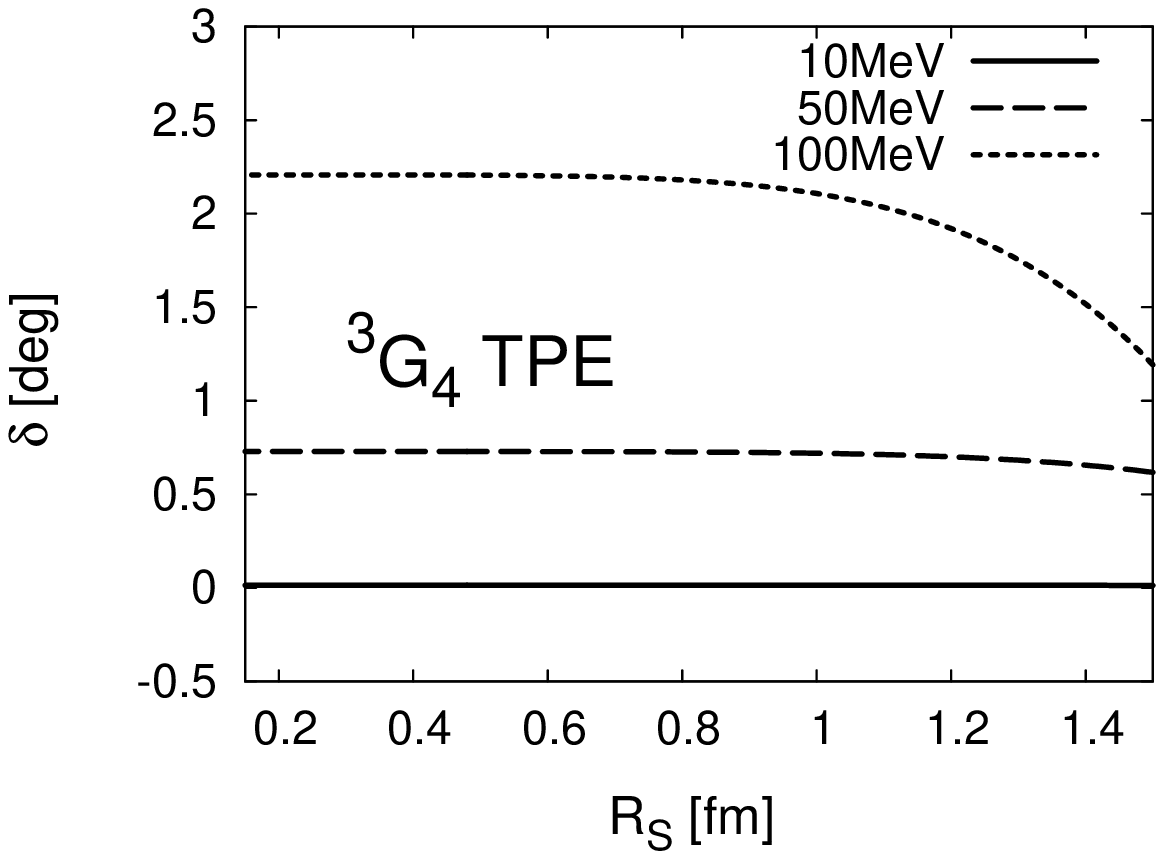,height=4cm,width=5cm}
\epsfig{figure=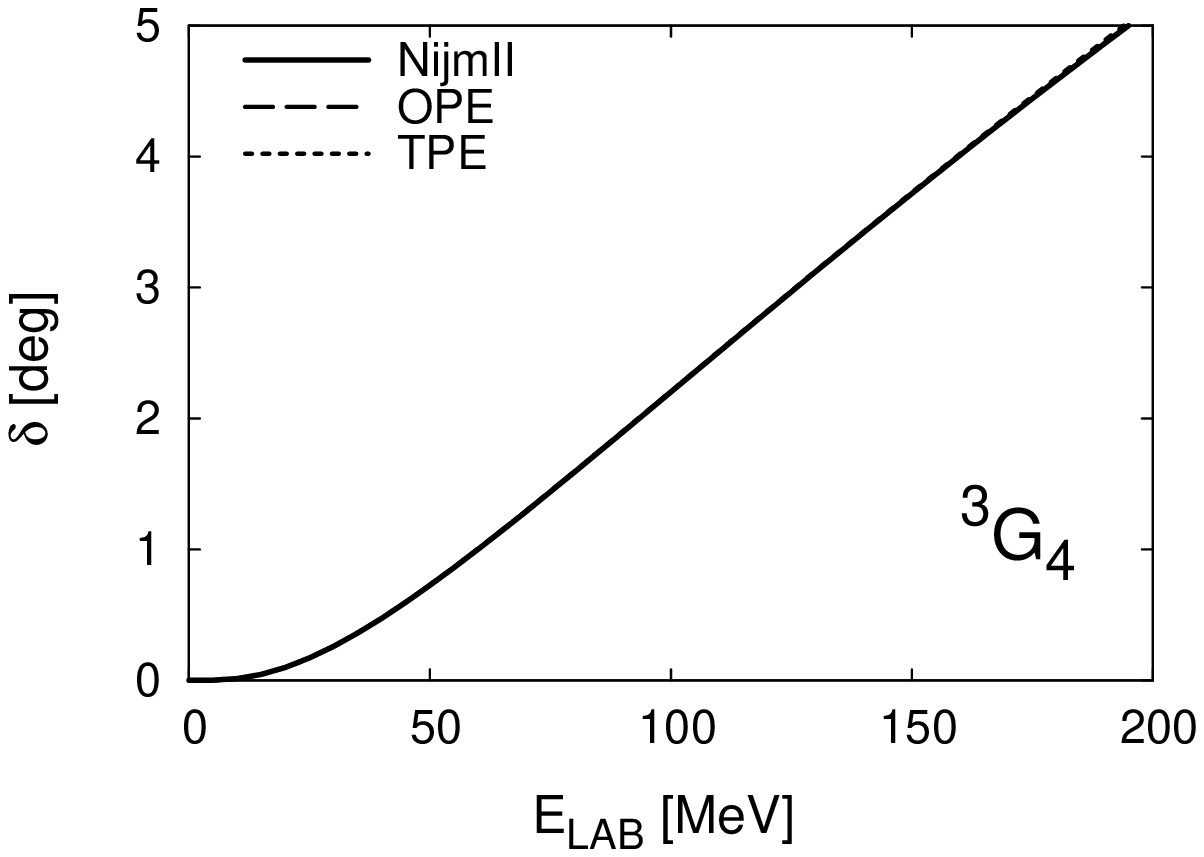,height=4cm,width=5cm} \\
\epsfig{figure=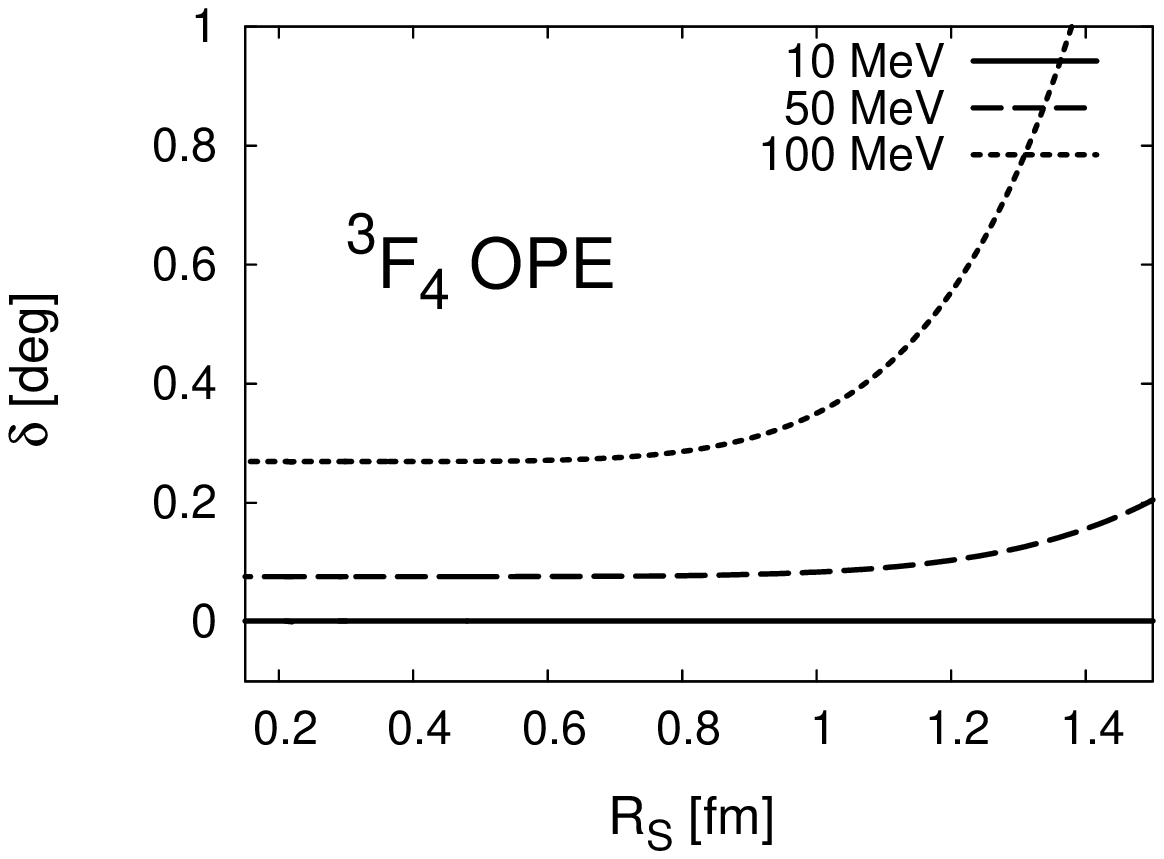,height=4cm,width=5cm}
\epsfig{figure=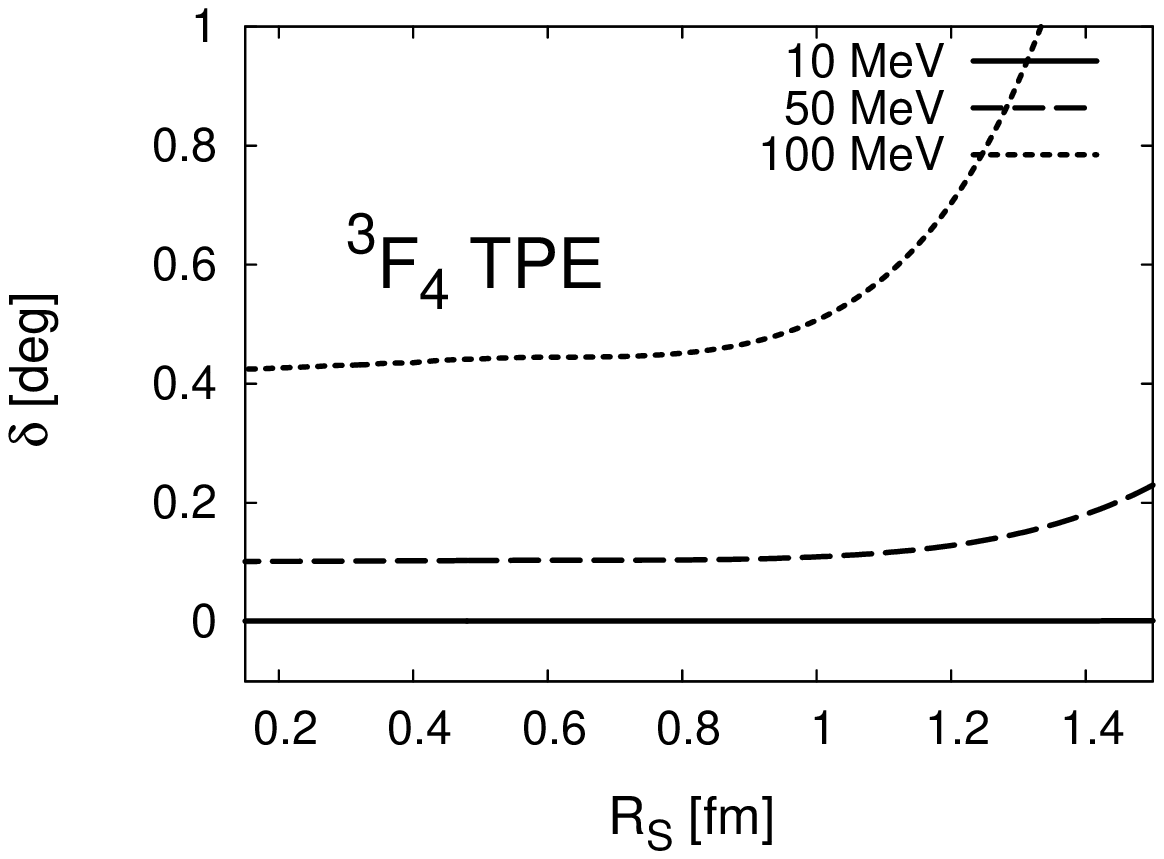,height=4cm,width=5cm}
\epsfig{figure=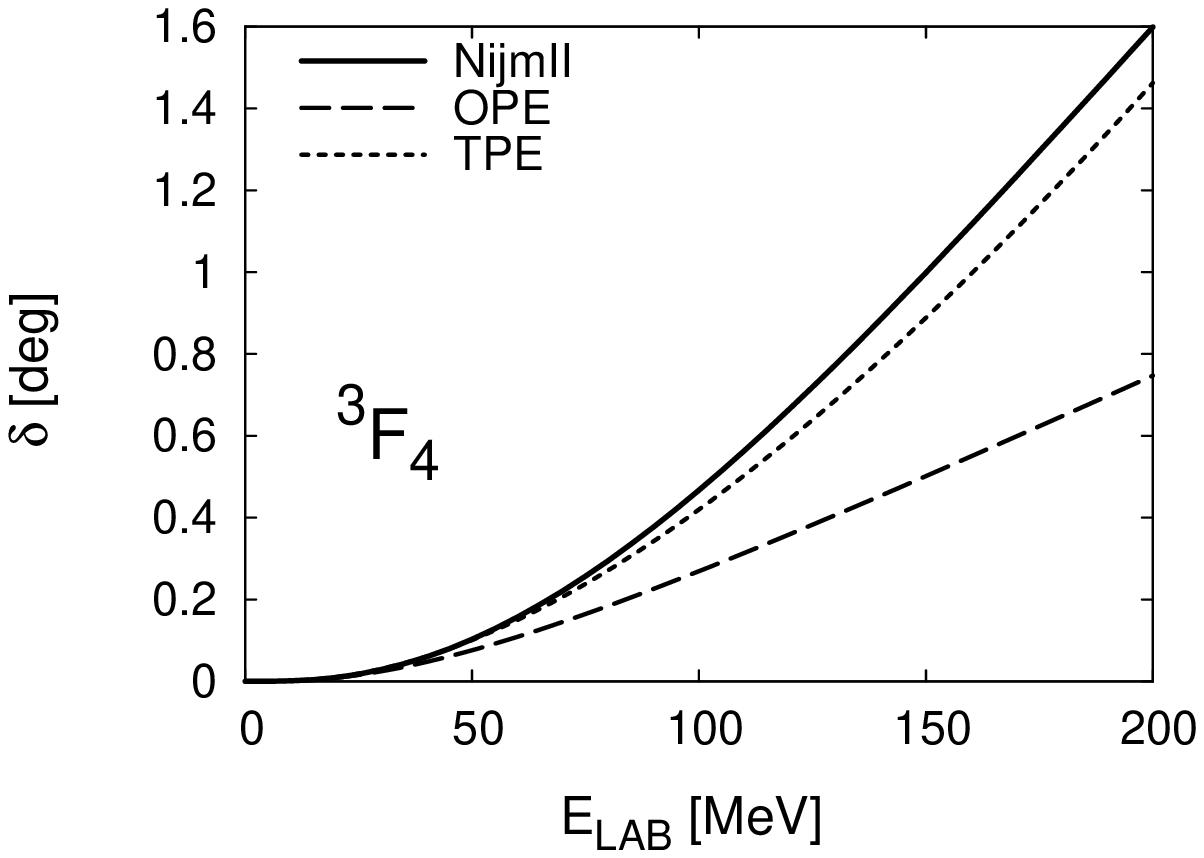,height=4cm,width=5cm} \\ 
\epsfig{figure=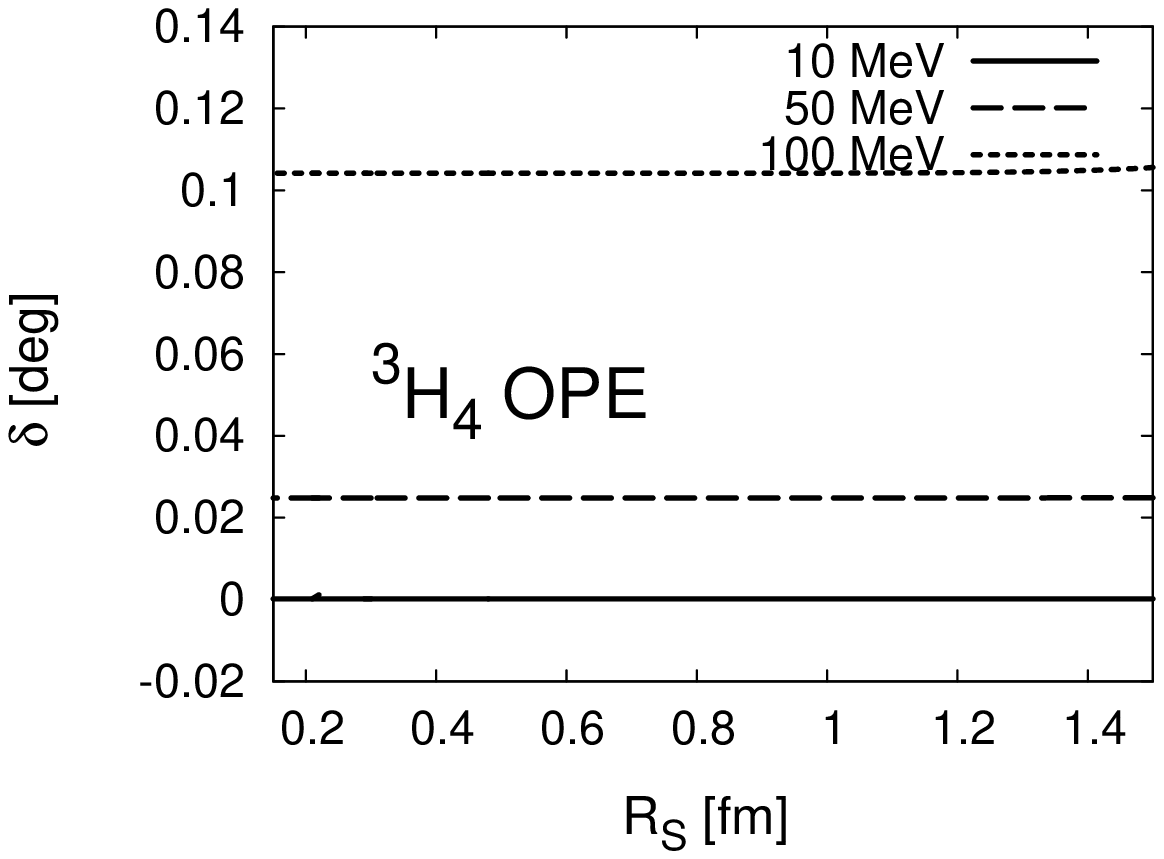,height=4cm,width=5cm}
\epsfig{figure=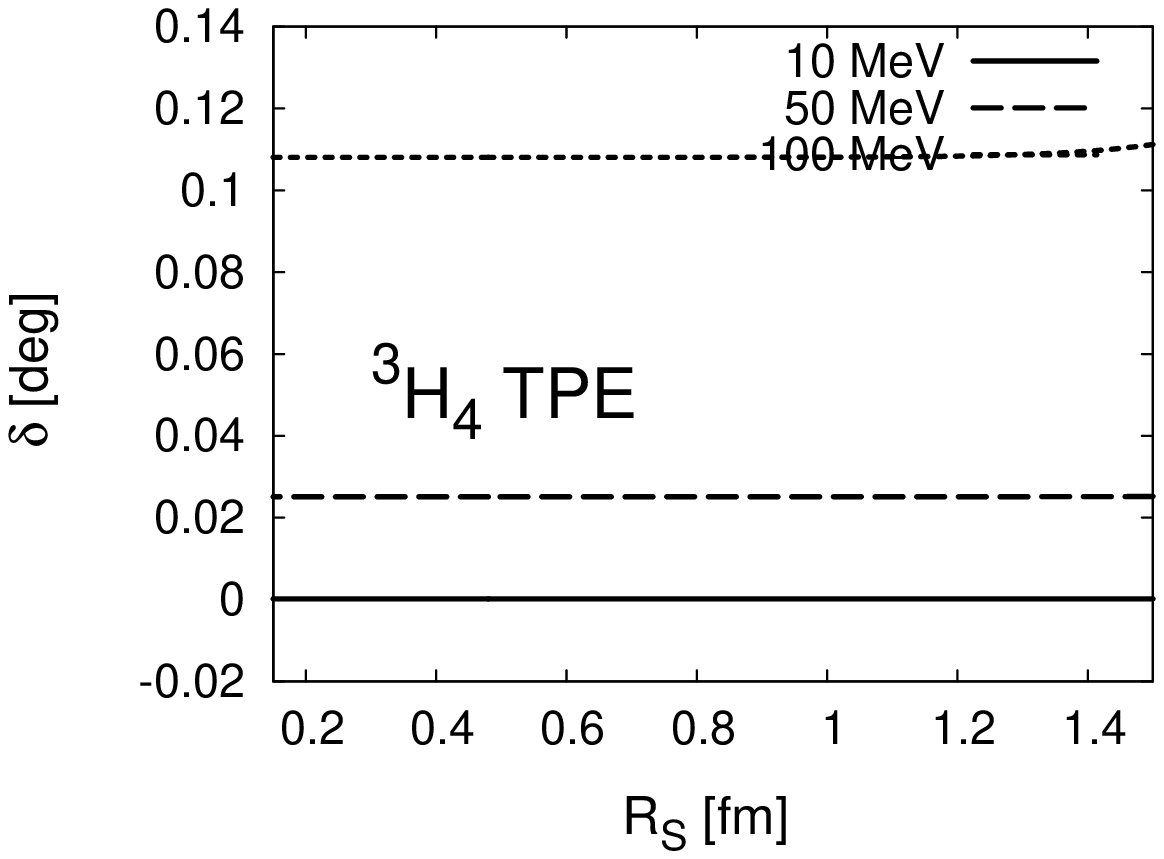,height=4cm,width=5cm}
\epsfig{figure=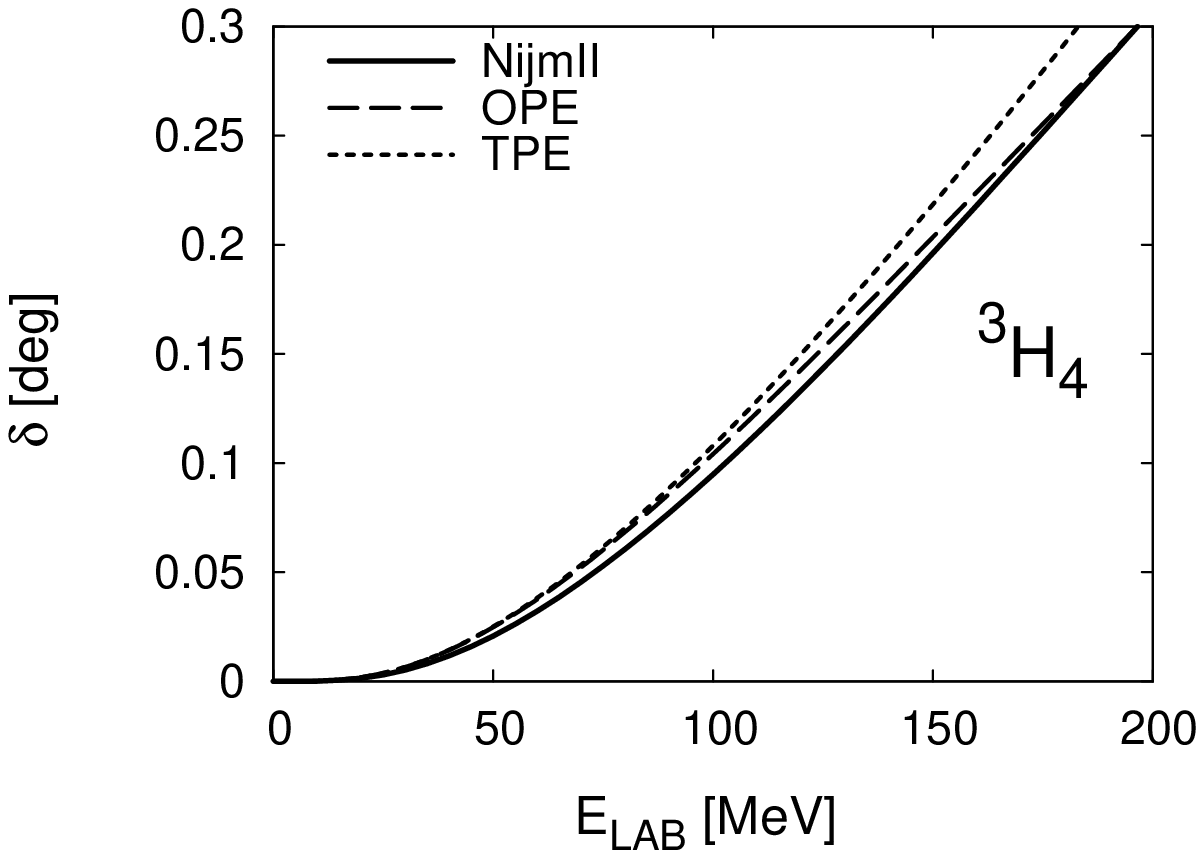,height=4cm,width=5cm} \\ 
\epsfig{figure=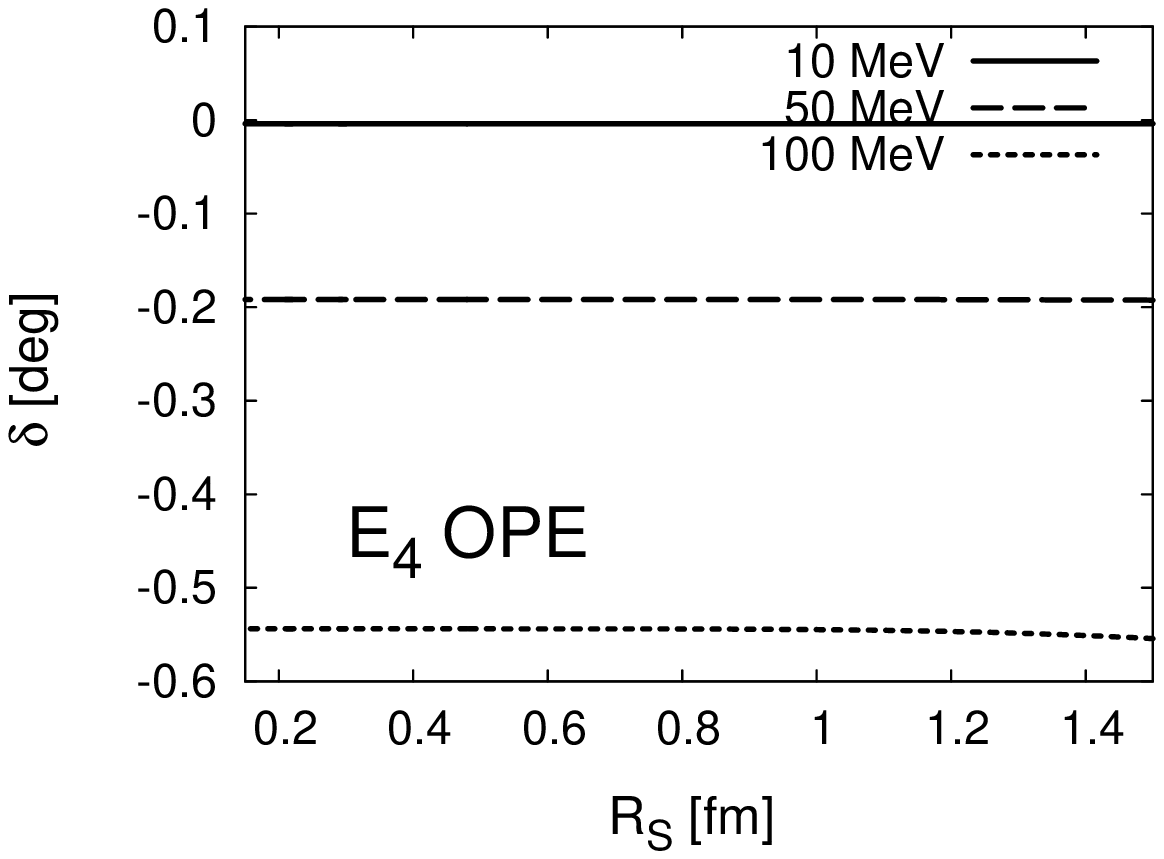,height=4cm,width=5cm}
\epsfig{figure=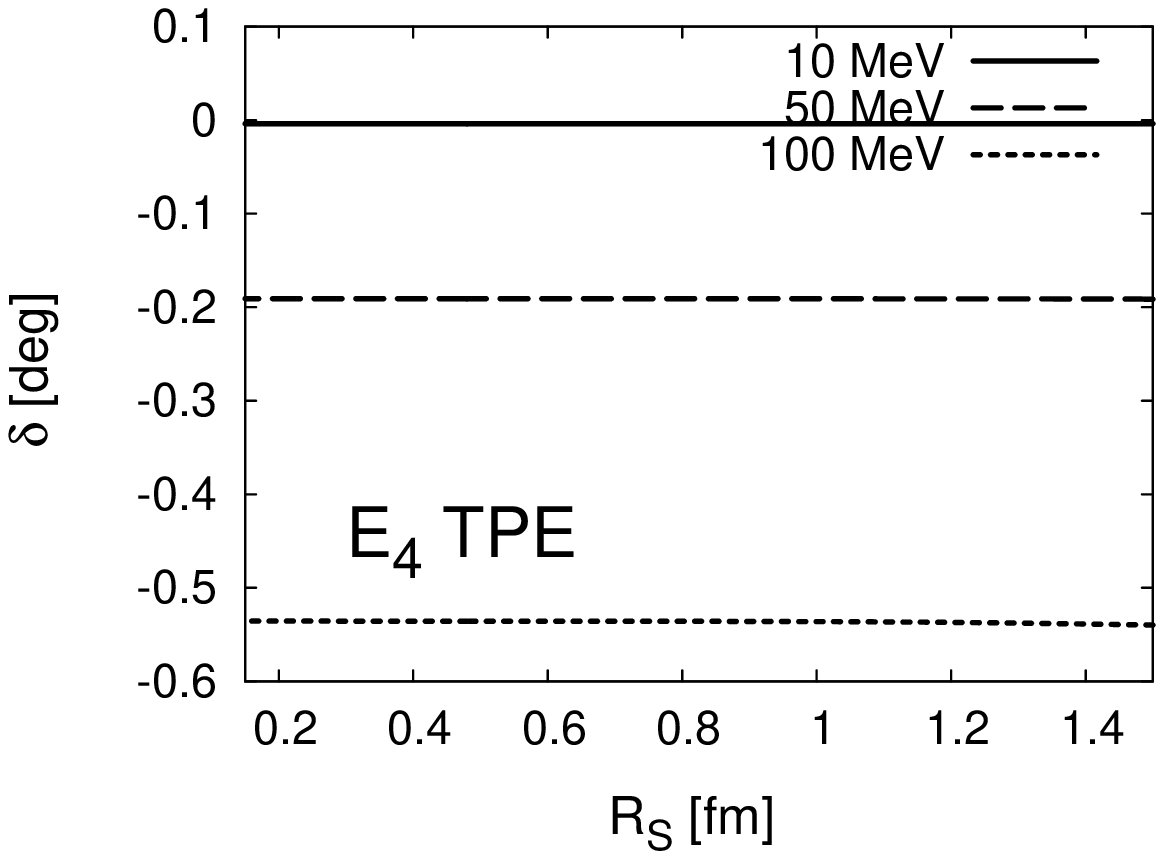,height=4cm,width=5cm}
\epsfig{figure=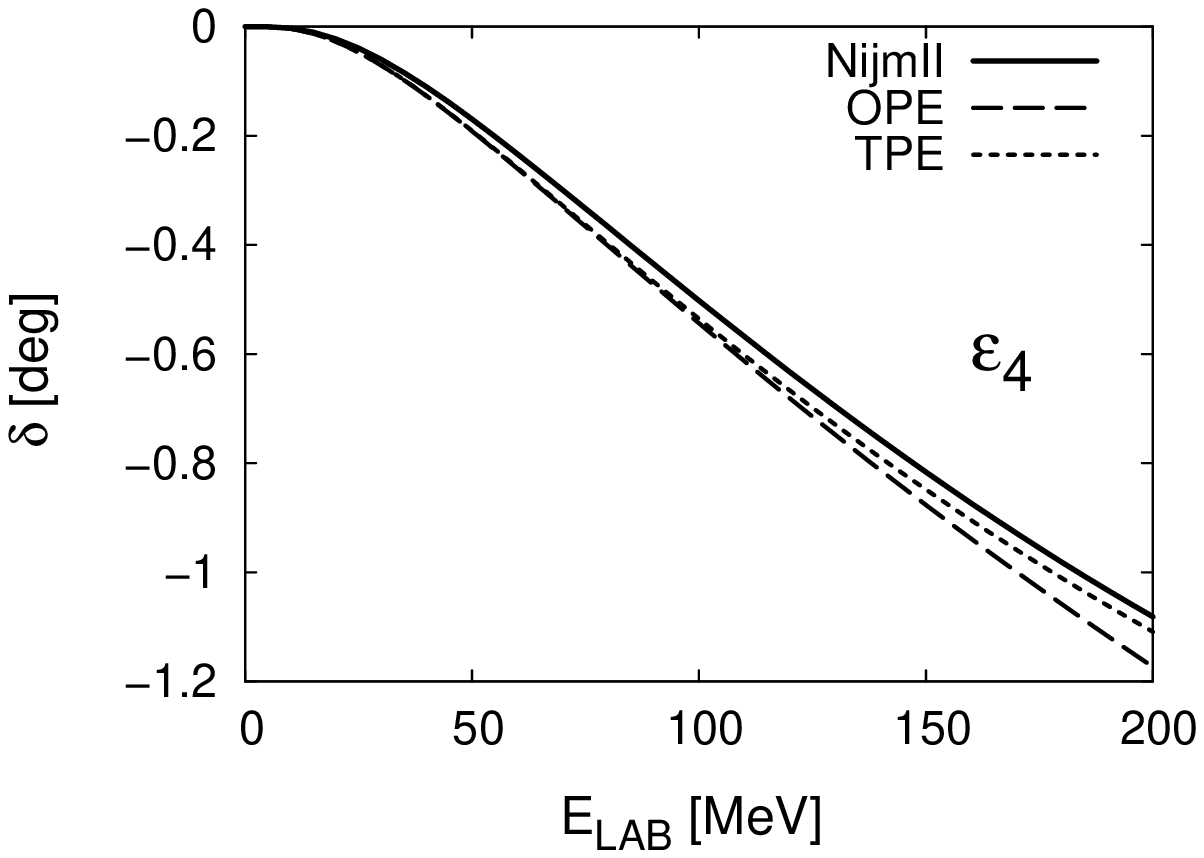,height=4cm,width=5cm} 
\end{center}
\caption{Same as Fig.~\ref{fig:fig-j=0} but for $j=4$.}
\label{fig:fig-j=4}
\end{figure*}

\begin{figure*}[]
\begin{center}
\epsfig{figure=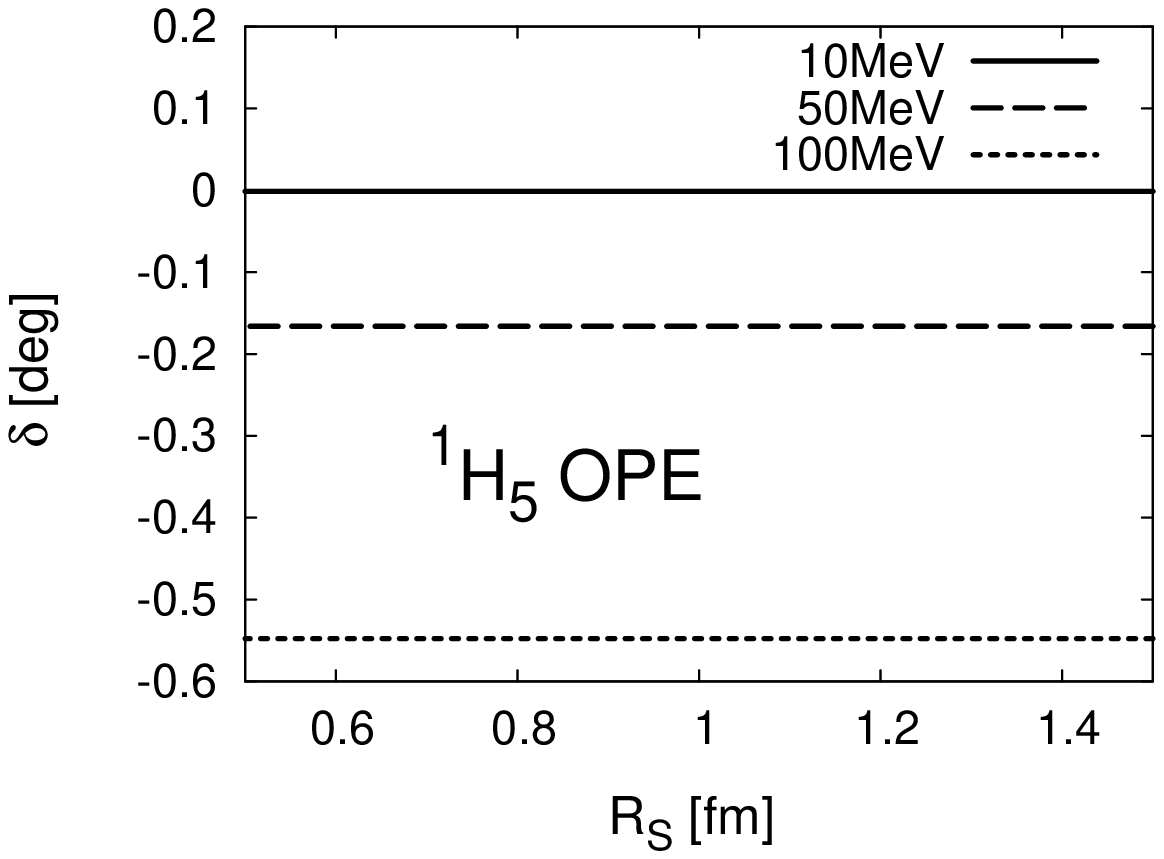,height=4cm,width=5cm}
\epsfig{figure=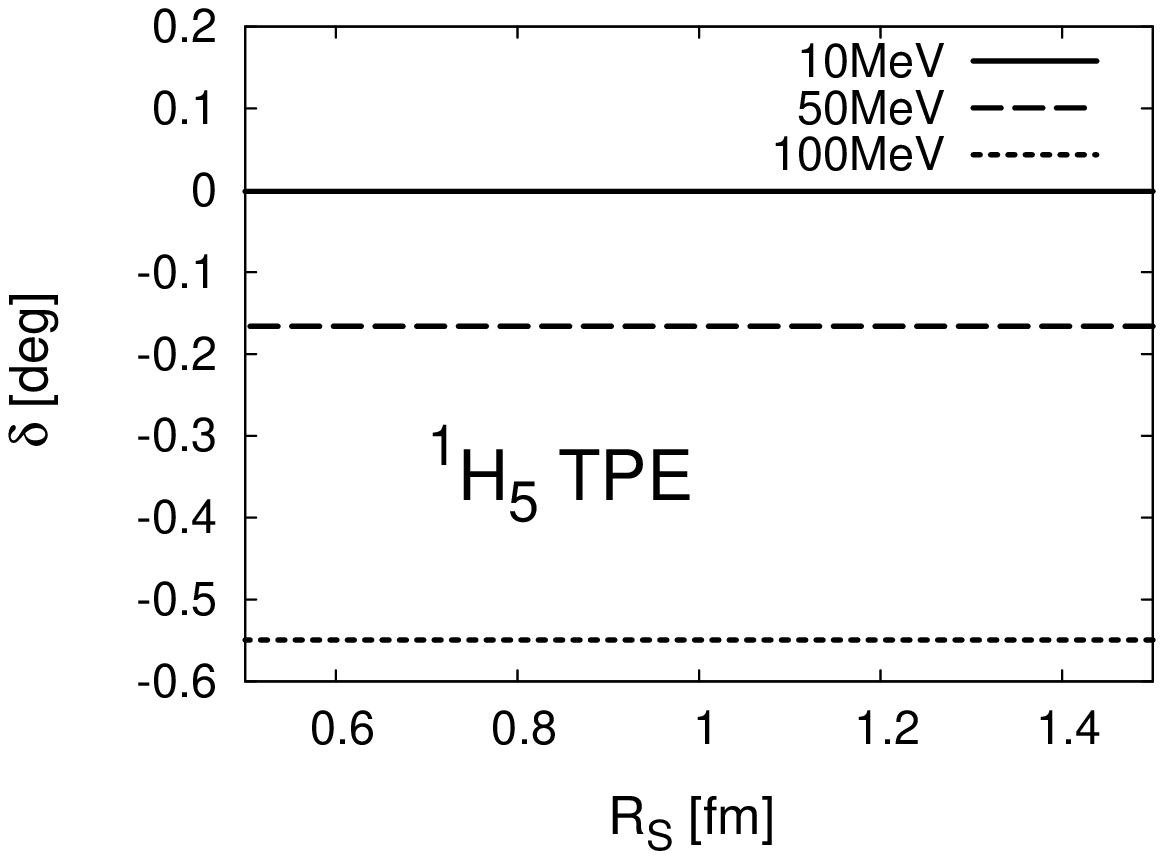,height=4cm,width=5cm}
\epsfig{figure=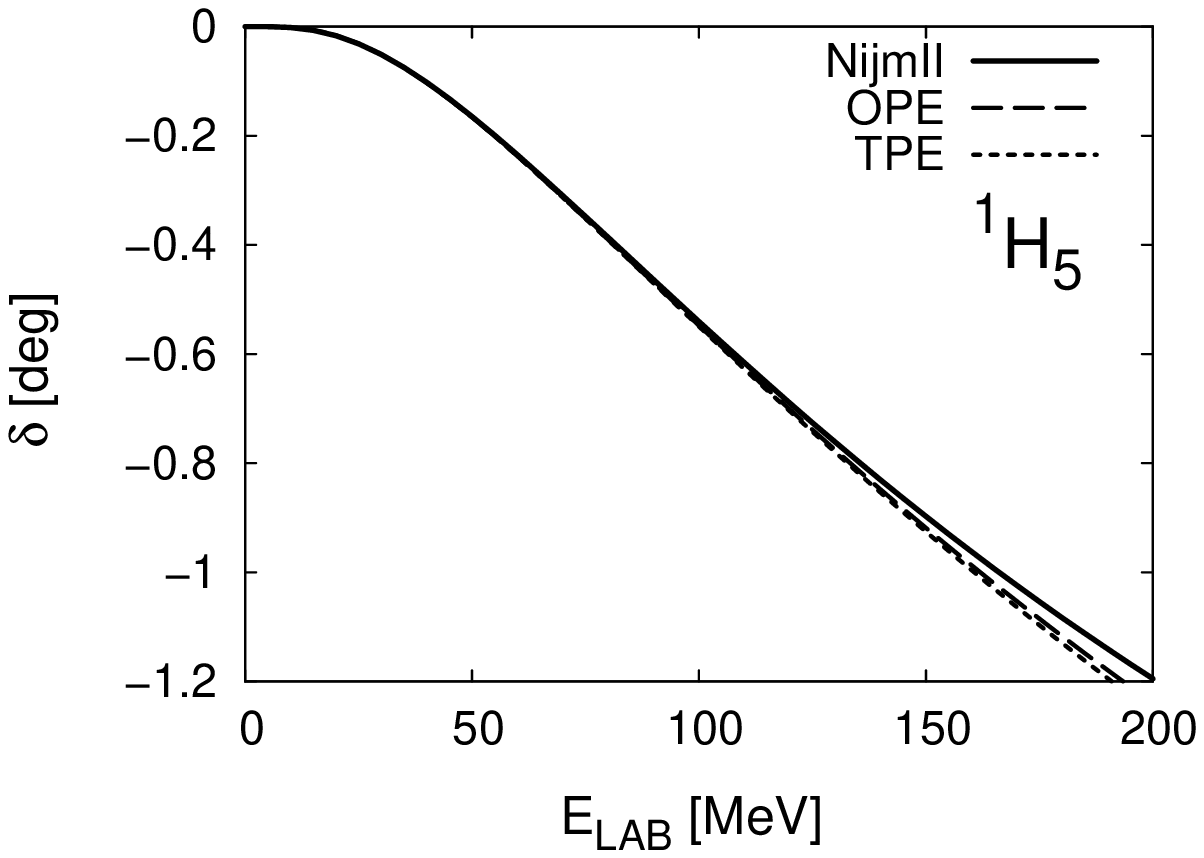,height=4cm,width=5cm} \\ 
\epsfig{figure=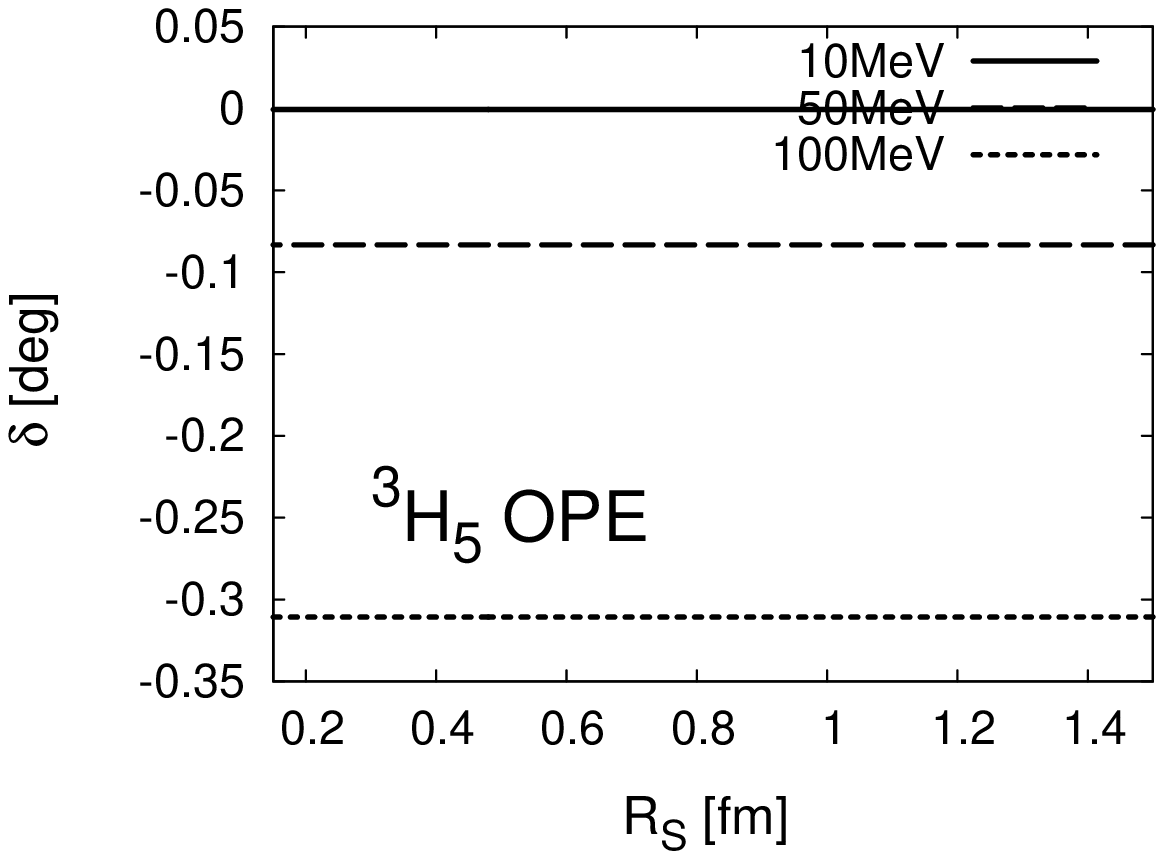,height=4cm,width=5cm}
\epsfig{figure=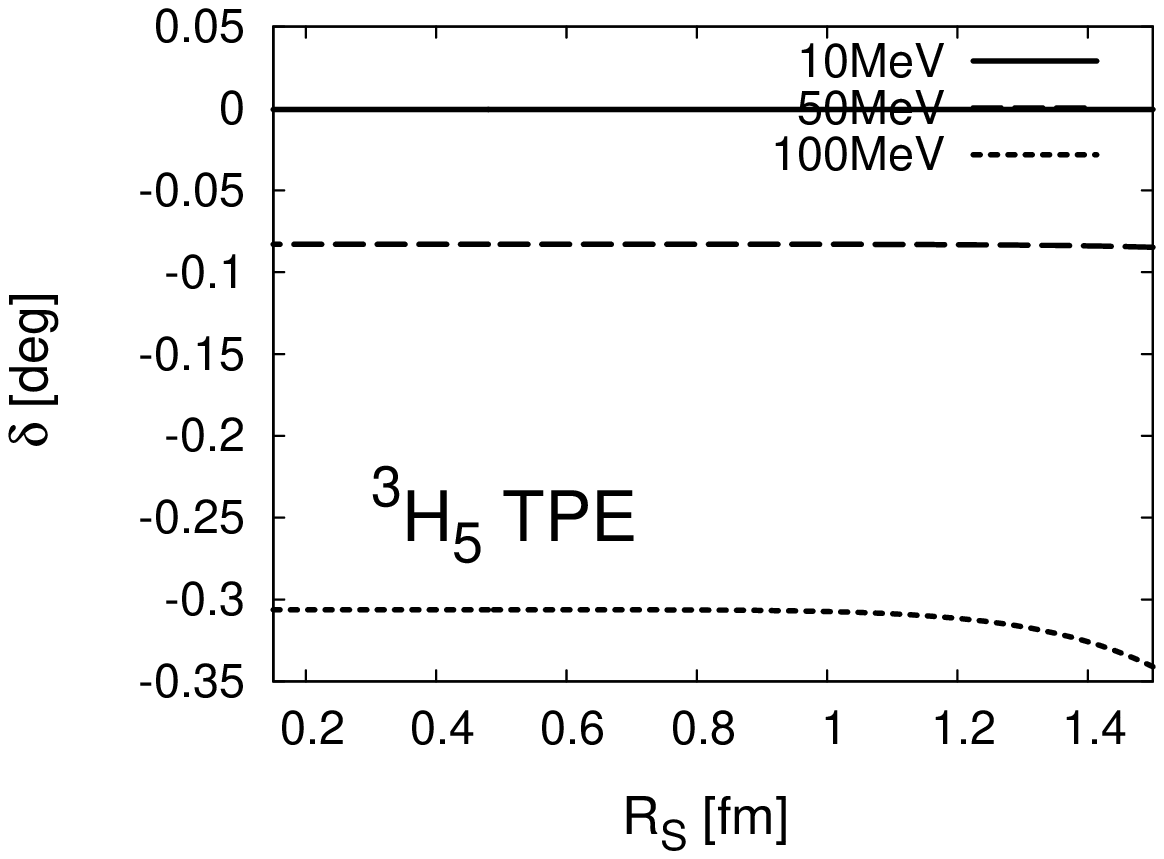,height=4cm,width=5cm}
\epsfig{figure=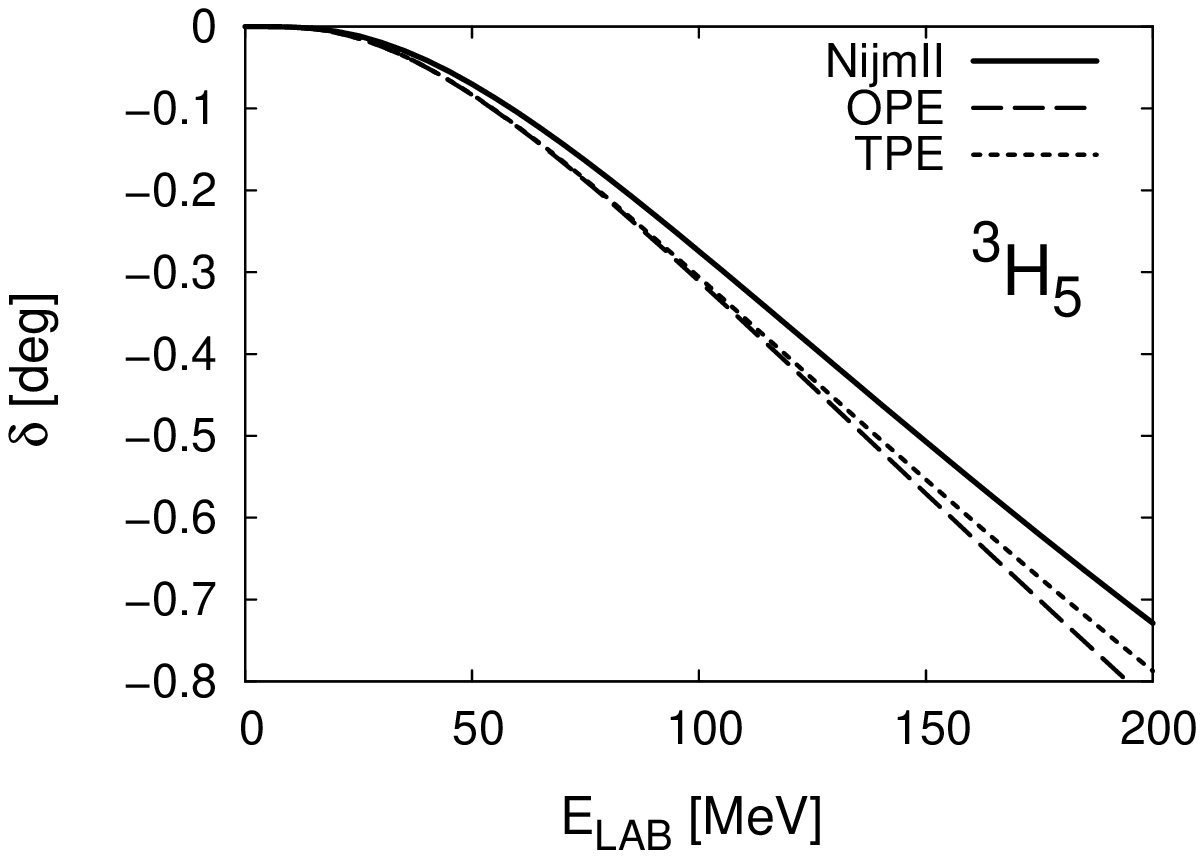,height=4cm,width=5cm} \\
\epsfig{figure=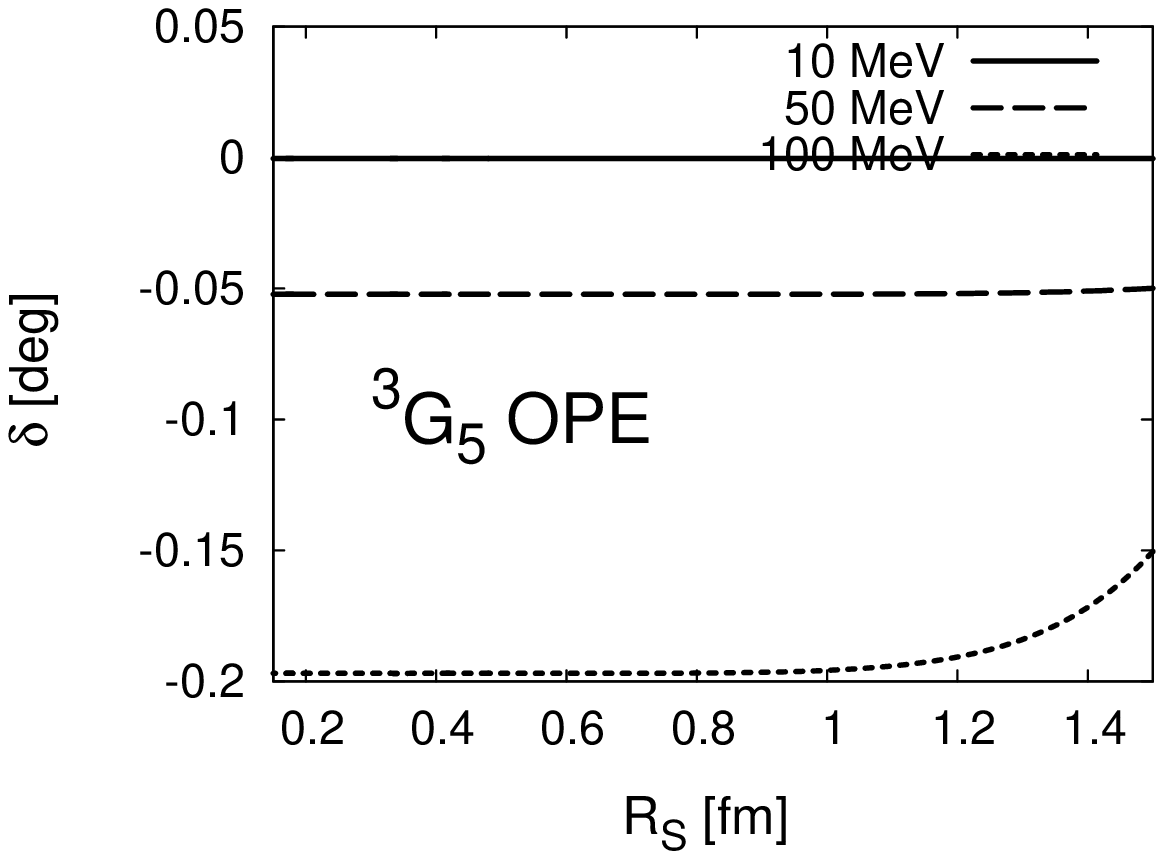,height=4cm,width=5cm}
\epsfig{figure=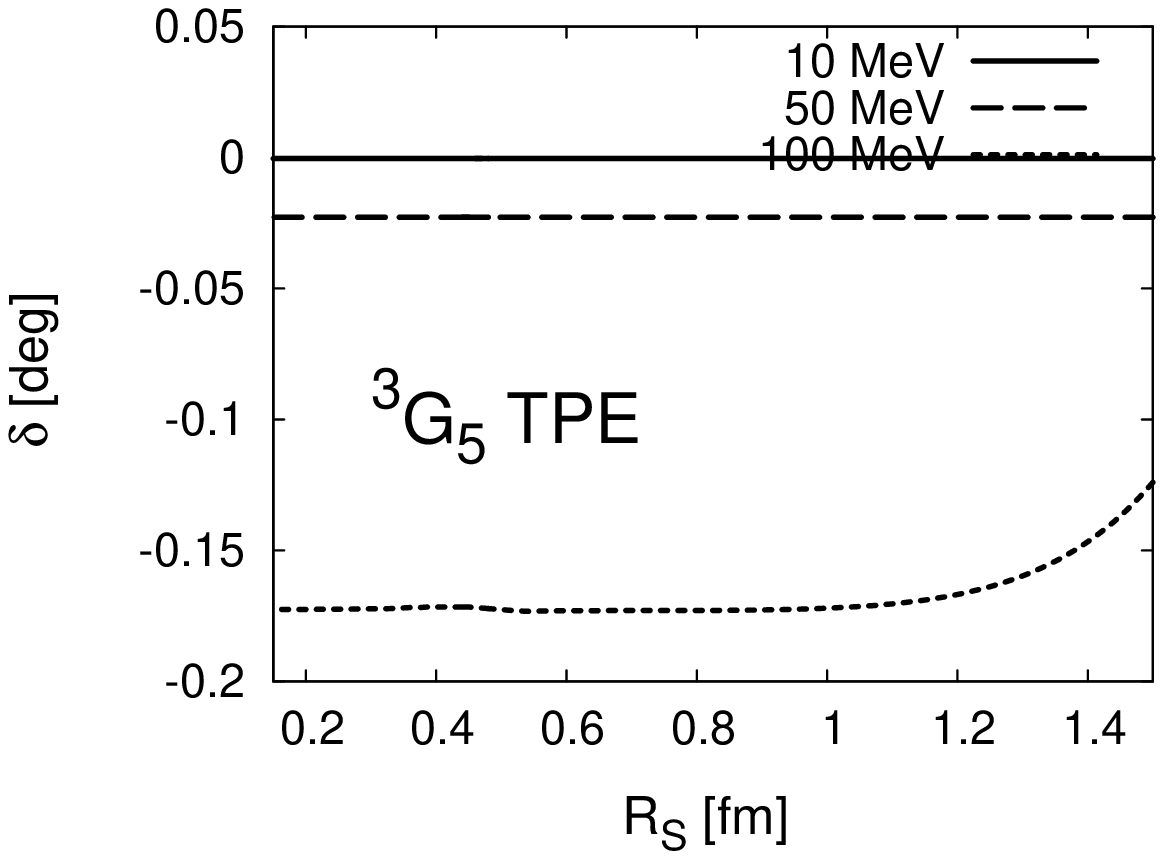,height=4cm,width=5cm}
\epsfig{figure=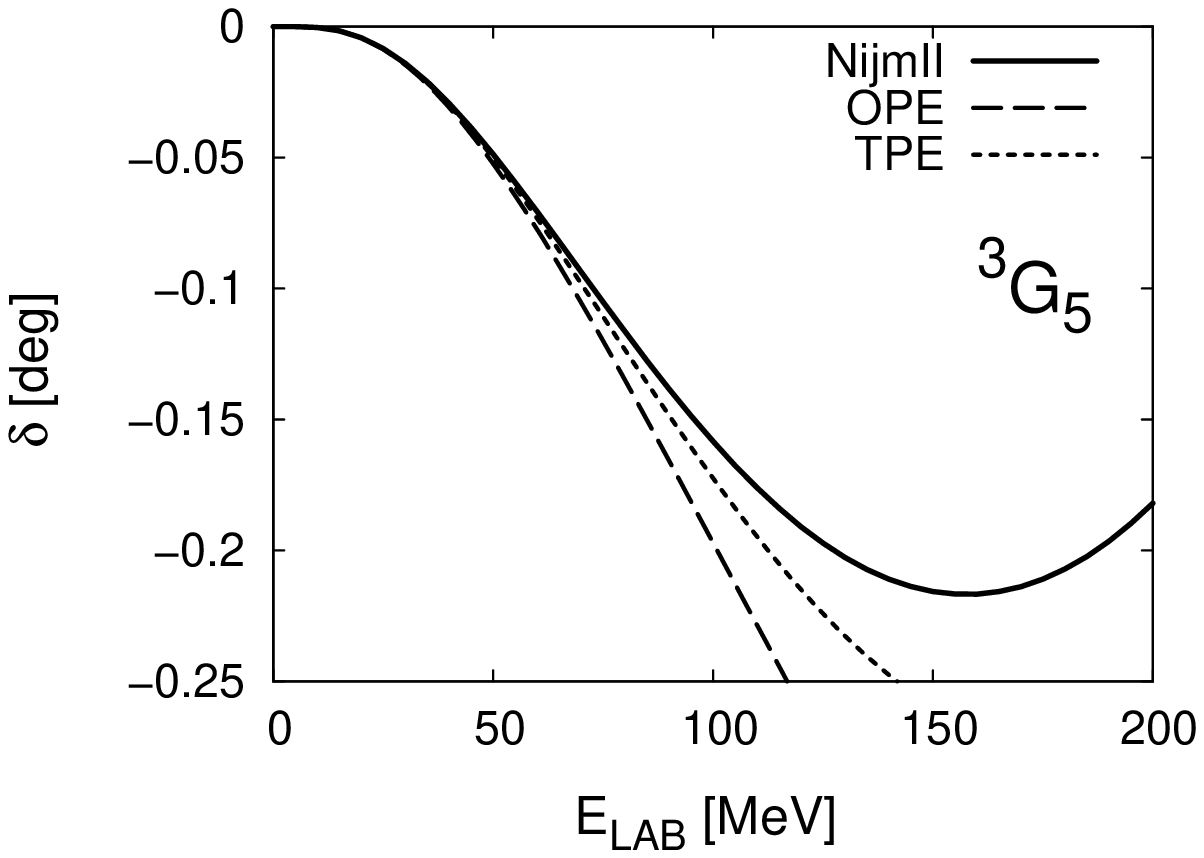,height=4cm,width=5cm} \\ 
\epsfig{figure=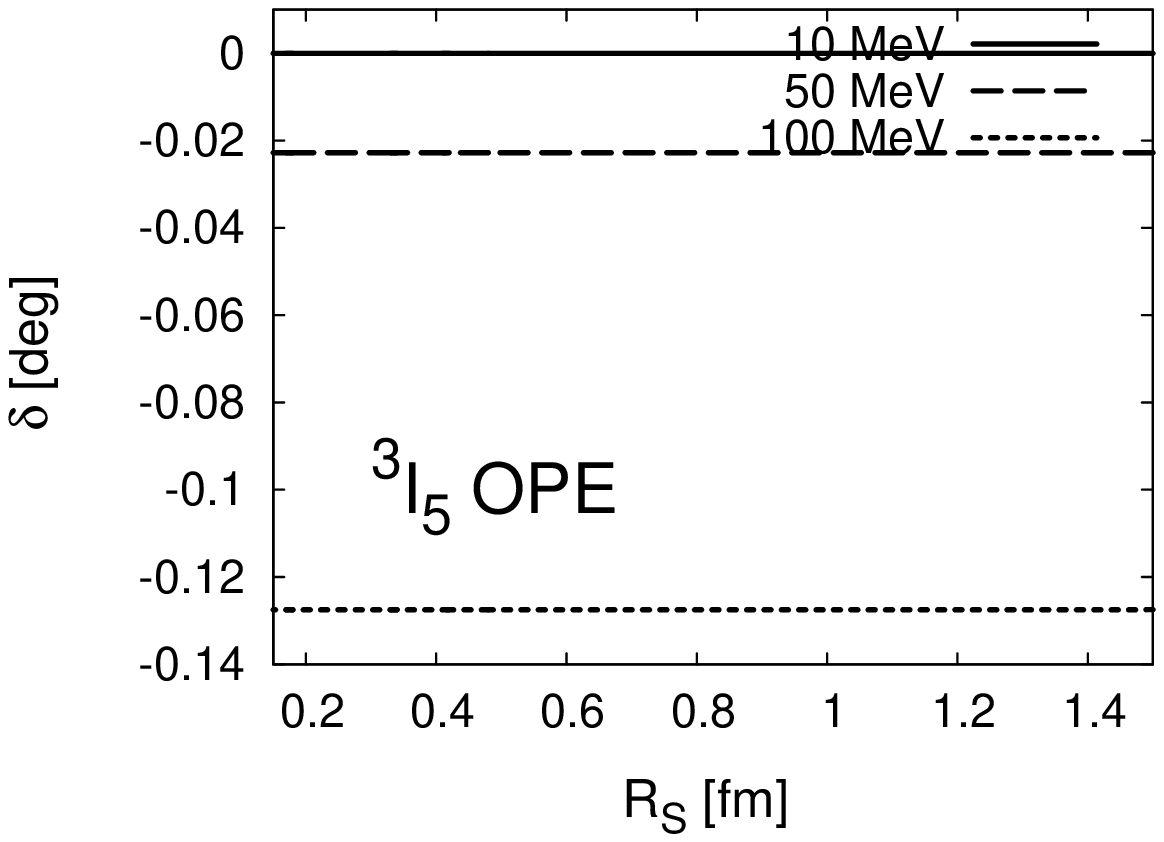,height=4cm,width=5cm}
\epsfig{figure=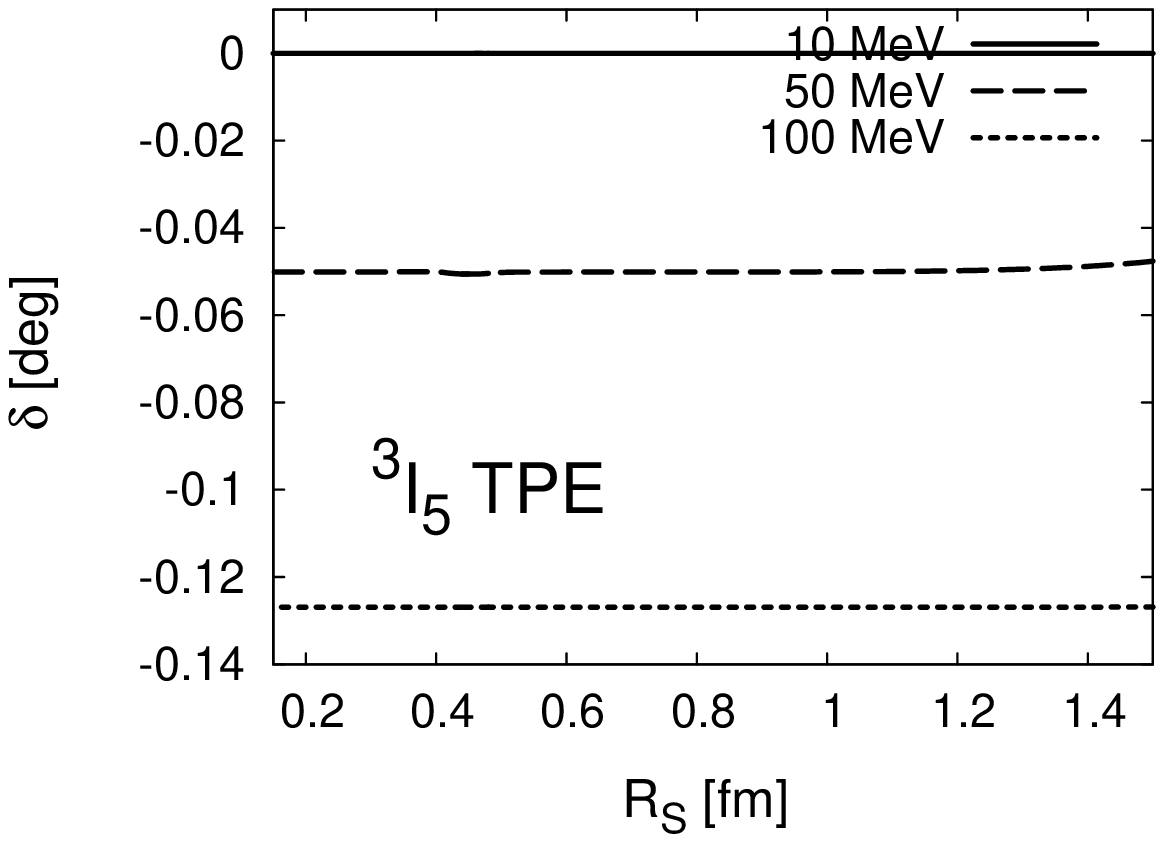,height=4cm,width=5cm}
\epsfig{figure=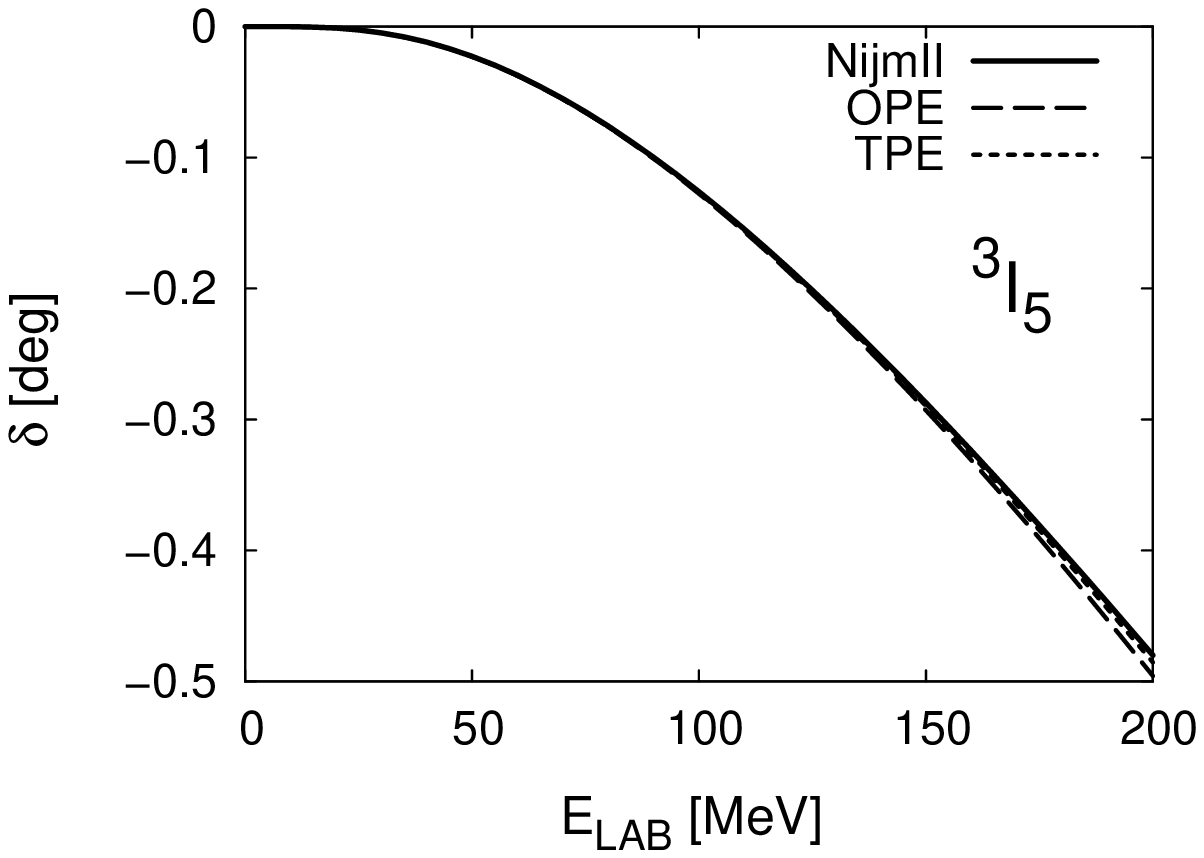,height=4cm,width=5cm} \\ 
\epsfig{figure=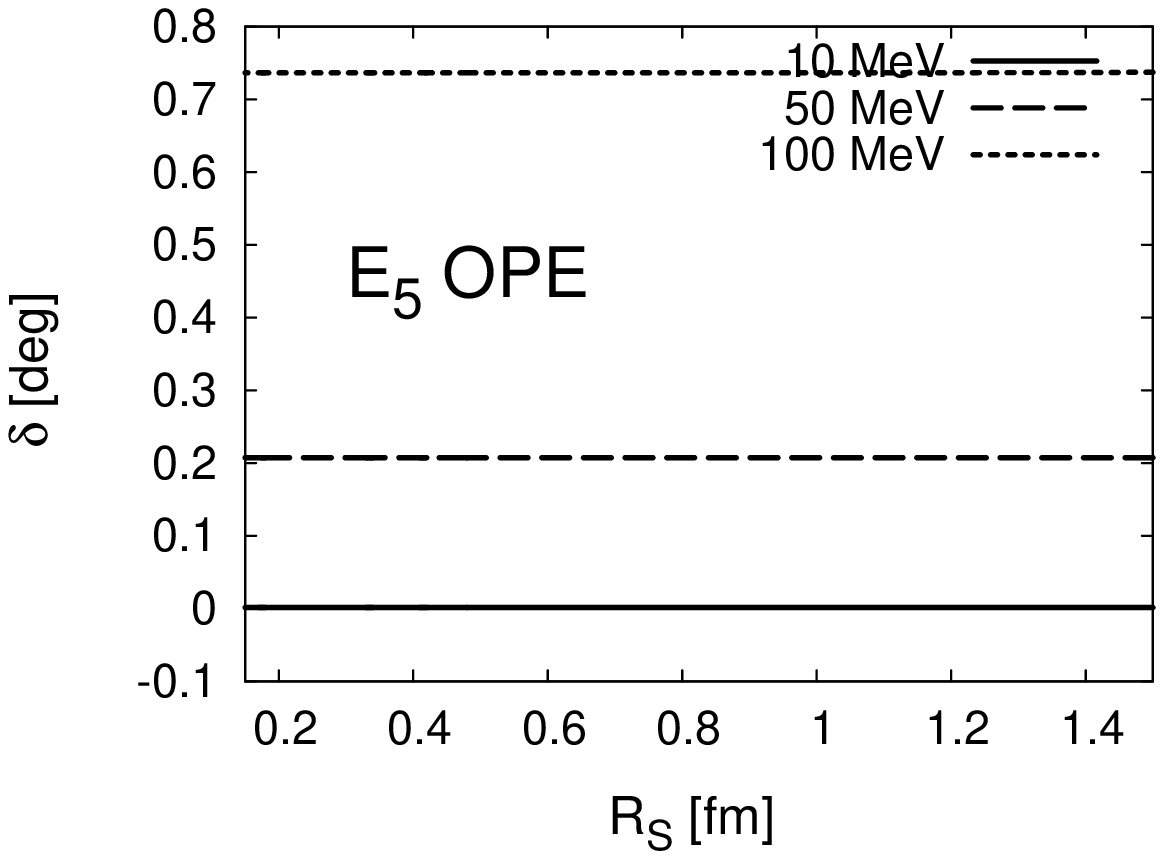,height=4cm,width=5cm}
\epsfig{figure=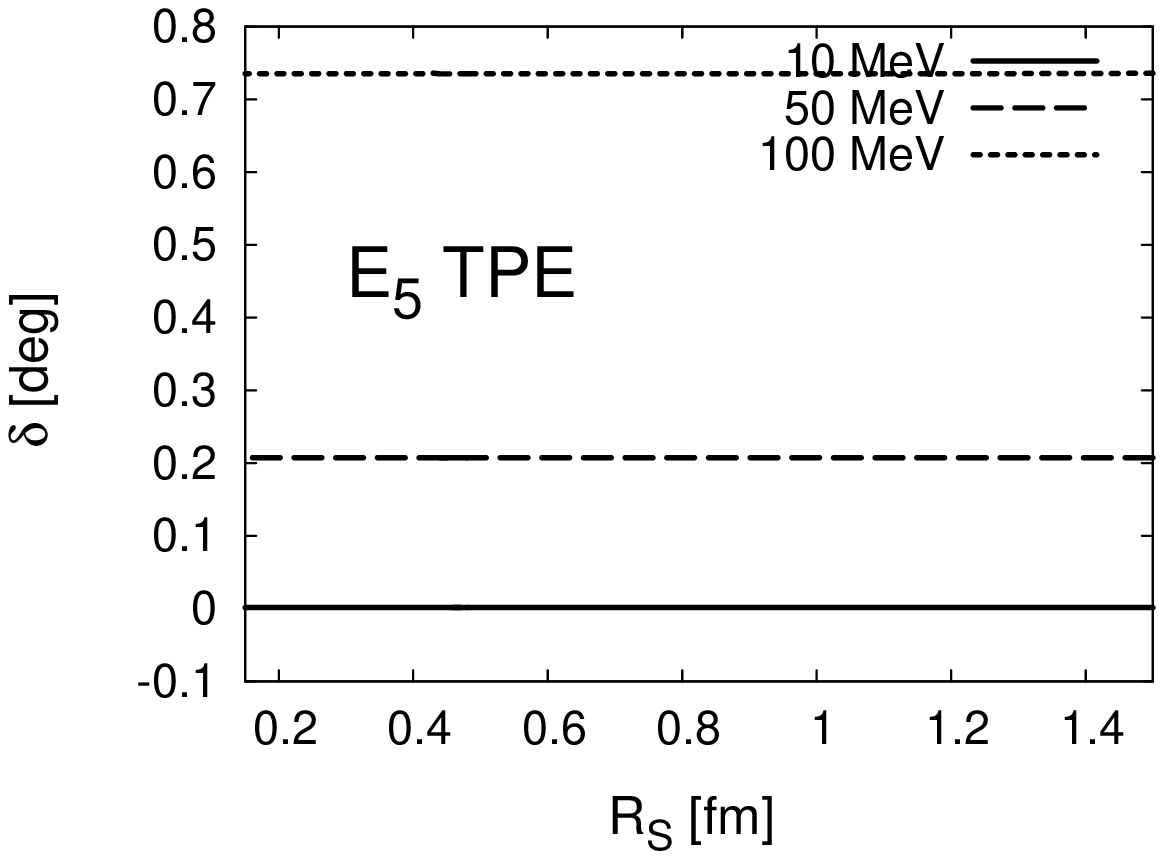,height=4cm,width=5cm}
\epsfig{figure=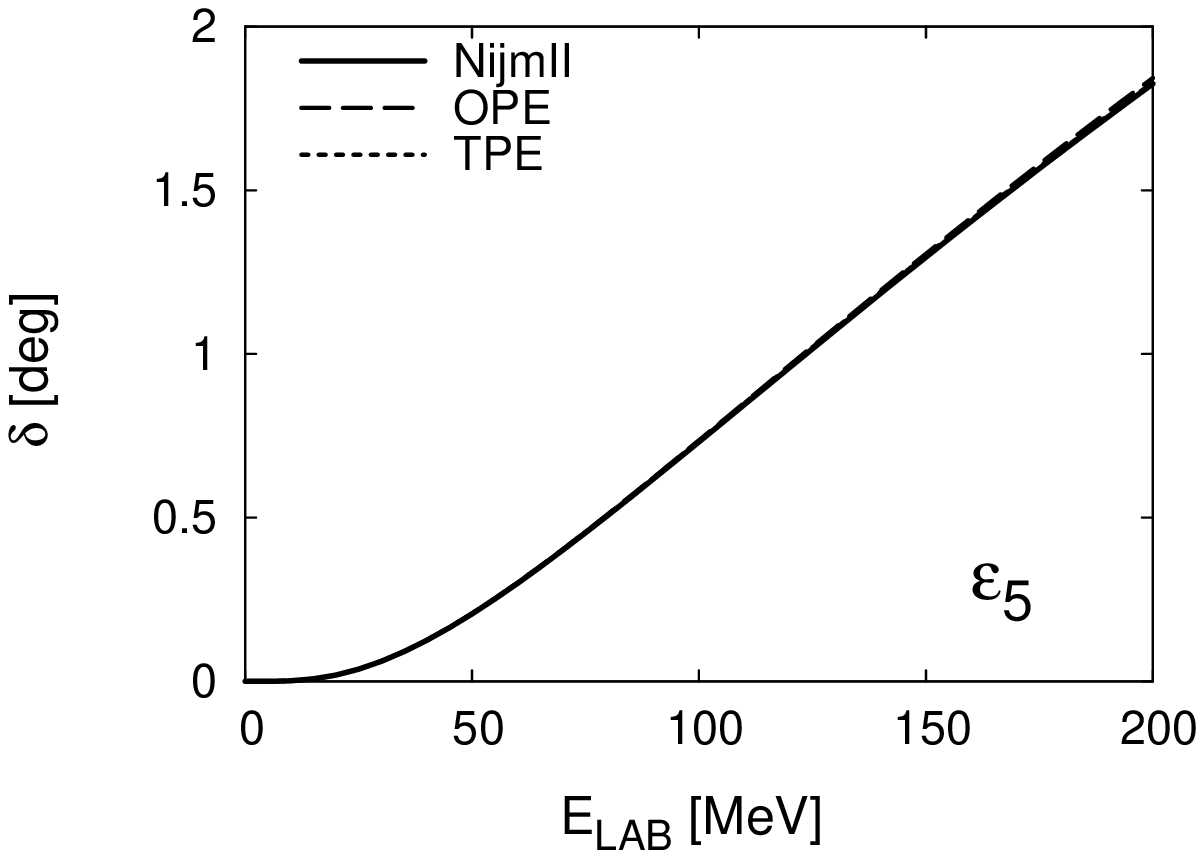,height=4cm,width=5cm} 
\end{center}
\caption{Same as Fig.~\ref{fig:fig-j=0} but for $j=5$.}
\label{fig:fig-j=5}
\end{figure*}

\section{\bf Results for the Phase Shifts} 
\label{sec:phases}

\subsection{Numerical parameters}

For our numerical calculations we take $f_\pi=92.4 {\rm MeV}$,
$m=138.03 {\rm MeV}$, $ 2 \mu_{np}= M = M_p M_n /(M_p+M_n) = 938.918
{\rm MeV}$, $ g_A =1.29 $ in the OPE piece to account for the
Goldberger-Treimann discrepancy and $g_A=1.26 $ in the TPE piece of
the potential. The corresponding pion nucleon coupling constant takes
then the value $ g_{\pi NN}=13.083$ (i.e. $g_A=1.29$) according to the
Nijmegen phase shift analysis of NN
scattering~\cite{deSwart:1997ep}. The values of the coefficients
$c_1$, $c_3$ and $c_4$ used along this paper can be looked up in Table
\ref{tab:table1} for completeness. The potentials in configuration
space used in this paper are exactly those provided in
Ref.~\cite{Ordonez:1995rz,Kaiser:1997mw,Rentmeester:1999vw} but
disregarding relativistic corrections, $M/E \to 1$~\footnote{As
mentioned in our previous work~\cite{Valderrama:2005wv} these effects
are tiny for the deuteron. For central waves they are about 0.2 $^0$
at the maximum CM momentum $p=400 {\rm MeV}$. This trend is general 
also for peripheral waves.}. The potentials are
listed in Appendix~\ref{sec:potentials} for completeness. The short
distance van der Waals coefficients for all channels studied in the
present work are presented in Appendix~\ref{sec:short}. The output of
such a channel by channel analysis is briefly summarized in
Table~\ref{tab:table2} where we indicate which scattering lengths are
used as input parameters according to the discussion given in
Sect.~\ref{sec:form}. Low energy parameters for the high quality
potentials~\cite{Stoks:1993tb,Stoks:1994wp} have been obtained in
Ref.~\cite{PavonValderrama:2004se}. We will use the Nijm II values,
but to give an idea on the expected lower uncertainties on those
parameters we also list the Reid93 values. Probably the real
uncertainties are much larger since the actual value of these low
energy parameter will depend upon which long range physics is included
in the high quality potentials where explicit TPE effects have not
been included, as we do in the present work~\footnote{This will generate slight
inconsistencies in the TPE results of Sect.~\ref{sec:NNLO-TPE} which
will be amended by a small modification of the threshold parameters,
yet larger than the discrepancies between the threshold parameters for
NijmII and Reid93 potentials obtained in
Ref.~\cite{PavonValderrama:2004se}.}.

\subsection{The deuteron channel revisited}

Before embarking in the full fledged discussion of all partial waves,
it is interesting to reanalyze first the $^3S_1-^3D_1$ channel already
studied in our previous work on the
deuteron~\cite{PavonValderrama:2005gu,Valderrama:2005wv}. There, we
used the orthogonality to the deuteron bound state. The scattering
lengths $\alpha_{02}= 1.67\,{\rm fm}^3$ and $\alpha_2= 6.6\,{\rm fm}^5$
were deduced from the experimental deuteron binding energy, the
asymptotic $D/S$ ratio and the S-wave scattering length $\alpha_0$.
These values turned out to be a bit off the values deduced from the
NijmII and Reid93 potentials~\cite{PavonValderrama:2004td} (see
Table~\ref{tab:table2}). Nevertheless, the intermediate energy region
turned out to be better described than the low energy behaviour
suggested. In the present work we choose instead to build scattering
states which are orthogonal to the zero energy states, so deuteron
properties can be deduced, as done in table~\ref{tab:deut_prop}.
In Fig.~\ref{fig:3C1} we show the results
when either the zero energy or the deuteron bound state are used as
reference states. One obvious lesson from this comparison is that
phase shifts, particularly the $E_1$ channel, may be better described
in the intermediate energy region if the deuteron is used as a
reference state, despite the fact that the threshold behaviour is a
bit off.
This is maybe explained by the observation that $\alpha_{02}$
and $\alpha_2$ encode higher energy information about the system
than $\alpha_0$ or $\gamma$~\footnote{It should be note that $\alpha_{02}$ 
and $\alpha_2$ are related with the behaviour of the scattering amplitude 
at order $k^2$ and $k^4$ respectively, relative to $\alpha_0$.}, 
so the latter parameters are more suited
to obtain an effective description of the system.
This feature will become evident in other partial waves.

\subsection{Cut-off dependence}

In Figs.~\ref{fig:fig-j=0}, \ref{fig:fig-j=1}, \ref{fig:fig-j=2}, 
\ref{fig:fig-j=3}, \ref{fig:fig-j=4} and \ref{fig:fig-j=5} we show the
results of our calculation for all partial waves with $j \le 5 $ as a
function of the nucleon LAB energy. For definiteness we use the chiral
constants $c_1$, $c_3$ and $c_4$ of Ref.~\cite{Entem:2003ft} (Set IV)
which already provided a good description of deuteron properties after
renormalization~\cite{Valderrama:2005wv} at NNLO. This choice allows a
more straightforward comparison to the N$^3$LO calculation of
Ref.~\cite{Entem:2003ft} with finite cut-offs. Unless otherwise
stated, the needed low energy parameters for these figures are {\it
always} taken to be those of Ref.~\cite{PavonValderrama:2004se} for
the NijmII potential (see Table~\ref{tab:table2}).

In order to test the stability of the phase-shifts against changes in
the short distance cut-off parameter, $R_S$, we show in
Figs.~\ref{fig:fig-j=0}, \ref{fig:fig-j=1}, \ref{fig:fig-j=2},
\ref{fig:fig-j=3}, \ref{fig:fig-j=4} and \ref{fig:fig-j=5}, and
similarly to the OPE study in momentum space of
Ref.~\cite{Nogga:2005hy}, the cut-off dependence for fixed values of
the lab energy both for the OPE as well as for the TPE
potentials. This is done in the range $0.15 {\rm fm} \le R_S \le 1.5
{\rm fm}$. If we identify this short distance cut-off with the sharp
momentum cut-off $\Lambda= \pi/2 R_S$~\cite{PavonValderrama:2004td},
the smallest boundary radius, $\sim 0.15 {\rm fm}$, corresponds to a
maximum cut-off $\Lambda \sim 2 {\rm GeV} $. This is much larger than
the cut-offs used in
Refs.~\cite{Rentmeester:1999vw,Epelbaum:1999dj,Epelbaum:2003gr,Epelbaum:2003xx,Entem:2003ft,Rentmeester:2003mf}
but comparable to the exponential cut-off used in
Ref.~\cite{Nogga:2005hy} for the renormalization of the OPE potential~
\footnote{\label{footnote:gauss} 
There, a cut-off has been introduced according to the rule in
the potential $ V(k',k) \to e^{-{k'}^4 / {\Lambda}^4} V(k'k) e^{-k^4 /
\Lambda^4} $ and counterterms have been added. To get an
order of magnitude of the equivalent sharp cut-off $\tilde \Lambda $
we estimate the linear divergence at zero energy in the contact
theory,
$$ 
\tilde \Lambda= \int_0^\infty e^{-2 q^4 / \Lambda^2 } dq =
\frac{\Gamma(\frac54)}{2^{\frac14}} \Lambda = 0.762 \Lambda \, ,
$$ 
and also using $\tilde \Lambda = \pi /2 R_S $ ~\cite{PavonValderrama:2004td}, 
we get $\Lambda = 1 / (0.48 R_S)$ .}. 
Note that the limit $R_S \to 0$ may be taken independently 
for any different channel. 

The evolution of the increasingly oscillating wave function in the
attractive case can be identified with the cycles (improperly called
limit-cycles, see footnote 5 in Ref.~\cite{PavonValderrama:2004nb})
described in Refs.~\cite{Beane:2000wh,Beane:2001bc,PavonValderrama:2004nb,
PavonValderrama:2004td}
by looking at suitable logarithmic combinations of the wave
functions. The cycles documented in Ref.~\cite{Nogga:2005hy} in
momentum space can be mapped into the coordinate space cycles by
relating the coordinate and momentum space cut-offs. 

Generally speaking, the inclusion of chiral TPE effects generates
smoother limits as compared to the OPE results, as one would
expect. We have checked that for short distance repulsive
(eigen)channels results are not very sensitive to the choice of the
regulator for small values of $R_S$.  As we also see from the figures,
the convergence depends both on the partial wave as well as on the
energy. As expected, the needed value of the short distance cut-off
$R_S$ for which stability is achieved is rather high for peripheral
waves, $R_S \sim 1/m_\pi$. Another feature of the calculation are the
observed stability plateaus for a number of partial waves. This trend
has also been noted in previous works with finite
cut-offs~\cite{Epelbaum:2004fk} where there appear sequential cut-off
windows. In coordinate space this is originated by the almost
self-similar pattern of the short distance oscillations of the wave
function which suggest a sequential and faster convergence modulo
cycles~\cite{PavonValderrama:2004nb}.

Let us remark at this point that the existence of an $R_S \to 0$ limit
does not necessarily mean a plateau-like approach to it.  This is the
case, for example, of the $^1S_0$ wave, which for OPE shows a linear
dependence on the cut-off due to the mild $1/r$ singularity of the
potential, generating a linear-like behaviour which corresponds to the
ratio of regular ($\sim r$) and irregular ($\sim 1$) solutions at the
origin~\footnote{Anyway, the lack of a clear plateau in this wave
becomes obvious in the coordinate space treatment. Assuming the
relationship $\Lambda \sim 1/(0.48R_S)$ for a gaussian cut-off ( see
footnote \ref{footnote:gauss}) between the momentum and coordinate
space cut-offs, a linear dependence of the phase shifts on the $R_S$
coordinate cut-off would map into a $1/\Lambda$ dependence in momentum
space, which might be regarded as a plateau in a sufficiently thin
cut-off window.. Note that going from $R_s=0.2{\rm fm} $ to $R_s =
0.1{\rm fm}$ corresponds equivalently to double the momentum space
cut-off from $\sim 2 {\rm GeV}$ to $\sim 4 {\rm GeV}$.}.  A
similar behaviour can be found on other singlet waves in which the OPE
potential also behaves as $1/r$, but highly attenuated by the
influence of the centrifugal barrier.

Finally, let us note that there are some channels where the phase
shifts exhibit a very strong dependence on the regulator~\footnote{The
jump in the evolution of the OPE potential in the $^3D_3$ channel
around $R_S = 0.3 {\rm fm}$, Fig.~\ref{fig:fig-j=3} resembles a
coupled channel resonance, corresponding to tunneling across the
centrifugal barrier into the short distance attractive singularity.}.

\subsection{Renormalized phase shifts}

\subsubsection{LO (OPE)}

In Figs.~\ref{fig:fig-j=0}, \ref{fig:fig-j=1}, \ref{fig:fig-j=2},
\ref{fig:fig-j=3}, \ref{fig:fig-j=4} and \ref{fig:fig-j=5} we also
compare the OPE (LO), the NNLO TPE and the Nijmegen phase shift
analysis~\cite{Stoks:1993tb,Stoks:1994wp}. As noted in
Tab.~\ref{tab:table2} in some cases with attractive singular
potentials some scattering lengths must be specified, in order to
determine the phase shifts, but for repulsive singular potentials the
scattering lengths and hence the phase shifts are fully determined
from the potential. In the coupled channel case where only one
parameter should be fixed we have chosen, as indicated in
Table~\ref{tab:table2}, to take the scattering length of the
corresponding partial wave with the lower orbital angular momentum. As
we see from
Figs.~\ref{fig:fig-j=0}, \ref{fig:fig-j=1}, \ref{fig:fig-j=2},
\ref{fig:fig-j=3}, \ref{fig:fig-j=4} and \ref{fig:fig-j=5}, OPE does a
relatively good job for the phases when compared to the NijmII
results, up to a reasonable energy. This calculation extends our
previous results~\cite{PavonValderrama:2005gu} using the same
regularization for the singlet $^1S_0$ and triplet $^3S_1-^3D_1$
channels.

The LO results corresponding to static OPE potential have also been
obtained recently in momentum space by a solution of the
Lippmann-Schwinger equation in Ref.~\cite{Nogga:2005hy} for $j \le 3$.
These authors see that in the limit $\Lambda \to
\infty $ (in practice $\Lambda = 4 {\rm GeV} ) $ it is always possible
to adjust a counterterm in such a way that the phase shifts are
cut-off independent. They also find that the needed counterterm does
not correspond to the expectations based on Weinberg's dimensional
power counting argument, so that one is forced to promote 
counterterms which are of higher order in Weinberg's counting 
to make the theory free of short distance ambiguities. 
This proposal not only fits quite naturally into our
analysis of short distance boundary conditions, but can also be
anticipated by just looking at the short distance behaviour of the
potential. In general, we reproduce their results for the phase shifts
using our boundary condition regularization (our shortest distance
cut-off is typically $a=0.1 {\rm fm} $ for OPE). This is precisely one
of the points of renormalization; different regularization methods
should yield identical results when the regulator is removed provided
the same renormalization conditions are imposed. Note that in our case
whenever a scattering length must be provided we exactly construct the
phase shift as to reproduce the threshold behaviour of the Nijmegen
phases~\cite{Stoks:1993tb,Stoks:1994wp} by exactly fixing the
scattering length (the renormalization condition). This requires
solving the zero energy problem by integrating in with the given
scattering length, and matching at short distances the finite energy
problem to finally determine the phase shift by integrating out. In
this approach we never make a fit.  In the approach of
Ref.~\cite{Nogga:2005hy} counterterms are adjusted to fit the phases
in the region around threshold. Although this is in spirit the same
renormalization condition to fix the counterterms, we expect some
numerical discrepancies, due to the fact that the threshold parameters
in Ref.~\cite{Nogga:2005hy} may be slightly different to ours.

\begin{figure*}[ttt]
\begin{center}
\epsfig{figure=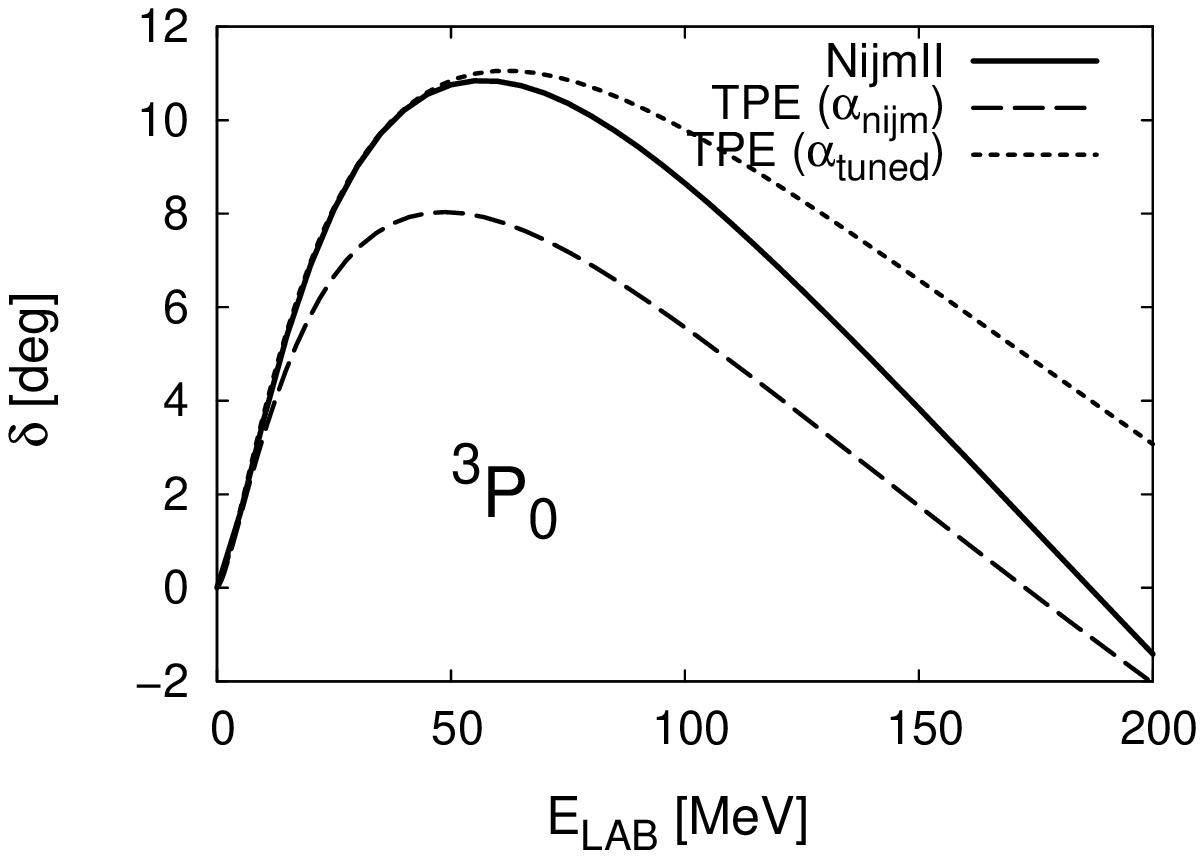,height=4cm,width=5cm}
\epsfig{figure=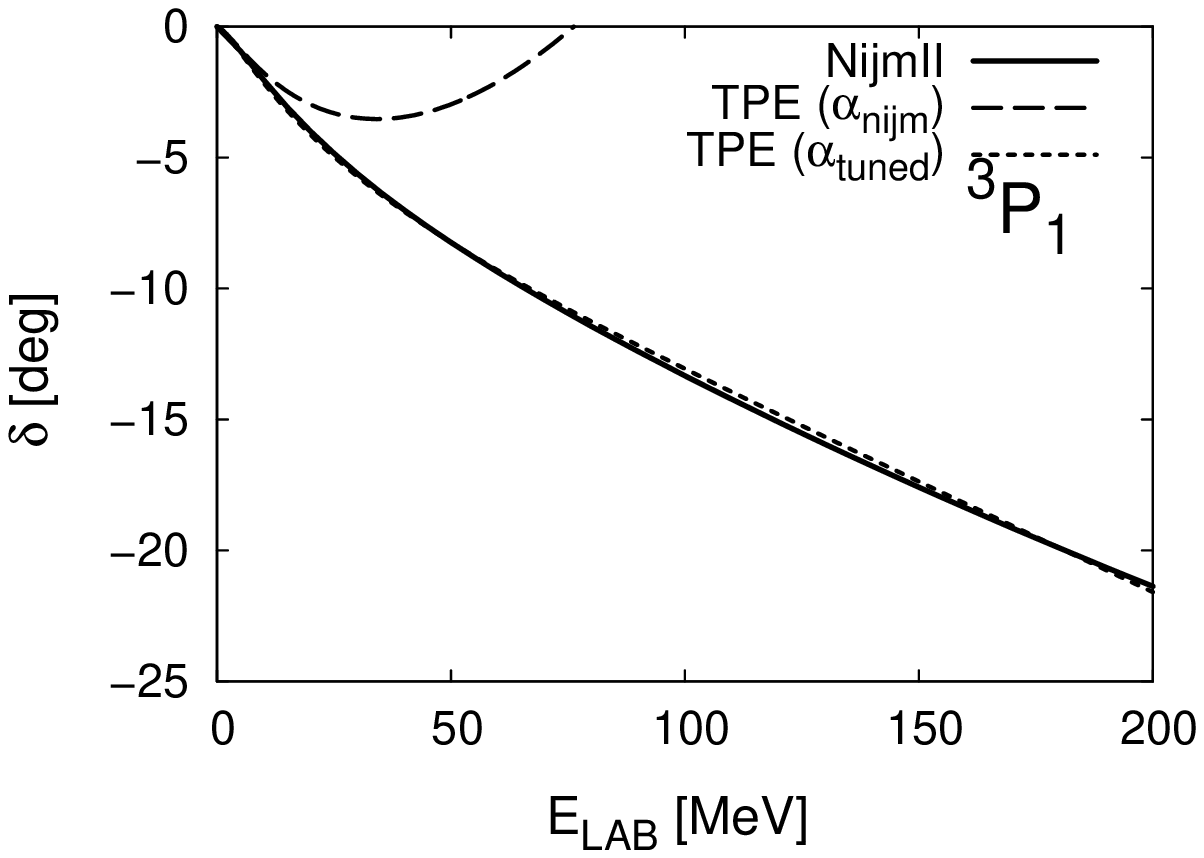,height=4cm,width=5cm}
\epsfig{figure=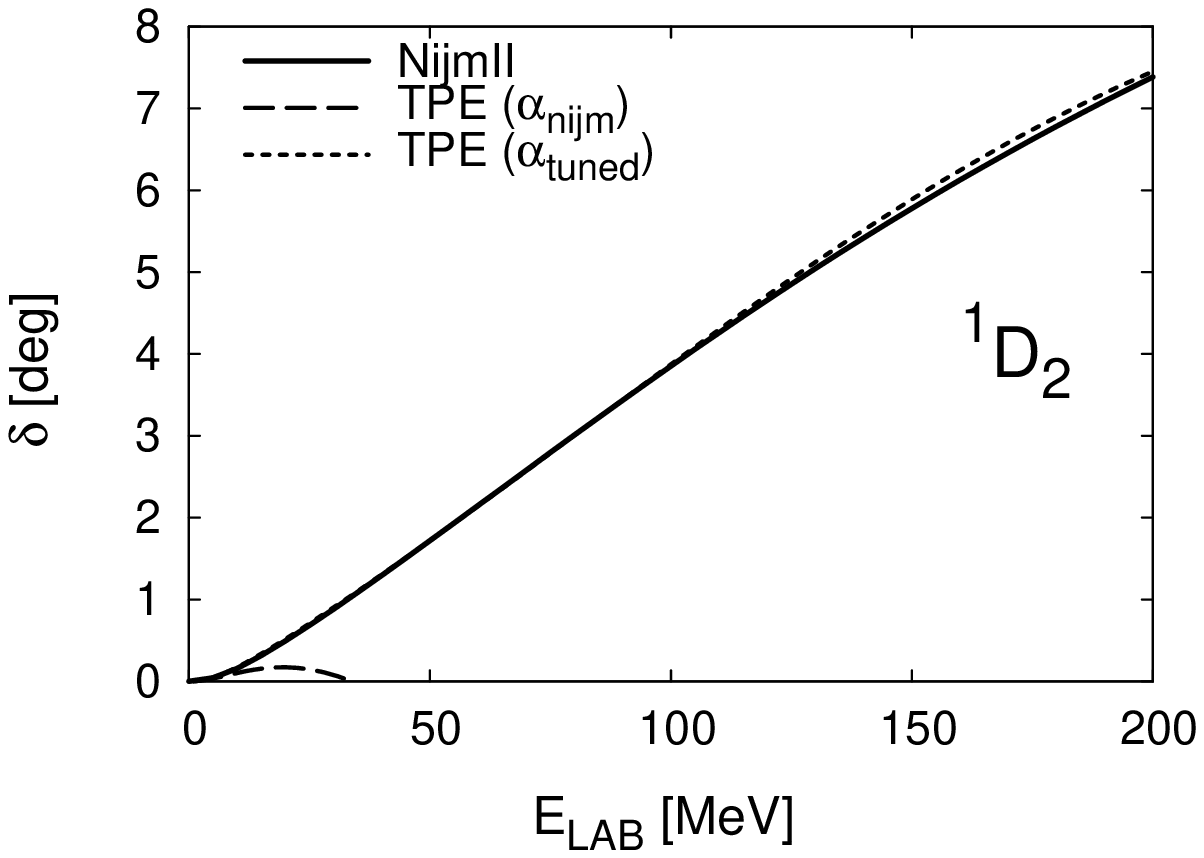,height=4cm,width=5cm} 
\end{center}
\caption{Dependence of some phases for the chiral TPE potential on the
scattering lengths compared to the NijmII
phases~\cite{Stoks:1993tb,Stoks:1994wp}. We use Set IV of chiral
constants.  Label TPE ($\alpha_{\rm nijm}$) means scattering lengths
of Table~\ref{tab:table2} are taken as deduced in
Ref.~\cite{PavonValderrama:2004se}, while TPE ($\alpha_{\rm tuned}$)
stands for the values tuned to fit the phases in the intermediate
energy region. For $^3P_0 $ we take $\alpha_1=-2.670 {\rm fm}^3$, for
$^3P_1$ we take $\alpha_1=1.692 {\rm fm}^3 $ and for $^1D_2$ we take
$\alpha_2=-1.666 {\rm fm}^5$.}
\label{fig:alphas}
\end{figure*}

\subsubsection{NLO (TPE)} 

Regarding NLO we do not show the results as they fail completely to
describe the data in triplet $^3S_1-^3D_1$ channel. The problem we
found already~\cite{Valderrama:2005wv} in the triplet $^3S_1-^3D_1$
channel persists in other channels; the short distance behaviour of
the NLO potential corresponds to $1/r^5$ repulsive
eigenpotentials. This feature explains the relatively small maximal
cut-offs allowed in NLO calculations in momentum space.  As stressed
on our previous work there are at least two scenarios where the
problem may be overcome. One possibility appeals to the role of the
$\Delta$ resonance and the fact that its contribution to $c_3$ and
$c_4$ scales as the inverse of the $N \Delta$ splitting $\Delta \sim 2
m_\pi $ as found in Ref.~\cite{Ordonez:1995rz,vanKolck:1994yi,Kaiser:1998wa}. 
In the $\Delta$
counting the $c_3$ and $c_4 $ contributions to the NNLO deltaless
potential become actually NLO contributions, and the short distance
behaviour becomes a $1/r^6$ attractive singularity. The second scenario
has to do with the influence of relativity beyond a truncated heavy
baryon expansion, since according to
Refs.~\cite{Higa:2003jk,Higa:2003sz,Higa:2004cr} one has a
relativistic $1/r^7$ van der Waals short distance behaviour with
attractive-repulsive eigen potential meaning that as in the OPE case
one has one free parameter. Calculations taking into account these
effects in all partial waves are currently underway~\cite{HPR2005}.

\subsubsection{NNLO (chiral-TPE)} 
\label{sec:NNLO-TPE}

We turn now to the NNLO calculation which are the genuine predictions
of ChPT because they contain the chiral constants $c_1$, $c_3$ and
$c_4$ (see e.g. Table~\ref{tab:table1}) and for definiteness we will
use mainly Set IV~\cite{Entem:2003ft} in our analysis~\footnote{By
genuine we mean that the NNLO potential contains parameters which are
relating $\pi N$ and $NN$ data in some intricate way. The fact that we
are using the parameter Set IV~\cite{Entem:2003ft} is because it
nicely reproduces the deuteron properties. One could, of course
improve on this by a large scale fit to the data}. Results for the TPE
renormalized phase shifts are presented in Figs.~\ref{fig:fig-j=0},
\ref{fig:fig-j=1}, \ref{fig:fig-j=2}, \ref{fig:fig-j=3},
\ref{fig:fig-j=4} and \ref{fig:fig-j=5}.  Some expected features do
indeed occur. Peripheral waves are slightly modified by going from OPE
to the chiral TPE potential. On the other hand, low partial waves are
also improved in the low energy region. For instance, the $^1S_0 $
phase has an attractive singular interaction, requiring fixing the
scattering length. The difference in the curves is mainly related to
the difference in the effective range which improves when going from
OPE to TPE~\cite{Valderrama:2005wv}. This is a rather general feature,
the error at low energies is controlled by the low energy threshold
parameters, like the effective range and others. If one looks at the
$^3P_0$ channel, we see that there is improvement but not as dramatic
as in the $^1S_0$ channel.

\begin{figure*}[ttt]
\begin{center}
\epsfig{figure=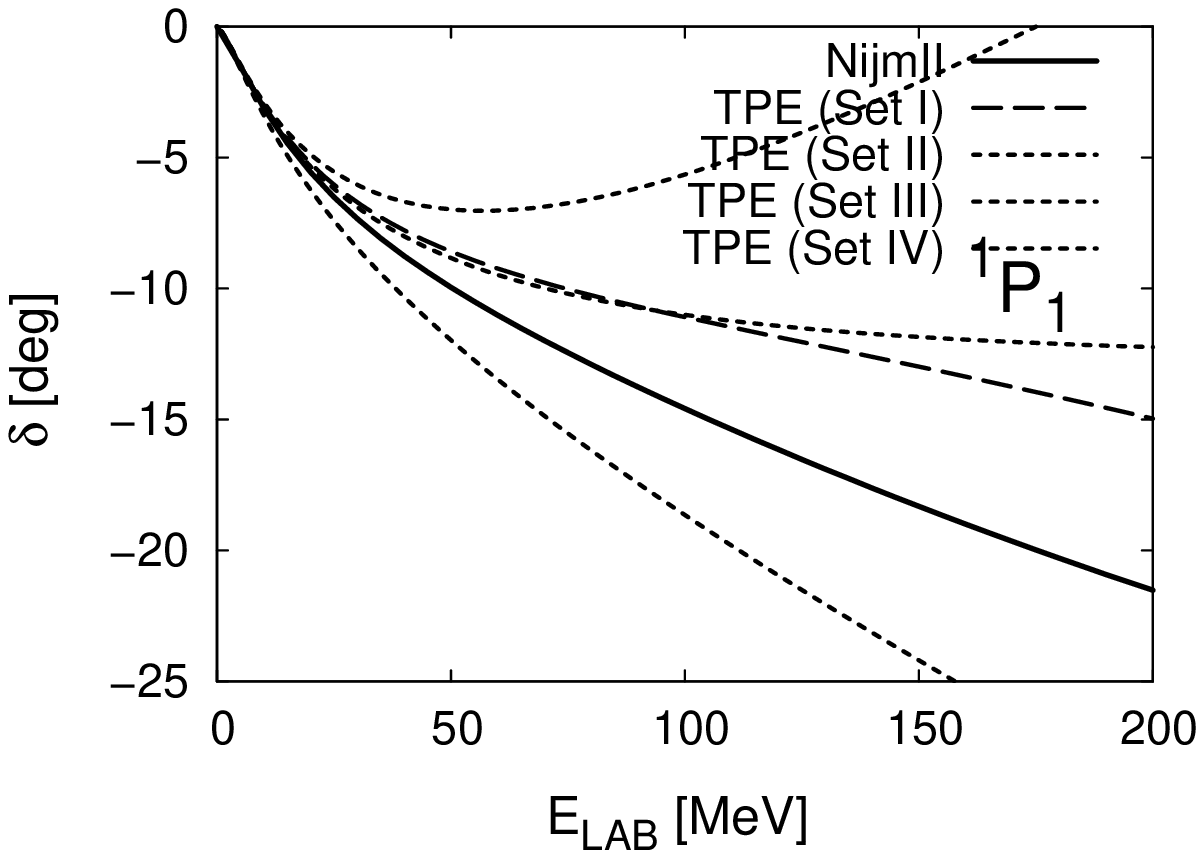,height=4cm,width=5cm}
\epsfig{figure=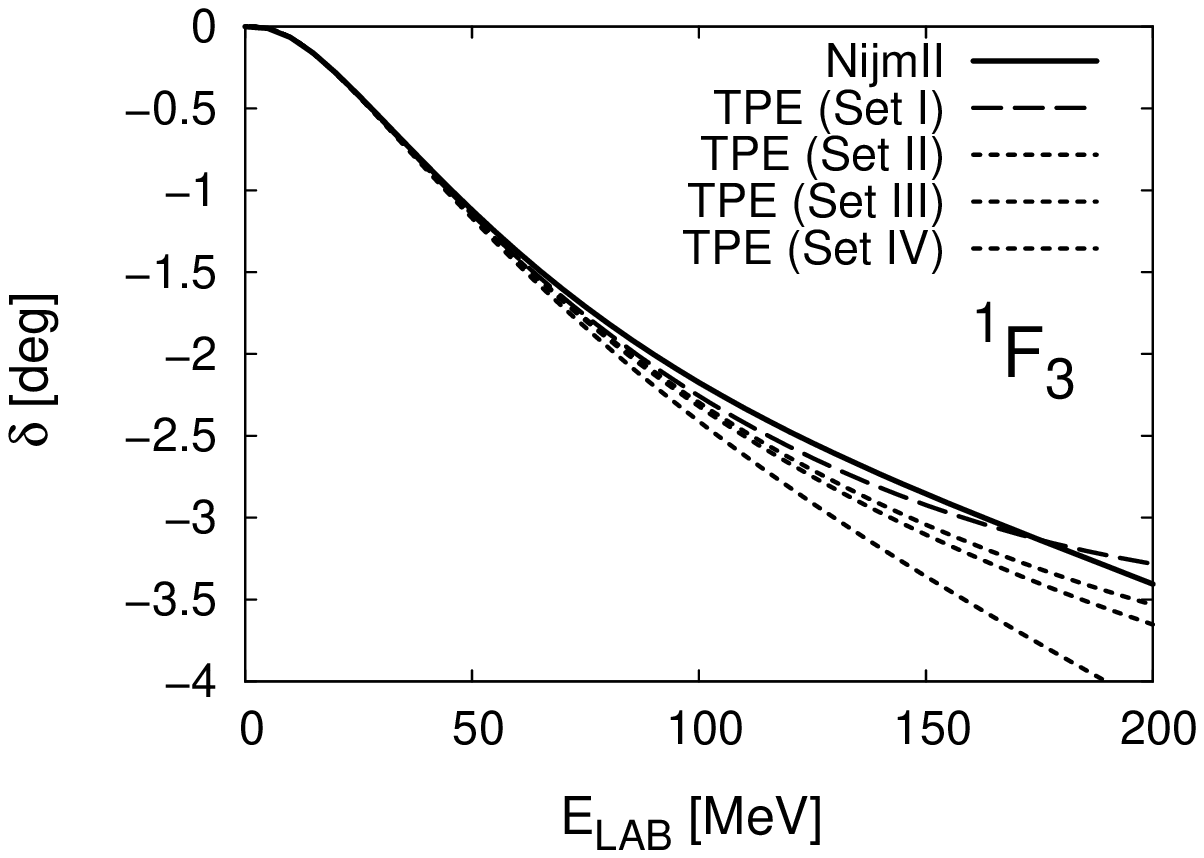,height=4cm,width=5cm}
\epsfig{figure=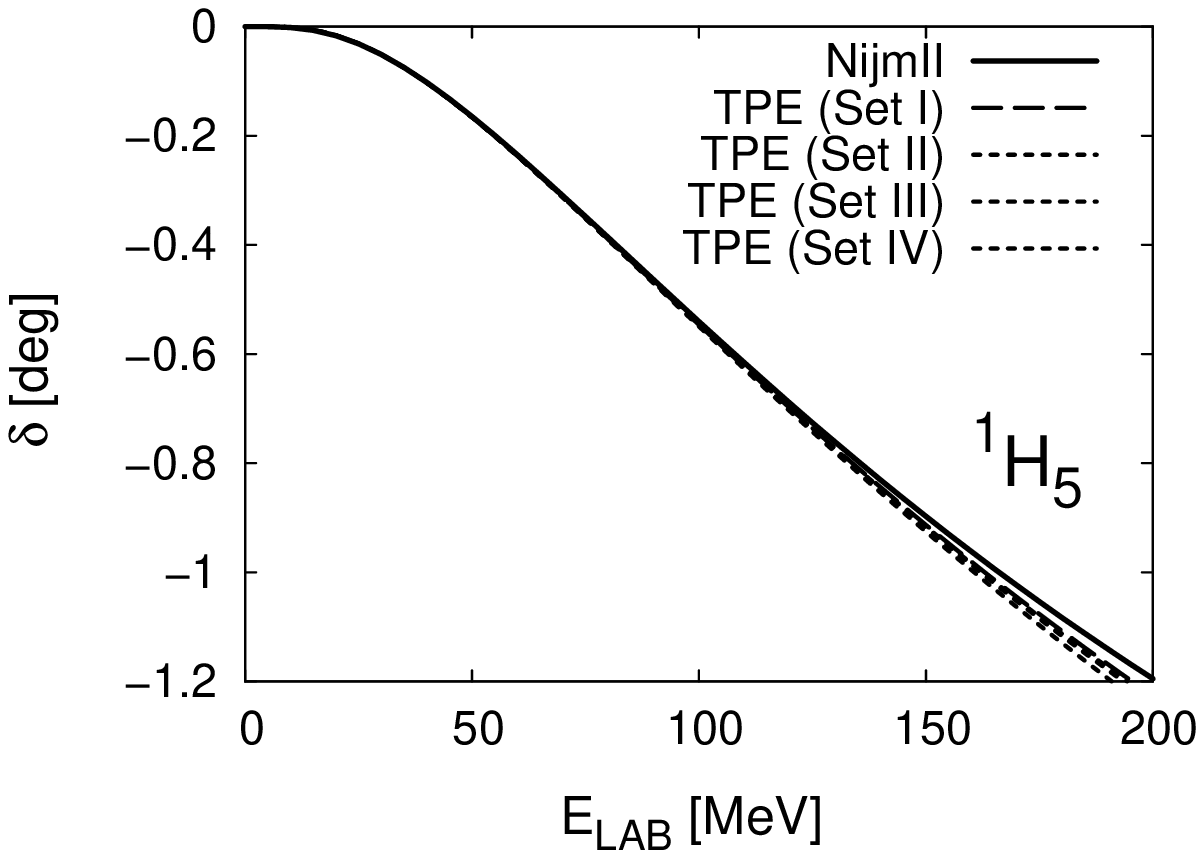,height=4cm,width=5cm} 
\end{center}
\caption{Dependence of the NNLO (SYM-nuclear bar) phases on the chiral
constants. These are the only channels where the potential is singular
repulsive.}
\label{fig:sing_rep}
\end{figure*}

As we have said, in singular repulsive channels, which at NNLO
correspond to the $^1P_1$, $^1F_3$ and $^1H_5$ singlet states, and to
the $^3P_2-{}^3F_2$ and $^3F_4-{}^3H_4$ triplet states, the phase
shift and also the scattering length are entirely determined by the
potential. So, these phases are a good place to study the influence of
different values for the chiral constants, $c_1$, $c_3$ and $c_4$,
presented in Table~\ref{tab:table1}. In Fig.~\ref{fig:sing_rep} we
show this dependence for these special partial waves. As we see, the
$^1P_1$ phase exhibits a strong dependence on the parameter set, while
$^1F_3$ and $^1H_5$ are less sensitive to this particular choice. The
strong dependence in the $^1P_1$ channel suggests that this may be an
ideal place to fit the chiral constants, since the scattering lengths
are fixed. We will not attempt such determination of the chiral
constants here because that would require realistic error estimates of
the phase shifts.

If we restrict to the spin singlet channels we see that there is very
good agreement for higher peripheral waves, $^1H_5$, $^1G_4$ and
$^1F_3$.  This is expected from perturbative calculations.  Note,
however, that unlike perturbation theory we fix by construction the
scattering lengths for the case of singular and attractive potentials.
Some intermediate waves, such as $^1D_2$, which potential is singular
attractive, are badly reproduced despite the fact that the threshold
behaviour is in theory reproduced since we use the corresponding
scattering length as input. Actually, for these waves the TPE result
seems to worsen the OPE prediction. Presumably this is an indication
either on the inadequacy of the (NijmII) scattering lengths used as
input for NNLO or on the importance of N$^3$LO contributions. Let us
note that the NijmII potential does not incorporate explicit TPE
effects in their long range part. In fact, if we take a slightly
different scattering length, $\alpha_2 = -1.666 {\rm fm}^5 $ instead
of the values deduced in Ref.~\cite{PavonValderrama:2004se},
$\alpha_2=-1.389 {\rm fm}^5$ for the NijmII potential a rather good
agreement with the Nijmegen analysis is obtained for the $^1D_2$ phase
shift (see similar results for $^3P_0$ and $^3P_1$ waves in
Fig.~\ref{fig:alphas}).  Although the small difference between the
fitted and experimental values for the scattering length could also be
explained by N$^3$LO corrections, suggesting that they are not large,
a definite conclusion cannot be drawn in the absence of a large scale
fit~\footnote{This also applies to the non-static OPE corrections
which account for about $0.1^o$ at $E_{\rm LAB}=200 {\rm MeV}$. The
effect can be mocked up by even tinier readjustments of both the
scattering lengths as well as the chiral couplings $c_1$, $c_3$ and
$c_4$ than deduced from inaccuracies in the NijmII potentials.}.

These general trends are confirmed in the triplet channels, where in
high partial waves there is an overall improvement when going from OPE
to TPE. In some cases, like in the $^3D_2$ , $\epsilon_2 $, $^3P_2$
and $^3F_2$ the improvement is rather satisfactory all over the energy
range. However, the theory has notorious problems in the $^3P_1$ and
$\epsilon_1$ and to a lesser extent in the $^3D_3 $ and $E_1$ channels
if one insists on keeping the scattering lengths of the NijmII
potential. As before, small changes in the scattering lengths allow
for an overall improved description as can be deduced from
Fig.~\ref{fig:alphas} in some particular cases (see also
Fig~\ref{fig:3C1}). This suggests that higher orders in the potential
may be needed. This fact was pointed out in our previous work on the
central phases, where the NNLO potential made {\it almost} the
effective range although there was a statistically significant
discrepancy to the experimental number, which called for the inclusion
of N$^3$LO terms. This may possibly happen also in some higher partial
waves and it would be interesting to see whether improved long
distance potentials might account for the observed discrepancies to
phase shifts provided the scattering lengths are kept to their
physical values.

As we have shown (see Fig.~\ref{fig:alphas}), small changes in the
scattering lengths indeed allow for a better description of the phases
in the intermediate energy region. On the other hand, we would expect
our description to become increasingly better for lower energies. This
situation is a bit disconcerting. Given the similarity between the
scattering lengths computed in Ref.~\cite{PavonValderrama:2004se} for
the NijmII and Reid93 potentials it seems unlikely that potential
models yield a completely off value for the $\alpha$'s in non-central
waves, but one must admit that errors will in general be larger than
the difference between these two potential values suggests as already
argued above.  If one takes into account the fact that both potentials
include similar long range physics, this means that the true error
could be larger due to systematic uncertainties in their short range
part (non OPE). 

Nevertheless, let us mention that current calculations involving
chiral potentials not only ignore this possible disagreement at
threshold but they in fact modify the corresponding scattering lengths
since the counterterms are determined by a fit to the phase shifts in
the region above threshold with no obvious control on the low energy
parameters (see e.g. Ref.~\cite{Nogga:2005hy}).  The arguments above
do not prove that taking slightly different scattering lengths than
those suggested by the high quality potentials is a legitimate
operation but at least shows that no more assumptions are made. From
this viewpoint it might be profitable to study the impact on those
calculations of either imposing exact threshold behaviour or
alternatively evaluating the threshold parameters themselves.

 \subsection{Remarks on the perturbative nature of peripheral waves}

The numerical coincidence of our non-perturbative calculations with
perturbation theory expectations~\cite{Kaiser:1997mw,Entem:2002sf},
although quite natural on physical grounds, deserves some explanation
on the basis of the formalism and the relevance of short distance
singularities.  Indeed, the attractive character of the singular NNLO
potentials at the origin implies a non-trivial boundary condition of
the form of Eq.~(\ref{eq:uA}), which cannot be reproduced to any given
order in perturbation theory, at least without the inclusion of extra
counterterms in the perturbative expansion, a point which will be
further discussed at the end of this section.  This point was
previously illustrated in Ref.~\cite{Beane:2000wh} for s-waves and
also in our previous work on the renormalization of the
OPE~\cite{PavonValderrama:2005gu} by comparing the exact deuteron wave
functions with the perturbative ones. There, one observes that the
first order perturbative calculation provides finite results, but the
expansion at second order produces divergent results due to the short
distance non-normalizable $D-$wave component. Thus, observables
cannot, strictly speaking, be analytical functions of the coupling
(for the purpose of discussion we could visualize the problem by
thinking of singularities of the sort $g^2 + g^4 \log g^2 $). This
does not mean that for the physical range of couplings the
non-analytical contribution is necessarily large numerically. For
instance, in the deuteron channel the residual non-analytical higher
order terms happens to be numerically sizeable even for a weakly bound
deuteron.

Based on the results of Ref.~\cite{PavonValderrama:2005gu}, there is
no reason to expect that higher partial waves will not exhibit this
failure of perturbation theory at some finite order. Nevertheless, the
perturbative short distance behaviour of higher partial waves tames
the singularity due to the kinematical $r^l$ suppression. This is a
perturbative long distance feature where the centrifugal barrier
dominates. The point is that this short distance behaviour is not
invariant order by order in strict perturbation theory for a singular
potential and, actually, one finds a short distance enhancement of the
wave function even in perturbation theory. So, one expects that the
perturbation theory on a singular potential will diverge at some
finite order also for high partial waves. In Appendix~\ref{sec:pert}
we show that this is indeed the case; for a singular potential
diverging like $1/r^n$ ($n > 2$) and a partial wave with angular
momentum $l$, the perturbative expansion diverges at $k-$th order in
perturbation theory provided $k > (2l+1)/(n-2) $. This estimate
provides the order at which, if desired, a long distance perturbation
theory on boundary conditions might be applied as discussed previously
for the deuteron channel~\cite{PavonValderrama:2005gu}. Using the
techniques developed in Ref.~\cite{Valderrama:2005wv} to make
perturbation theory on distorted OPE central waves it would be
interesting to see, as claimed by renormalization arguments on the
OPE~\cite{Birse:2005um}, whether such an expansion is indeed possible.

Having established that perturbation theory will diverge at some
finite order, we would like now to understand why it still can
accurately represent the full non-perturbative solutions obtained
numerically. The reason can be found in the very efficient way how the
short distance singularity of the potential makes short distances to
be inessential in the wave function for the regular non perturbative
solution. For high angular momenta and attractive singular potential
the wave function senses the singularity {\it after} tunneling through
the barrier, an exponentially suppressed effect. In perturbation
theory the effect is just substituted by the core provided by the
centrifugal barrier.

\section{Conclusions} 
\label{sec:concl}

In the present paper we have analyzed the renormalization of
non-central waves for NN scattering for the OPE and chiral TPE
potentials. This calculation extends our previous studies on central
phases and the deuteron for OPE and TPE potentials presented in
Refs.~\cite{PavonValderrama:2005gu,Valderrama:2005wv} respectively. As
already stressed in those works the requirement of finiteness of the
scattering amplitude as well as the orthogonality of wave functions
impose tight constraints on the allowed structure of counterterms for
a given potential. Using the standard Weinberg counting for the
potential, the counterterm structure is deduced and does not generally
coincide with the naive expectations. In some cases forbidden
counterterms in the Weinberg counting must be
allowed~\cite{PavonValderrama:2005gu,Nogga:2005hy} whereas in some
other cases allowed counterterms must be
excluded~\cite{Valderrama:2005wv}. Finite cut-off calculations based
on the Weinberg counting allow to introduce counterterms which are
usually readjusted to globally fit the data but are forbidden by
finiteness and orthogonality, in renormalized calculations.  The
success of the original counting relies heavily on keeping finite the
cut-off, while at the same time it is usually emphasized that low
energy physics does not depend crucially on short distance details. As
we have argued, these two facts are mutually contradicting; the
standard Weinberg counting is incompatible with exact renormalization,
i.e. removing the cut-off, as was suggested in
Ref.~\cite{Kaplan:1998tg} within a perturbative set up and shown in
Ref.~\cite{Nogga:2005hy} non-perturbatively, at least in the heavy
baryon expansion and when only nucleons and pions are taken into
account.  This feature changes when relativistic effects and $\Delta$
degrees of freedom are taken into account, showing that perhaps
renormalization, i.e. the independence on short distance details may
be a strong condition on admissible potentials. In this regard we find
that, as one would expect, the cut-off dependence is milder for the
chiral TPE potential than for OPE potential. This suggests that higher
order corrections become even more cut-off independent. Indeed, the
finite cut-off N$^3$LO calculations of Ref.~\cite{Epelbaum:2004fk} do
exhibit this feature in spite of the strong cut-off dependence
observed at lower orders.

Using this modified Weinberg counting, the quality of the agreement
and improvement depends on the particular partial wave. High partial
peripheral waves, when treated non-perturbatively reproduce the data
fairly well and deviations from OPE to TPE are small, as one would
expect in a perturbative treatment. Nevertheless, we have also shown
that regardless on the orbital angular momentum, there is always a
limit to the order in perturbation theory for which finite results are
obtained. The divergence is related to an indiscriminate use of the
perturbative expansion, and not to an intrinsic deficiency in the
definition of the scattering amplitude. Thus, also for peripheral
waves the phase-shifts are perturbatively non renormalizable while
they are non-perturbatively renormalizable. This result extends a similar 
observation for the
deuteron~\cite{PavonValderrama:2004nb,PavonValderrama:2005gu}. Nevertheless,
we have also argued why convergent perturbative calculations to finite
order are useful and may even provide accurate descriptions when
compared to the non-perturbative result.

Unlike naive expectations, it is not always true that after
renormalization the NNLO TPE phases improve over OPE ones if one {\it
insists} on keeping the scattering lengths required by finiteness to
the same physical values as those
extracted~\cite{PavonValderrama:2004se} from the high quality Nijmegen
potentials\cite{Stoks:1994wp}. This renormalization condition at zero
energy has been adopted to highlight the difference between these
potentials~\cite{Stoks:1994wp} and the chiral NNLO singular
potentials~\cite{Kaiser:1997mw}. Remarkably, using zero energy to fix
the parameters has never been considered before within the chiral
potentials approach to NN scattering, thus some of the problems we
find and discuss have not even been identified so far. Actually, we
find that some partial waves such as $^1D_2$ and $^3P_1$ are
particularly sensitive to the value of the scattering length. In fact,
it is found that small deviations of the scattering lengths at the few
percent level in these partial waves improve dramatically the
description in the intermediate energy region. The improvement can
also be achieved in other partial waves by suitably tuning the
scattering lengths in all the channels characterized by singular
attractive interactions. This means that the absolute error is small
up to $E_{\rm LAB} \sim 100 {\rm MeV }$. Three pion exchange effects
should become relevant at about CM momentum of $k= 3m /2 $ which
corresponds approximately to this LAB energy. The modification
corresponds to change the renormalization condition to some finite
energy, or maximizing the overlapp between the chiral phase shifts and
the fitted ones in a given energy window, very much along the lines
pursued in previous works. However, changing the scattering lengths
produces large relative errors near the threshold. At this point the
discussion on errors on the phase shifts becomes a crucial matter,
particularly in the low energy region.  In this regard, it seems
likely that the difference in low energy threshold parameters
determined in Ref.~\cite{PavonValderrama:2004se} for the Reid93 and
NijmII in all partial waves with $j \le 5 $ provides a lower bound for
the true error. Obviously, a meticulous error analysis of these
threshold parameters would be very helpful. 

We have also found that some partial waves, with repulsive singular
interactions and where no free scattering lengths are allowed, are
particularly sensitive to the choice of chiral constants $c_1$, $c_3$
and $c_4$. This suggests that a fit of the chiral constants to these
partial waves may be possible. To do so, and again, a realistic
estimate on the errors of the phase shifts would be
mandatory. According to our findings on the deuteron for the chiral
TPE potential~\cite{Valderrama:2005wv} it is quite likely that, if
such error estimate was reliably done, theoretical determinations for
deuteron observables with unprecedented precision based on chiral
potentials might be achieved. This issue is being currently under
consideration and is left for future research~\cite{Pavon2005}.

From a practical viewpoint there is a potential disadvantage in
requiring exact renormalization for the approximated long distance
chiral potentials, due to the tight constraints imposed by finiteness
on the short distance behaviour of the wave functions. To some extent,
although the chiral potentials are motivated by the Effective Field
Theory idea, these additional conditions remind also aspects of
renormalization of fundamental theories. This is not entirely
surprising since we expect the chirally based potentials to resemble
the {\it true} NN potential, at least at sufficiently long distances.
For instance OPE is a true long distance contribution. Full TPE would
also be a true long distance part, which is known
in an approximate manner within the current ChPT schemes based on
dimensional power counting.  Nevertheless, the essential difference is
that non perturbative dimensional transmutation, i.e. the generation
of dimension-full parameters not encoded in the potential, occurs due
to the singular and attractive nature of long distance interactions
already at the lowest order approximation consisting of OPE. This
non-perturbative renormalizability is the essential feature that makes
this problem particularly tough and so distinct from the previous
experience of perturbative renormalization on Effective Field Theories
or finite cut-off representations of the problem.

The present work not only shows that the theoretical requirement of
renormalizability can be implemented as a matter of principle and as a
practical way of controlling short distance ambiguities in the
predictions of Chiral Perturbation Theory for the study of NN
scattering, but also that interesting physical and phenomenological
insights are gathered from such an investigation. We have shown under
what conditions such a program can successfully be carried out as a
possible alternative and model independent way of describing the data
by using very indirect, but essential, information on the implications
of chiral symmetry for the NN problem below the pion production
threshold.

\begin{acknowledgments}

We have profited from lively and stimulating discussions with the
participants at ``Nuclear Forces and QCD: Never the Twain Shall Meet ?
`` at ECT$^*$ in Trento.  We would also like to thank Andreas Nogga
for pointing out an error in the $^1S_0$ phase shift plot.  This work
is supported in part by funds provided by the Spanish DGI and FEDER
funds with grant no. FIS2005-00810, Junta de Andaluc\'{\i}a grant
No. FQM-225, and EU RTN Contract CT2002-0311 (EURIDICE).

\end{acknowledgments}

\appendix

\section{Potentials} 
\label{sec:potentials} 

For completeness we list here the potentials found in
Ref.~\cite{Kaiser:1997mw}, and used in this paper. In coordinate space
the general form of the potential is written as
\begin{eqnarray} {\cal V}_{NN} &=& V_C (r) + \vec \tau_1 \cdot \vec \tau_2  
W_C (r) \nonumber \\ &+&
\big[V_{S} (r)+ \vec \tau_1 \cdot \vec \tau_2  W_{S} (r)\big]\,\vec\sigma_1
\cdot \vec \sigma_2 \nonumber \\ &+& \big[ V_T (r)+ \vec \tau_1 \cdot \vec \tau_2 W_T (r)
\big]\, \left( 3 \vec \sigma_1 \cdot \hat r \vec \sigma_2 \cdot \hat r - \vec\sigma_1
\cdot \vec \sigma_2  
 \right) \nonumber \\ &+&  \big[ V_{LS}(r) +\vec\tau_1 \cdot \vec \tau_2 W_{LS} (r)\big] \,\vec L \cdot \vec S  \, ,
\end{eqnarray} 
For states with good total angular momentum one obtains 
\begin{eqnarray}
U_{jj}^{0j} (r) &=& M \left[ (V_C - 3 V_S )+ \tau (W_C - 3 W_S )
\right] \, ,\\ 
U_{jj}^{1j} (r) &=& M \big[ (V_C + V_S - V_{LS})  \nonumber \\ 
&+& \tau (W_C + W_S- W_{LS}) + 2 (V_T + \tau W_T ) \big] \, , \nonumber \\  \\ 
U_{j-1,j-1}^{1j} &=&  M \big[ (V_C + \tau W_C + V_S + \tau W_S ) \nonumber\\ 
&+& (j-1) \left( V_{LS}+ \tau W_{LS} \right) \nonumber  
\\ &+&\frac{2(j-1)}{2j+1}  \left( V_T +\tau W_T \right) \big] \, ,  
\\ U_{j-1,j+1}^{1j} &=&  - \frac{6\sqrt{j(j+1)}}{2j+1} M 
\left( V_T +\tau W_T\right) \, ,
\\ U_{j+1,j+1}^{1j} &=& M \big[ (V_C + \tau W_C + V_S + \tau W_S ) \nonumber  
\\ &+& 2(j+2) \left( V_{LS}+ \tau W_{LS} \right) \nonumber 
\\ &+& \frac{2(j+2)}{2j+1}  \left( V_T +\tau W_T \right) \big] \, ,
\end{eqnarray} 
with $\tau= 2 T(T+1) -3 $. Remember that Fermi-Dirac statistics
requires $(-1)^{L+S+T}=-1$. 

The LO (OPE) potentials read ($x= m_\pi r $ )
\begin{eqnarray} 
W_S^{OPE} &=& \frac{g^2 m^3 }{ 48 \pi f^2 }\frac{e^{-x}}{x} \, ,\\ 
W_T^{OPE} &=& \frac{g^2 m^3 }{ 48 \pi f^2 } \frac{e^{-x}}{x}\left( 3 +
\frac{3}{x} + \frac1{x^2} \right) \, ,
\end{eqnarray} 
all others being zero.

The non-vanishing NNLO (TPE) potentials are given by 
\begin{widetext} 
\begin{eqnarray}
V_C^{TPE} (r) &=& \frac{3 g^2 m^6 }{32 \pi^2 f^4 }\frac{e^{-2 x}}{x^6} \Big\{ 
\left( 2 c_1 + \frac{3 g^2}{16 M} \right) x^2 (1+x)^2 + \frac{g^5 x^5}{32 M} 
+  \left(c_3 + \frac{3 g^2}{16 M} \right) \left( 6 + 12 x + 10 x^2 + 4 x^3 + x^4 \right) \Big\}  \nonumber \, , \\
W_T^{TPE} (r) &=& \frac{g^2 m^6 }{48 \pi^2 f^4 }\frac{e^{-2 x}}{x^6} \Big\{ -
\left( c_4 + \frac{1}{4 M} \right) (1+x) (3 + 3 x +x^2)   
+ \frac{ g^2}{32 M}  \left( 36 + 72 x + 52 x^2 + 17 x^3 + 2 x^4 \right) \Big\} \nonumber \, , \\  
V_T^{TPE} (r) &=& \frac{g^4 m^5 }{128 \pi^3 f^4 x^4 }  \Big\{ -12 K_0 ( 2x) - (15 + 4 x^2 ) K_1 (2 x) 
+  \frac{3 \pi m e^{-2x}}{8 M x }  \left( 12 x^{-1} + 24 + 20 x + 9 x^2 + 2 x^3  \right) \Big\} \nonumber \, , \\ 
W_C^{TPE} (r) &=& \frac{g^4 m^5 }{128 \pi^3 f^4 x^4 } \Big\{ \left[ 1 + 2
g^2 ( 5 + 2 x^2 ) - g^4 ( 23 + 12 x^2 ) \right] K_1 (2 x ) 
+  x \left[ 1+ 10 g^2 - g^4 ( 23 + 4 x^2 ) \right] K_0 ( 2 x)
\nonumber \, , \\ &+&  \frac{g^2 m \pi e^{-2x}}{4 M x} \left[ 2 ( 3g^2 - 2 )
\left( 6 x^{-1} + 12 + 10 x + 4 x^2 + x^3 \right) \right] + g^2 x \left( 2 + 4
x + 2 x^2 + 3 x^2 \right) \Big\} \nonumber \, ,\\ 
V_S^{TPE} (r) &=& \frac{g^4 m^5 }{32 \pi^3 f^4} \Big\{ 3 x K_0 (2 x) + ( 3 + 2 x^2 ) 
K_1 (2 x) - \frac{3 \pi m e^{-2x}}{16 M x} \left( 6 x^{-1} + 12 + 11 x + 6 x^2 + 2 x^3 \right) \Big\}  \nonumber \, ,  \\
W_S^{TPE} (r) &=& \frac{g^2 m^6 }{48 \pi^2 f^4 }\frac{e^{-2x}}{x^6} \Big\{ 
\left(c_4 + \frac1{4M} \right) (1+x) ( 3 + 3 x + 2 x^2 ) 
- \frac{g^2 }{16 M }\left( 18 + 36 x + 31 x^2 + 14 x^3 + 2 x^4 \right) \Big\} \nonumber \, , \\
V_{LS}^{TPE} (r) &=& - \frac{3 g^4 m^6 }{64 \pi^2 M f^4 } \frac{e^{-2 x}}{x^6} (1+x) \left( 2 + 2 x + x^2 \right) \nonumber \, , \\ 
W_{LS}^{TPE} (r) &=& \frac{g^2 ( g^2 -1) m^6 }{32 \pi^2 M f^4 } \frac{e^{-2x}}{x^6}(1+x)^2  \, ,
\end{eqnarray} 
\end{widetext} 
where $K_0$ and $K_1$ are modified Bessel functions. The NLO terms are
obtained by dropping all terms in $1/M$ and $c_1$, $c_3$ and $c_4$.

\section{The divergence of perturbation theory for peripheral waves} 
\label{sec:pert} 

In this appendix we show that for a singular, attractive or repulsive,
potential at the origin which diverges like $1/r^n $, there is always
a finite order in perturbation theory where the phase shift diverges,
regardless on the particular value of the angular momentum. Let us
consider for simplicity the single channel case. The radial equation
can be transformed into the integral equation
\begin{eqnarray} 
u_l (r) = \hat j_l (k r ) + \int_0^\infty G_{k,l} (r,r') U(r') u_l (r') dr' 
\, , \label{eq:int_eq}
\end{eqnarray} 
 where $G_{k,l}$ is the Green function given by
\begin{eqnarray}
k G_{k,l}(r,r') &=& \hat j_l (k r) \hat y_l (kr') \theta (r'-r) \nonumber
\\ &+& \hat j_l (k r') \hat y_l (kr) \theta (r-r') \, ,
\end{eqnarray} 
where $\theta (x)$ is the Heavyside step function, $\theta(x) = 1 $
for $x \ge 0 $ and $\theta(x)=0$ for $x < 0 $ and $\hat j_l (x) = x
j_l (x) $ and $\hat y_l (x) = x y_l (x) $ are the regular and singular
reduced spherical Bessel functions respectively.  To regularize the
lower limit of integration in Eq.~(\ref{eq:int_eq}) one may assume a
short distance regulator which will eventually be removed. The phase
shift is given by
\begin{eqnarray}
\tan \delta_l &=& - \frac1{k} \int_0^\infty \hat j_l (k r) U(r) u_l (r) \, .
\end{eqnarray} 
In perturbation theory by successive iteration of
Eq.~(\ref{eq:int_eq}) the Born series 
\begin{eqnarray}
\tan \delta_l &=& - \frac1{k} \int_0^\infty dr \left[\hat j_l (k
r)\right]^2 U(r) \nonumber \\ &-& \frac1{k} \int_0^\infty dr dr' \hat
j_l (k r) U(r) U(r') G_{k,l} (r,r') j_l (k r') + \dots  \nonumber \, ,\\
\end{eqnarray} 
is obtained. For our purposes of proving the divergence of
perturbation theory it is sufficient to analyze the low energy
limit. Using $\delta_l \to - \alpha_l k^{2l +1} $ and using known
properties of the Bessel functions 
\begin{eqnarray}
\hat j_l (x) \to \frac{x^{l+1}}{(2l+1)!!} \qquad 
\hat y_l (x) \to -\frac{(2l-1)!!}{x^{l}} \, .
\end{eqnarray} 
The Green's function becomes 
\begin{eqnarray}
-(2l+1) G_{0,l}(r,r') = \frac{r^{l+1}}{{r'}^l} \theta (r'-r) +
\frac{{r'}^{l+1}}{r^l} \theta (r-r') \, , \nonumber \\ 
\end{eqnarray} 
we get
\begin{eqnarray}
(2l+1)!!^2 \alpha_l &=& \int_0^\infty dr r^{2l+2} U(r) \nonumber \\
&+& \frac{2}{2l+1} \int_0^\infty dr r \int_0^r dr'  (r') ^{2l+2}
U(r) U(r') + \dots \nonumber \, .\\
\end{eqnarray} 
Since we only want to analyze the short distance behaviour we can
estimate the convergence of integrals by using the finite range and
singular potential $U(r) = (R/r)^n /R^2 \theta (a-r)$. Thus, we see
that in the first Born approximation the integral converges for $ 2 l
+ 1 > n-2 $, whereas the second Born approximation requires $2 l + 1 >
2 (n-2) $. This is obviously a more stringent condition. In general,
at $k-$th order convergence at the origin is determined by the
integral
\begin{eqnarray}
\int_0^\infty dr_1 r_1 U(r_1 ) \int_0^{r_1} dr_2 r_2 U(r_2) \dots
\int_0^{r_{k-1}} dr_k r_k^{2l+2} U(r_k) \, ,\nonumber \\ 
\end{eqnarray} 
which is finite only for $2 l +1 > k (n-2) $, a condition violated for sufficiently high $k$ when $n> 2$.  So, for $n > 2 $ there
will always occur a divergent contribution at a given finite order,
even if the Born approximation was finite due to a high value of the
angular momentum, $l$.

\section{Leading singularities in the Short distance expansion} 
\label{sec:short}

The determination of the short distance behaviour from the full
potentials is straightforward, but it is necessary to determine the
number of independent parameters in every channel and at any level of
approximation. For a quick reference we list the leading singularity
behaviour in Table~\ref{tab:table3}


\begin{table*}
\caption{\label{tab:table3} The leading short distance singularity of
the NN reduced potentials, $U= 2 \mu V $, Eq.~(\ref{eq:pot_chpt}) to
LO, NLO and NNLO for all channels considered in this work. The sign of
the coefficients for the one channel case (singlet and triplet) or the
eigenvalues for the triplet coupled channel case determines the number
of independent parameters. $ \bar c_3 = c_3 M $ and $ \bar c_4 = M c_4
$ are the dimensionless chiral constants.}
\begin{ruledtabular}
\begin{tabular}{|c|c|c|c|}
\hline 
Wave  & LO & NLO & NNLO \\ 
\hline 
\hline $^1S_0 $ &  $ - \frac{g^2 m^2 M}{16 \pi f^2} \frac{1}{r}  $ & 
$\frac{(1+10 g^2 -59 g^4)M}{256 \pi^3 f^4}\frac{1}{r^5}  $ & $  \frac{3 g^2 (-4+24 \bar c_3 - 8 \bar c_4+ 15 g^2)}{128 \pi^2
f^4 }\frac{1}{r^6}  $   \\
\hline $^3P_0 $ & $ - \frac{g^2  M }{4 \pi f^2}\frac{1}{r^3} $  & 
$ \frac{(1+10 g^2 +49 g^4)M}{256 \pi^3 f^4 }\frac{1}{r^5}  $  &  $ \frac{g^2 (12+72 \bar c_3 +40 \bar
c_4+g^2)}{128 \pi^2 f^4}\frac{1}{r^6} $   \\
\hline
\hline 
$^1P_1 $ & $ \frac{3 g^2 m^2 M }{16 \pi f^2}\frac{1}{r}  $ & $ \frac{3(-1-10 g^2 +11 g^4)M}{256 \pi^3 f^4 }\frac{1}{r^5}  $ & $   \frac{9 g^2 (4+8 \bar c_3 +8 \bar
c_4-3 g^2)}{128 \pi^2 f^4 }\frac{1}{r^6} $ \\
\hline $^3P_1 $ & $ \frac{g^2  M }{8 \pi f^2 }\frac{1}{r^3}  $  & $ \frac{(1+10 g^2 -41 g^4)M}{256 \pi^3 f^4 }\frac{1}{r^5} $  & $ \frac{g^2 (-2+36 \bar c_3 -4 \bar
c_4+ 19 g^2)}{64 \pi^2 f^4 }\frac{1}{r^6}  $   \\
\hline 
$^3S_1 $ & $ 0$   & 
$ \frac{3(-1-10 g^2 + 27 g^4) M}{256 \pi^3 f^4}\frac{1}{r^5}   $   &  $ 
-\frac{3 g^2 (-4-24 \bar c_3 + 8 \bar c_4 + 3 g^2 )}{128 \pi^2 f^4 }\frac{1}{r^6} $
\\
$^3D_1 $  & $ \frac{3 g^2}{8 f^2 \pi }\frac{1}{r^3} $ & 
$  \frac{3 (-1 -10 g^2 + 37 g^4)M }{256 \pi^3 f^4 }\frac{1}{r^5} $
 & $ \frac{9 g^2 (-1 +2 \bar c_3 - 2 \bar c_4 + 2 g^2 )}{32 \pi^2 f^4 }\frac{1}{r^6} $ 
 \\
$E_1 $   & $ -\frac{3g^2}{4 \sqrt{2} f^2 \pi}  \frac{1}{r^3}$    & $-\frac{15 g^4 M }{64 \sqrt{2} f^4 \pi^3 } \frac{1}{r^5} $  & $   \frac{- 3g^2 (-4 -16 \bar c_4 + 3 g^2 ) }{64 \sqrt{2} \pi^2 f^4 } \frac{1}{r^6} $   \\
\hline 
\hline 
$^1D_2 $  &  $ - \frac{g^2 m^2 M}{16 \pi f^2}\frac{1}{r}   $ & 
$\frac{(1+10 g^2 -59 g^4)M}{256 \pi^3 f^4}\frac{1}{r^5}  $ & $  \frac{3 g^2 (-4+24 \bar c_3 - 8 \bar c_4+ 15 g^2)}{128 \pi^2
f^4 }\frac{1}{r^6} $   \\ 
\hline $^3D_2 $ & $ - \frac{3 g^2 M}{8  \pi f^2}\frac{1}{r^3}   $  &  $\frac{(1+10 g^2 -89 g^4)M}{256 \pi^3 f^4}\frac{1}{r^5}  $  & $  \frac{ g^2 (-4+18 \bar c_3 - 10 \bar c_4+15 g^2)}{32 \pi^2
f^4 }\frac{1}{r^6} $ \\
\hline $^3P_2 $  & $ - \frac{g^2 M }{40 f^2 \pi } \frac{1}{r^3}$  & $ \frac{(1+10 g^2 - 5 g^4)M }{256 \pi^3 f^4 }\frac{1}{r^5} $& $ \frac{g^2 (-9 + 90 \bar c_3 + 14 \bar c_4 + 5 g^2 )}{160 \pi^2 f^4 }\frac{1}{r^6}$  \\
$^3F_2 $  & $-\frac{g^2 M}{10 \pi f^2 }\frac{1}{r^3}$ & $ \frac{(1+10 g^2 + 13 g^4 ) M }{256 \pi^3 f^4 }\frac{1}{r^5} $& $ \frac{g^2(76 + 360 \bar c_3 + 104 \bar c_4 + 175 g^2 )}{640 \pi^2 f^4 }\frac{1}{r^6}$  \\
$E_2 $  & $\frac{3\sqrt{3} }{20 \sqrt{2} \pi f^2 }\frac{1}{r^3}$ & $-\frac{9 \sqrt{3} g^4 M }{64 \sqrt{2} f^4 \pi^3 }\frac{1}{r^5}$  & $\frac{3 \sqrt{3}g^2 (-4 -16 \bar c_4 + 15 g^2)}{320 \sqrt{2} \pi^2 f^4 }\frac{1}{r^6} $  \\
\hline
\hline 
$^1F_3 $ & $ \frac{3 g^2 m^2 M }{16 \pi f^2 } \frac{1}{r}$  
& $ \frac{3(-1-10 g^2 + 11 g^4)M }{256 \pi^3 f^4 } \frac{1}{r^5}$ & $-\frac{9 g^2 (-4 - 8 \bar c_3 - 8 \bar c_4 + 3 g^2 )}{128 \pi^2 f^4 }\frac{1}{r^6}$  \\
\hline $^3F_3 $&  $ \frac{g^2  M }{8 \pi f^2 }\frac{1}{r^3}$  
& $ \frac{(1+10 g^2 - 41 g^4 )M}{256 \pi^3 f^4 }\frac{1}{r^5}$ & $\frac{g^2 (-2+36 \bar c_3 - 4\bar c_4 + 19 g^2) }{64 \pi^2 f^4 }\frac{1}{r^6}$  \\ 
\hline 
$^3D_3 $ & $ -\frac{g^2  M }{28 \pi f^2 }\frac{1}{r^3}$  
& $ \frac{(7+70 g^2 - 17 g^4 ) M}{1792 \pi^3 f^4 }\frac{1}{r^5}$ & $-\frac{g^2 (76 - 504 \bar c_3 - 88 \bar c_4 + 37 g^2 }{896 \pi^2 f^4 }\frac{1}{r^6}$  \\
$^3G_3 $  & $ -\frac{5 g^2  M }{56 \pi f^2 } \frac{1}{r^3}$  
& $ \frac{(7+70 g^2 +  73 g^4 ) M}{1792 \pi^3 f^4 }\frac{1}{r^5}$ & $ \frac{g^2 (66 +252\bar c_3 + 68 \bar c_4 + 155 g^2 )}{448 \pi^2 f^4 }\frac{1}{r^6} $  \\
$E_3 $  & $ \frac{3 \sqrt{3}  g^2  M }{28 \pi f^2 }\frac{1}{r^3}$  
& $ -\frac{45 \sqrt{3} g^4  M}{448 \pi^3 f^4 }\frac{1}{r^5}$ & $ \frac{3 \sqrt{3} g^2 (-4 -16\bar c_4 + 15 g^2 )}{448 \pi^2 f^4 }\frac{1}{r^6}$  \\
\hline 
\hline 
$^1G_4 $  & $ -\frac{ g^2 m^2 M }{16 \pi f^2}\frac{1}{r}  $ & $ \frac{(1+10 g^2 -59 g^4)M}{256 \pi^3 f^4 }\frac{1}{r^5}  $ & $   \frac{3 g^2 (-4+24 \bar c_3 -8 \bar
c_4+15 g^2)}{128 \pi^2 f^4 } \frac{1}{r^6}$ \\ 
\hline $^3G_4 $ & $ -\frac{3 g^2 M }{8 \pi f^2}\frac{1}{r^3}  $ & $ \frac{3(-1-10 g^2 + 17 g^4)M}{256 \pi^3 f^4 } \frac{1}{r^5} $ & $   \frac{3 g^2 (2+12   \bar c_3 +4  \bar
c_4+ g^2)}{64 \pi^2 f^4 } \frac{1}{r^6} $ \\ 
\hline 
$^3F_4 $ & $ \frac{3 g^2 M }{28 \pi f^2} \frac{1}{r^3} $ & $ \frac{3(-7-70 g^2 + 209 g^4)M}{1792 \pi^3 f^4 }\frac{1}{r^5}  $ & $  \frac{3 g^2 (76 +168 \bar c_3 -88  \bar
c_4- 127 g^2)}{896 \pi^2 f^4 }  \frac{1}{r^6}$\\
$^3H_4 $  & $ \frac{15 g^2 M }{56 \pi f^2}\frac{1}{r^3}  $ & $ \frac{3(-7-70 g^2 + 239 g^4)M}{1792 \pi^3 f^4 }\frac{1}{r^5}  $ & $  \frac{3 g^2 (-66 + 84 \bar c_3 -68 \bar
c_4+ 137 g^2)}{448 \pi^2 f^4 } \frac{1}{r^6}$ \\
$E_4 $ & $ -\frac{9 \sqrt{3} g^2 M }{28 \pi f^2}\frac{1}{r^3}  $ & $ -\frac{45 \sqrt{3} g^4 M}{448 \pi^3 f^4 }\frac{1}{r^5}  $ & $  -\frac{9 \sqrt{3} g^2 (-4 - 16 \bar c_4 + 3 g^2)}{448 \pi^2 f^4 } \frac{1}{r^6}$ \\
\hline
\hline 
$^1H_5 $  & $\frac{3 g^2 m^2 M }{16  \pi f2 }\frac{1}{r}$ & $\frac{3(-1-10 g^2 + 11 g^4 )M}{256 \pi^3 f^4 }\frac{1}{r^5} $ & $\frac{9 g^2 ( 4 + 8 \bar c_3 + 8 \bar c_4 - 3 g^2 )}{128 \pi^2 f^4 }\frac{1}{r^6}$ \\
\hline 
$^3H_5 $  & $\frac{g^2 M }{8 \pi f^2}\frac{1}{r^3}$ & $\frac{1+10g^2 - 41 g^4 )M}{256 \pi^3 f^4 }\frac{1}{r^5}$ & $\frac{g^2(-2 + 36 \bar c_3 - 4 \bar c_4 + 19 g^2 )}{64 \pi^2 f^4 }\frac{1}{r^6}$ \\
\hline 
$^3G_5 $  & $\frac{3 g^2 M }{22 \pi f^2 }\frac{1}{r^3}$ & 
$\frac{3(-11-110 g^2 + 337 g^4 )M}{2816 \pi^3 f^4 }\frac{1}{r^5}$  & $\frac{3 g^2 (204 + 264 \bar c_3 - 152 \bar c_4 - 373 g^2 )}{1408 \pi^2 f^4 } \frac{1}{r^6}$   \\
$^3I_5 $  & $\frac{21 g^2 M }{88 \pi f^2} \frac{1}{r}$ & $\frac{3(-11-110 g^2 + 367 g^4 )M}{2816 \pi^3 f^4 } \frac{1}{r^5}$  &  $ \frac{3 g^2 ( - 73 + 66 \bar c_3 - 50 \bar c_4 + 151 g^2 )}{352 \pi^2 f^4 } \frac{1}{r^6}$ \\
$E_5 $  & $- \frac{9 \sqrt{15}g^2 M }{44  \sqrt{2} \pi f^2 } \frac{1}{r} $  & 
-$\frac{45 \sqrt{15}g^4 M  }{704 \sqrt{2} \pi^3 f^4 } \frac{1}{r^5}$
 &  $-\frac{9 \sqrt{15}g^2 (- 4 - 16 \bar c_4 + 3 g^2 )}{704 \pi^2 f^4} \frac{1}{r^5}$
\end{tabular}
\end{ruledtabular}
\end{table*}


\end{document}